\documentclass{aa}  
\usepackage[varg]{txfonts}
\usepackage{graphicx}
\usepackage{siunitx}
\usepackage{xcolor}
\usepackage{natbib}
\usepackage{booktabs}
\usepackage{boldline}
\usepackage{nccmath}
\usepackage{amsmath}
\usepackage{tabularx}
\usepackage{algorithm2e}
\usepackage{csquotes}
\usepackage{comment}
\usepackage{longtable}
\usepackage{tabu}
\usepackage{gensymb}
\usepackage{enumitem}
\usepackage{float}
\usepackage{subcaption}
\usepackage[labelfont=bf]{caption}
\usepackage{mwe}
\usepackage[colorlinks,citecolor=blue,urlcolor=blue,bookmarks=false,allcolors=blue,hypertexnames=true]{hyperref} 
\usepackage{threeparttable}
\usepackage{color}
\usepackage[colorinlistoftodos]{todonotes}
\usepackage{tabularx}
\usepackage{adjustbox}
\usepackage{blindtext, rotating}
\usepackage{array,booktabs}

\newcommand\clearrow{\global\let\rowmac\relax}
\makeatletter

\DeclareRobustCommand{\OI}{%
  \mbox[{O\check@mathfonts\fontsize\sf@size\z@\selectfont I}]~%
}
\DeclareRobustCommand{\OII}{%
  \mbox[{O\check@mathfonts\fontsize\sf@size\z@\selectfont II}]~%
}
\DeclareRobustCommand{\OIII}{%
  \mbox[{O\check@mathfonts\fontsize\sf@size\z@\selectfont III}]~%
}
\makeatother
\def \kms         {\,km$\,$s$^\mathrm{-1}$}

\def \arcsec      {\text{$^\mathrm{\prime\prime}$}}

\def \whz         {\,W\,Hz$^{-1}$}

\def \ergpers {\,$\mathrm{erg s^{-1}}$}

\defcitealias{Gereb2014}{G14}
\defcitealias{Gereb2015}{G15}
\defcitealias{Maccagni2017a}{M17}

\newcommand{\Hi}{\text{H\textsc{i}}}
\newcommand{\hbeta}{\text{H\textsc{$\beta$}}}
\newcommand{\halpha}{\text{H\textsc{$\alpha$}}}
\newcommand{\hgamma}{\text{H\textsc{$\gamma$}}}
\newcommand{\hdelta}{\text{H\textsc{$\delta$}}}
\newcommand{\oiiihbeta}{\text{[O\textsc{iii}]}/\text{H\textsc{$\beta$}}}

\newcommand{\Nii}{\text{[N\textsc{ii}]}}
\newcommand{\Sii}{\text{[S\textsc{ii}]}}

\newlist{inlineroman}{enumerate*}{1}
\setlist[inlineroman]{itemjoin*={{ }},afterlabel=~,label=\roman*)}

\makeatletter
\renewcommand{\fnum@figure}{Figure \thefigure}
\makeatother

\begin{document} 
\title{Connecting the radio AGN life cycle to feedback}
\subtitle{Ionised gas is more disturbed in young radio AGN}
\author{Pranav Kukreti\inst{1,2,3}\thanks{\email{kukreti@uni-heidelberg.de}} and Raffaella Morganti\inst{2,3}}

\authorrunning{P. Kukreti and R. Morganti}

\institute{Astronomisches Rechen-Institut, Zentrum für Astronomie der Universität Heidelberg, Mönchhofstr. 12-14, 69120 Heidelberg,\\
Germany\and
ASTRON, the Netherlands Institute for Radio Astronomy, Oude Hoogeveensedijk 4, 7991 PD Dwingeloo, The Netherlands
\and
Kapteyn Astronomical Institute, University of Groningen, Postbus 800, 9700 AV Groningen, The Netherlands
}

  \abstract{In the host galaxies of radio active galactic nuclei (AGN), kinematically disturbed gas due to jet-driven feedback is a widely observed phenomenon. Simulations predict that the impact of jets on the surrounding gas changes as they grow. Useful insights into this phenomenon can be obtained by characterising radio AGN into different evolutionary stages and studying their impact on gas kinematics. We present a systematic study of the \OIII gas kinematics for a sample of 5\,720 radio AGN up to $z$\,$\sim$\,0.8 with a large 1.4\,GHz luminosity range of $\approx10^{22.5}-10^{28}$\,\whz, and 1\,693 \OIII detections. Our careful separation of radio emission from AGN and star formation allows us to isolate the impact of radio jets. Taking advantage of the wide frequency coverage of LOFAR and VLA surveys from $144-3000$\,MHz, we determine the radio spectral shapes, using them to characterise sources into different stages of the radio AGN life cycle. We determine the \OIII kinematics from SDSS spectra and link it to the life cycle. Our main conclusion is that the \OIII gas is $\sim$\,3 times more likely to be disturbed in the peaked spectrum (PS) sources (that represent a young phase of activity) than non-peaked spectrum (NPS) sources (that represent more evolved sources) at $z<0.4$. This changes to a factor of $\sim$\,2 at $z>0.4$. This shows that on average, the strong impact of jets is limited to the initial stages of the radio AGN life cycle. At later stages, the impact on gas is more gentle. We also determine the dependence of this trend on 1.4\,GHz and \OIII luminosities, and find that the difference between the two groups increases with 1.4\,GHz luminosity. Young radio AGN with $L_\mathrm{1.4GHz}>10^{25}$\whz have the most extreme impact on \OIII. Using a stacking analysis, we are further able to trace the changing impact on \OIII in the high frequency peaked spectrum (i.e. youngest), low frequency peaked spectrum (\enquote{less young}), and non-peaked spectrum (evolved) radio AGN.}

    \date{Received ** ** 2024  / Accepted ** ** 2024}    

    \maketitle
\section{Introduction}
\label{introduction}

Active galactic nuclei (AGN) are now understood to be linked to the evolution of massive galaxies in the Universe. This link manifests in the form of feedback that can quench star formation, releasing energy that can prevent cooling of the gas in hot halos, and even regulate the growth of the central supermassive black hole itself (for example \citealt{Silk1998,Fabian2012,McNamara2012}).
The mechanical energy released during a phase of activity can shock the host galaxy's gas, sometimes even sweeping it up in high-velocity outflows. This mechanical energy can be propagated by accretion-driven winds, radiation pressure, collimated jets of plasma, etc (see \citealt{Fabian2012, King2015, Harrison2024ObservationalInterpretation} for a review). Observational evidence of this form of feedback has been found in both radiatively efficient and inefficient AGN host galaxies. \par

Feedback driven by collimated jets has been widely observed in radio AGN host galaxies. These are systems where radio emission is dominated by the AGN, and cannot be entirely attributed to star formation. Although their energetic output is often dominated by mechanical energy in relativistic plasma jets, they can also release high levels of radiative energy. Over the years, many observational studies have established the presence of jet-driven feedback in radio AGN host galaxies (e.g. \citealt{Morganti2021a,Murthy2022,Nesvadba2021,Ruffa2022,Schulz2021} for some recent results). A limited number of statistical studies have confirmed a positive correlation between the presence of radio emission and disturbed ionised gas kinematics, in particular \citet{Mullaney2013}. An interesting aspect of this study is that their AGN sample was selected using optical emission line ratios, and these sources don't necessarily harbour radio jets. But it highlights the broader role that jets could play in feedback. Recently, similar results have also been found for quasars and broad-line AGN by \citealt{Escott2024UnveilingEmission}, who obtained a larger fraction of sources with outflows in radio-detected than non-radio-detected sources. However, it is worth noting that some studies have also found no significant relation between disturbed ionised gas and radio luminosity or morphology (for example \citealt{Woo2016,Ayubinia2023InvestigatingAGNs}).\par

Studies of radio AGN samples have also pointed towards a positive correlation between young radio AGN and more disturbed gas kinematics (ionised and neutral phase; e.g. \citealt{Holt2008,Gereb2015,Roche2016,Molyneux2019,Murthy2019,Kukreti2023}), proposing a picture where the impact of jet-driven feedback evolves throughout the active phase. This picture is broadly supported by simulations of jet-ISM interaction, which find a more extreme impact on the clumpy gas medium at early stages of jet growth \citep{Sutherland2007,Mukherjee2018}.\par

For radio AGN, observational evidence for the link between different stages of jet growth (young or evolved sources) and feedback, is still limited to powerful systems ($>10^{26}$\,\whz) or small samples. Although some trends have been found between radio emission and ionised gas kinematics for large samples (e.g. \citealt{Mullaney2013,Molyneux2019}, they have been focused on AGN selected using optical emission line ratios, and are not necessarily radio AGN. To understand the relation between jet growth and feedback using radio properties, selecting sources where the radio emission is dominated by jets is crucial. Our approach is to select such a sample, split it into different stages of life cycle and link it to its ionised gas kinematics.\par

The radio AGN life cycle has been established over the years using their spectral and morphological properties. In early stages of growth, young radio AGN are found to have a peaked radio spectrum and morphologically compact radio emission (for example \citealt{Callingham2017}, and see  \citealt{ODea1998} and \citealt{ODea2021a} for reviews). These properties have been observed in the compact steep spectrum (CSS) and gigahertz peaked spectrum (GPS) sources, and the period of this phase is found to be $\sim$\,$0.001-1$\,Myr from studies based on radio spectral modelling  \citep{Murgia1999SynchrotronSources,Murgia2003} and hot spot separation velocities \citep{Polatidis2002ProperObjects,Giroletti2003Ages31.04,Giroletti2009}. The peaked radio spectrum is either due to synchrotron self-absorption \citep{Tingay2003An1718-649,Snellen2000,DeVries2009FurtherAGN,Artyukh2008AstrophysicsB0108+388} or free-free absorption from an ionised medium \citep{Kameno2000,Kameno2005,Callingham2015a,Mhaskey2019GMRTSky,Keim2019ExtragalacticScales}. At later stages of the life cycle (a few tens of Myr), the radio spectrum is observed to be optically thin, due to the peak shifting to lower frequencies and eventually below the observing frequencies of most radio telescopes.\par

A single phase of activity can last up to $\sim$\,$10-100$\,Myr (see \citealt{Morganti2017b} for a review). After the activity stops and there is no more injection of fresh electrons, radiative losses dominate the emission leading to steep spectra and low surface brightness diffuse emission, called the remnant stage of the life cycle \citep{Parma2007a,Saripalli2012b,Brienza2017}. In some cases, remnant phase emission is found along with emission from a young phase at the centre, suggesting that the new phase of activity has started even before the radio plasma from the older phase has faded below detection limits (13-15\%; \citealt{Jurlin2020a}). These are called restarted radio AGN (for example, \citealt{Brienza2018,Morganti2021c,Morganti2021b,Kukreti2022a}). Therefore, the radio spectral properties change over the life cycle and can be used to characterise a sample of radio AGN into different stages of evolution. \par

This approach was used in \citet{Kukreti2023}, hereafter K23, where we already found some evidence for more disturbed ionised gas in young radio AGN than their evolved counterparts. We characterised the evolutionary stage using their radio spectral shape, and their ionised gas kinematics using the \OIII spectra. The main result of K23 was that on average, peaked radio spectrum sources had a broader \OIII profile than sources without a peak. Since peaked sources represent a younger phase of activity, as discussed above, this suggested a link between the life cycle and feedback on ionised gas.\par

Although interesting, this study only used a small sample of $\sim$\,130 sources up to $z\sim0.2$ covering a radio luminosity range of $\sim$\,$10^{23}-10^{25}$\,\whz. The small sample size also made it impossible to control for important AGN parameters like radio and optical luminosity, source sizes, etc. The purpose of the present paper is to investigate the link between radio AGN life cycle and feedback for a sample of 5\,720 radio AGN up to $z\sim0.8$ that cover a radio luminosity range from $\sim$\,$10^{22.5}-10^{28}$\,\whz. The large size and range of luminosities allow us to control for source properties and disentangle their role. \par  

The paper is structured in the following manner: Section 2 describes the approach for sample construction, Section 3 describes the modelling of the stellar continuum and emission line profiles of the sources, Section 4 describes the diagnostics used to select the radio AGN sample and Section 5 describes the radio properties of this radio AGN sample. We then present the results for different source groups in Section 6 and discuss them in Section 7. Throughout the paper, we have used the $\Lambda$CDM cosmological model, with H$_\mathrm{0}$ = 70 km s$^\mathrm{-1}$ Mpc$^\mathrm{-1}$, $\Omega_\mathrm{M}=0.3$ and $\Omega_\mathrm{vac}=0.7$.

\section{Sample construction}
\label{sample contruction}
This section outlines the approach used to construct the radio AGN sample. The sample was constructed using the Faint Images of the Radio Sky at Twenty-cm survey at 1400\,MHz with $\sim$\,5.4\arcsec~resolution (FIRST; \citealt{Becker1995TheCentimeters}), the LOFAR Two-metre Sky Survey at 144\,MHz with 6\arcsec~resolution (LoTSS; \citealt{Shimwell2017,Shimwell2022}), the Very Large Array Sky Survey at 3000\,MHz with $\sim$\,$2.5-3$\arcsec~resolution (VLASS; \citealt{Lacy2016,Lacy2020,Gordon2021}) and the Sloan Digital Sky Survey (SDSS) which provides spectra with a 3\arcsec~size fibre. These surveys were chosen as they provide radio and optical data over similar resolutions, allowing comparison of these properties over similar physical scales.

The parent sample was the \texttt{2014 December 17} version of the FIRST survey catalogue \citep{Helfand2015} which was cross-matched with the SDSS DR17 spectroscopic catalogue \citep{Abdurrouf2022} using a search radius of 3\arcsec (approximately half the FIRST beam size). The cross-matching was done down to a peak flux density of 5\,mJy in FIRST, which corresponds to $L_\mathrm{1.4GHz}\approx5\times10^{23}$\whz at $z=0.2$. This provided a sample of 29\,921 sources.\par

We then restricted the sample to $z=0.8$, after which the \OIII$\mathrm{\lambdaup\lambdaup}$4958,5007\AA~doublet moves outside the SDSS wavelength coverage. We removed sources below $z=0.02$, since below this redshift the SDSS fibres probe too small a region in a galaxy to determine reliable parameters. To obtain accurate redshifts, we only selected sources with a fractional error in redshift less than 0.01. This narrowed the sample down to 17\,405 sources.\par

Next, we cross-matched this sample with the LoTSS DR2 \citep{Shimwell2022} and VLASS Quick Look epoch 2\footnote{\url{https://cirada.ca/vlasscatalogueql0}} \citep{Lacy2016} catalogues. We used the recommendations provided in the VLASS catalogue user guide available online, to select sources with reliable detections. Cross-matching was done using TOPCAT \citep{Taylor2005}, with a radius of 6\arcsec, which is comparable to the FIRST and LoTSS resolution. This gave us a sample of 6\,556 sources, all of which have a LoTSS cross-match and out of which 6\,400 have a VLASS cross-match. 

 \begin{figure*}
 \centering
    \begin{subfigure}{0.6\columnwidth}
      \includegraphics[width=\columnwidth]{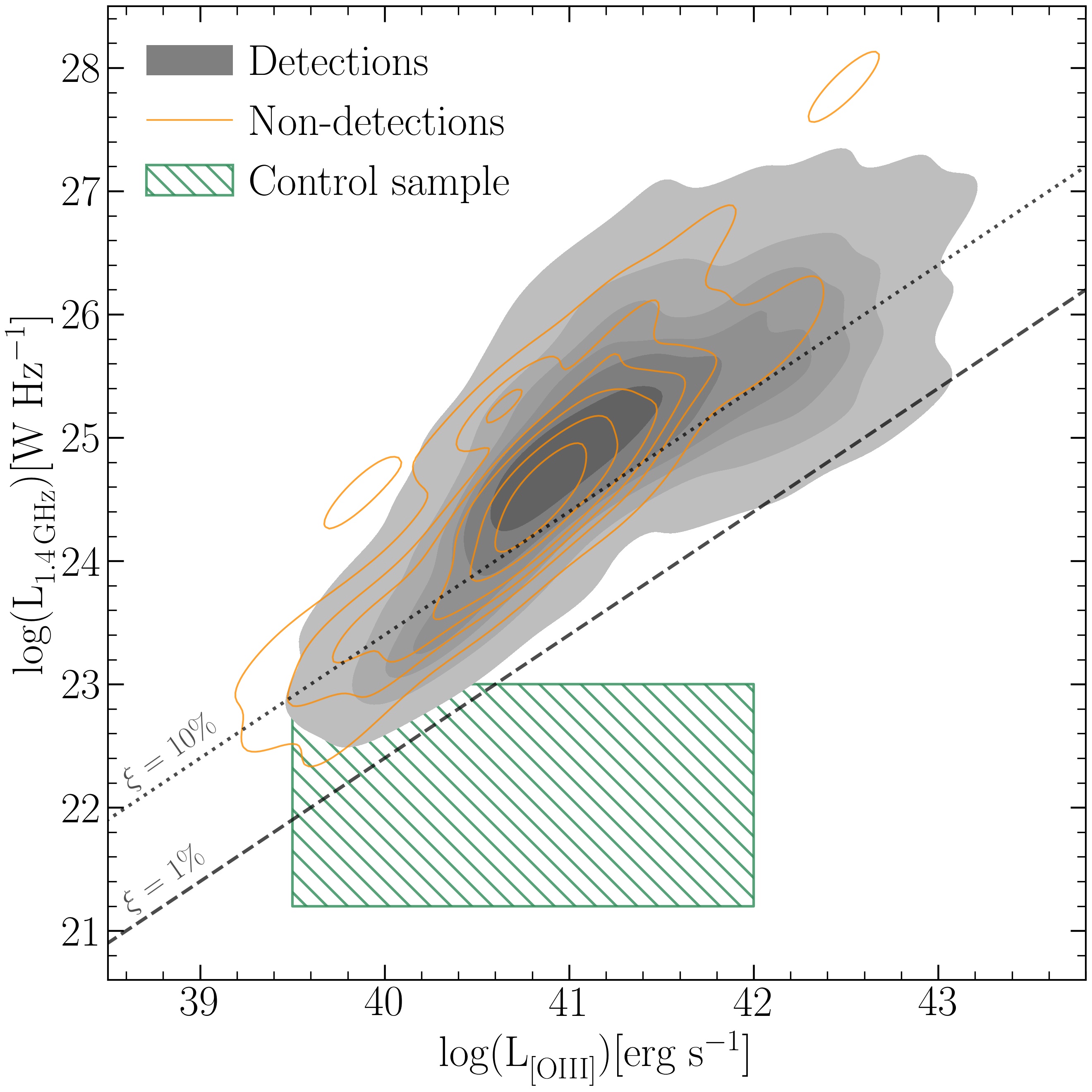}
      \caption{$L_\mathrm{1.4GHz}$ vs $L_{\textrm{\OIII}}$ for radio AGN}
       \label{radvsoptlum}   
    \end{subfigure}
    \hfill
    \begin{subfigure}{1.4\columnwidth}
      \includegraphics[width=1\columnwidth]{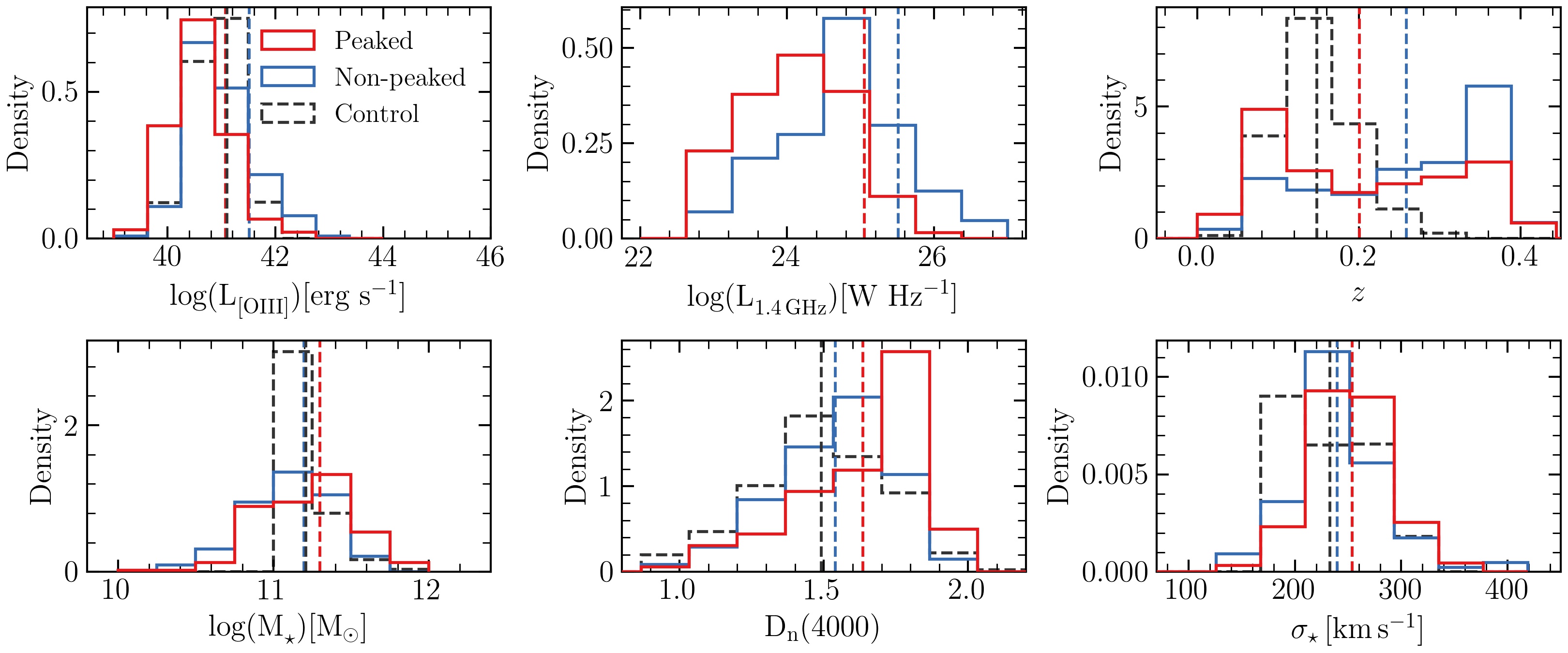}
      \caption{Properties of peaked and non-peaked sources in low redshift sample ($z<0.4$).}
      \label{sourceproperties}
    \end{subfigure}    
    \caption{Properties of the radio AGN sample. \textbf{(a)} $L_\mathrm{1.4GHz}$ vs $L_{\textrm{\OIII}}$ for the full radio AGN sample, with \OIII detections and non-detections shown separately. In the case of \OIII non-detections, the contour plot shows the upper limits for $L_{\textrm{\OIII}}$. The green hatched region marks the luminosity space occupied by the control sample selected from \citet{Mullaney2013}, described in Sect.~\ref{selectragn}. The dashed and dotted lines show the radio luminosity expected from the interaction of AGN-driven winds with the ISM from the model of \citet{Nims2015ObservationalNuclei}, for 1\% and 10\% of the shock energy going into relativistic electrons. \textbf{(b)} Distributions of host galaxy properties for the peaked and non-peaked radio AGN in the low redshift sample. Only the \OIII detections are shown. Vertical lines show the mean values of the distributions. The same plot for the high redshift sample is shown in Fig.~\ref{sampleproperties_highz}}
    \label{sample_properties_distribution}
 \end{figure*}
 
One of the diagnostics used for selecting radio AGN in this paper involves the Wide-field Infrared Survey Explorer (WISE) mid-infrared colours, which trace dust heated by AGN and stars. We added WISE data using the allWISE IPAC release from November 2013 \citep{Cutri2021}. We used three bands of the WISE catalogue - W1 at 3.4$\mu$m, W2 at 4.6$\mu$m and W3 at 12$\mu$m, that have an angular resolution of $6.1-6.5$\arcsec. We applied a signal-to-noise ratio (S/N) cut of 5 for bands W1 and W2, and 3 for band W3 due to its poorer sensitivity. However, we also kept sources with S/N less than 3 in W3 and used them to determine upper limits. We selected only sources with \texttt{cc\_flag=000} as suggested in the online user manual. To avoid performing $k$-corrections, we only cross-matched the 2\,310 sources up to $z=0.3$, and obtained a cross-match for 2\,308 sources.\par

Radio luminosity at 1.4\,GHz is commonly used in literature for radio AGN. To allow for comparison with other studies, we also estimated the total 1.4\,GHz luminosity of our sources. The low sensitivity of FIRST to extended emission means that it could underestimate the total luminosity. The $\sim$\,$5.4\arcsec$ resolution also means that a large source can be split up into several individual components, further underestimating the total luminosity. We therefore used the combined radio catalogue from \citet{Mingo2016}, who performed a detailed and accurate cross-match between FIRST and the NRAO VLA Sky Survey (NVSS; \citealt{Condon1998}) sources at 1.4\,GHz. The lower resolution ($\sim$\,45\arcsec) of NVSS makes it more reliable for estimating total radio luminosities. We used the combined flux densities from this catalogue to estimate the $k$ corrected 1.4\,GHz luminosities.\par

We aim to use the \OIII$\mathrm{\lambdaup\lambdaup}$4958,5007\AA~doublet to trace the ionised gas kinematics, and other emission line ratios to measure the ionisation properties. For this, we used the SDSS spectra. The details of the stellar continuum and emission line modelling, \OIII profile characterisation, and stacking analysis are described in Sect.~\ref{modelling optical spectra}. We remove 52 sources from this sample where either the spectra are corrupted at \OIII wavelength or an accurate characterisation of the \OIII profile is not possible. This gives a final sample of 6\,504 sources. A summary of the number of sources with different emission line detections is shown in Table.~\ref{emissionlinesummary}. \par

\begin{table}
\centering
\caption{Emission line detections summary}
         \label{emissionlinesummary}
\renewcommand{\arraystretch}{1.15}
\setlength{\tabcolsep}{4pt}
\begin{tabular}{cccccccc}
    \hlineB{3}
    \noalign{\vspace{0.05cm}}
    \hline
    \noalign{\smallskip}
      & Total & \OIII  & \hbeta  & \halpha & \Nii & \Sii  \\
    \noalign{\smallskip}
    \hlineB{3}
    \noalign{\smallskip}
    \noalign{\smallskip}
  Full sample & 6504 & 2333 & 1469 & 1574 & 1574 & 1111\\
  Radio AGN & 5720 & 1693 & 879 & 973 & 973 & 566\\
    \noalign{\smallskip}
    \hline
    \noalign{\vspace{0.05cm}}  
    \hlineB{3}
 \end{tabular}
    \flushleft
    \textbf{Note.} Summary of emission line detections in the full sample and selected radio AGN sample, discussed in Sect.~\ref{modelling optical spectra} and \ref{selectragn}.    
\end{table}

Finally, we identify 5\,720 radio AGN from this final sample, using a combination of diagnostics described in \citet{Best2005,Best2012} and \citet{Sabater2019}. These diagnostics help us construct a clean sample of radio AGN, where the radio emission is dominated by the AGN and not star formation. The procedure is described in Sect.~\ref{selectragn} in the appendix. Our radio AGN sample spans a wide range in total 1.4\,GHz luminosity from $\sim$\,$10^{22.5}-10^{27}$\,\whz, and has 1\,693 \OIII detections with an observed \OIII luminosity range of $\sim$\,$10^{39}-10^{44}$\,erg\,s$^{-1}$. This is shown in Fig.~\ref{radvsoptlum}. The host galaxies have stellar masses from $\sim$\,$10^{10}$ to $10^{12.4}$\,M$_{\odot}$. The host galaxies also span a wide range in $g-r$ colour, containing both red ($g-r>0.7$) and blue ($g-r<0.7$) objects, with a median $g-r=1.5$.\par

Radio emission in the so-called radio-loud AGN is attributed to jets, however, the origin is not completely clear for radio-quiet AGN. For these sources, radio emission can have multiple origins: star formation, innermost accretion disk coronal activity, low power jets and shocks due to interaction between AGN-driven winds and the ISM \citep{Panessa2019a,Zakamska2014,Zakamska2016}. Since we aim to study jet-driven feedback down to $L_\mathrm{1.4GHz}\sim10^{23}$\whz, it is important to confirm that the radio emission in our sample is dominated by jets even at low luminosities. We have selected our sources to have radio emission significantly more than expected from star formation, using the diagnostics mentioned before. Further, most of our sources also have radio luminosities $>$\,$10^{23}$\whz, which is more than expected from the corona \citep{Raginski2016AGNEmission}. In Fig.~\ref{radvsoptlum} we also plot the expected radio luminosity from AGN wind shocks, using the fiducial models of \citet{Nims2015ObservationalNuclei}. An overwhelming majority of the sources lie much above this model, confirming that the radio emission is dominated by jets in our sample.\par

\begin{figure}
\centering
  \includegraphics[width=0.55\columnwidth]{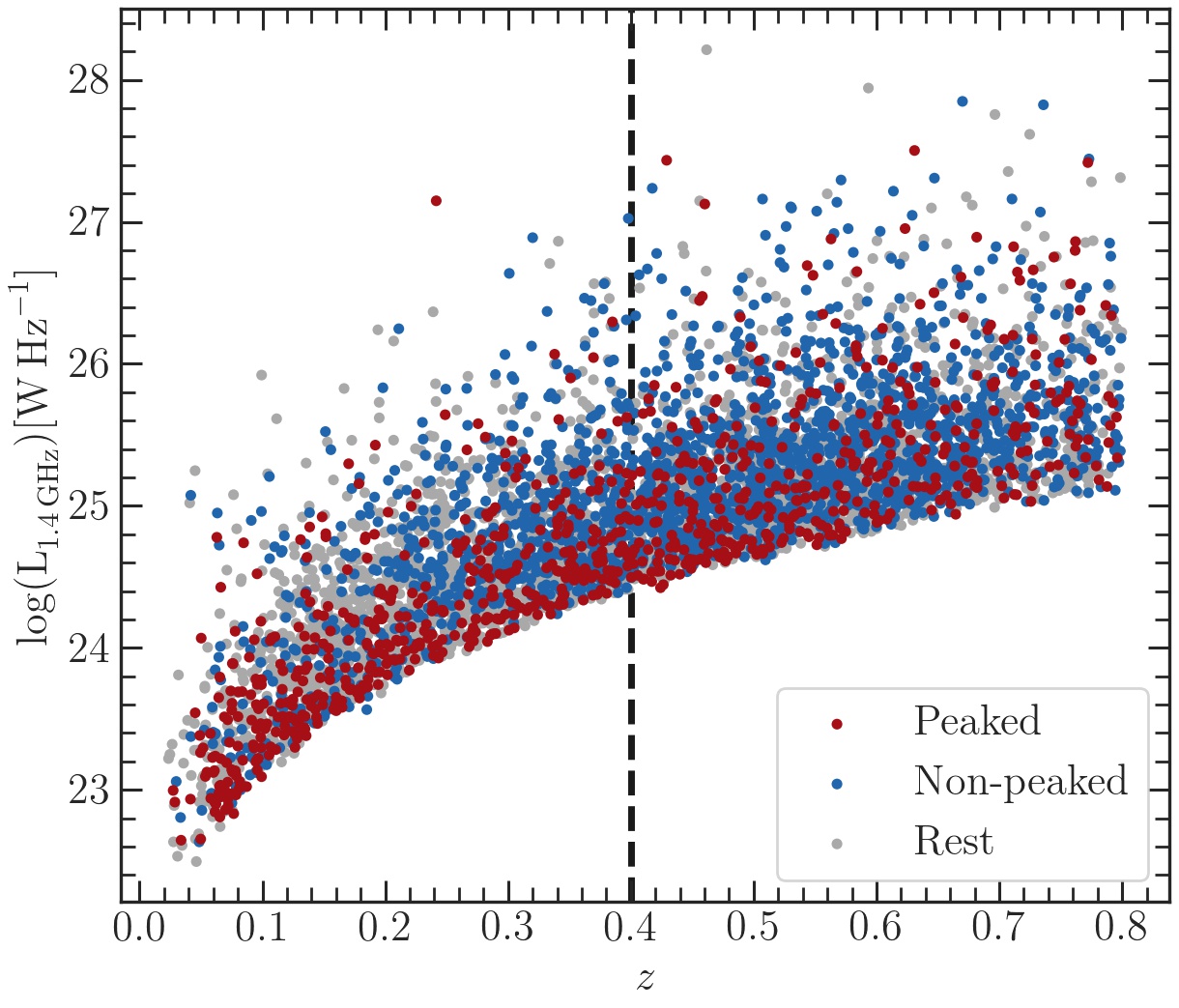}
  \caption{Redshift vs total $L_\mathrm{1.4GHz}$ for the selected from the sample in Sect.~\ref{selectragn}. Peaked and non-peaked sources are marked with different colours. The sample is split about the dashed line at $z=0.4$ into low and high redshift sources.}
  \label{lumvsz}    
\end{figure}

Our radio AGN sample spans a wide range in redshift from $z=0.02$ to $z=0.8$, leading to a large range of 1.4\,GHz luminosities. Figure~\ref{lumvsz} shows $L_\mathrm{1.4GHz}$ as a function of redshift. It can be seen that the typical radio luminosities being probed increase with redshift, as expected for a flux-limited sample. This means we are probing jets of intrinsically very different powers at the high and low redshift end. Therefore, we split the sample into two groups about $z=0.4$ (see source numbers in Table~\ref{spectralshapetable}). This allows a reasonable comparison, while also keeping enough sources in each sub-sample for a statistically significant analysis. The low redshift sample ($z<0.4$) probes a wide range of luminosities ($\sim$\,$10^{22.5}-10^{26}$\,\whz), while still being comparable to the sample used in K23. We discuss the results for the low and high redshift samples separately in Sect.~\ref{results}. \par

To determine the role of jets in driving feedback, it is also crucial to compare the \OIII kinematics of our radio AGN with a control sample of optical AGN sources that do not host radio jets. The diagnostics we use allow us to select a clean radio AGN sample, but the 743 sources classified as SF/radio-quiet AGN could still have low luminosity radio AGN. Thus they do not provide a clean control sample. Instead, we select a control sample from the $\approx$\,24\,000 type 1 and 2 AGN from \citet{Mullaney2013} up to $z=0.4$. This sample overlaps with the redshift range of our low redshift sample. These sources were classified as AGN using their optical emission line widths and ratios from SDSS. To select as `clean' a control sample as possible without radio jets, we only use sources with no radio detections in NVSS or FIRST. But at the sensitivity of NVSS and FIRST, a non-detection does not necessarily imply a lack of radio jets in the source. Therefore we restrict the sources to an upper limit of $L_\mathrm{1.4GHz}<10^{23}$\,\whz. Since LoTSS has higher sensitivity than FIRST and NVSS, we also removed any source with a LoTSS cross-match within a 6$\arcsec$ radius. This ensures that our control sample has no radio counterpart, and minimises any contamination from even low-luminosity radio jets. We further restrict the sample to $10^{39.5}<L_{\textrm{\OIII}}<10^{42}$\ergpers, $10^{11}<M_{\star}<10^{11.6}$\,M$_{\odot}$, and $180<\sigma_{\star}<300$\kms to match it to the low redshift sample. This gives us a control sample of 514 sources. The parameter space occupied by these sources is shown in Fig.~\ref{radvsoptlum}. \par

\section{Radio properties}
This section discusses the radio spectra and morphology of the sample. We first estimate the spectral indices using a combination of LoTSS, FIRST and VLASS data. We then divide the sources into different spectral shape groups and link them to their evolutionary stage. Finally, we also characterise the source sizes using LoTSS and FIRST data. These properties are then linked to the \OIII kinematics further in the paper.

\subsection{Characterising spectral shape}
\label{spectral properties of radio agn}

Using the high resolution of LoTSS, FIRST and VLASS we can probe the central $\sim$\,6$\arcsec$ region of the radio sources, over spatial scales similar to that covered by the SDSS fibre. This allows a direct comparison of the radio and optical properties. To do that, we estimate the spectral indices using the peak flux densities. For unresolved sources and sources with very weak extended emission, the peak and total flux densities are similar. However, for sources with bright extended emission, peak flux density can be significantly smaller than the total flux density. If the extended emission is optically thin and dominates the total flux density, using the total flux density to estimate the spectral index would lead to a source being classified as optically thin, even if there is a peaked spectrum source at the centre. This motivated our choice to use the peak flux density for estimating spectral indices of the radio AGN, to maximise the number of peaked spectrum sources and detect them even at the centre of large extended sources.\par

Estimating accurate spectral indices requires reliable flux density values and comparison over the same spatial scales. The flux densities in the VLASS epoch 1 quick look catalogue were found to be underestimated by a factor of $\sim$\,0.87 by \citet{Gordon2021}. They compared the VLASS flux densities with TGSS at 150\,MHz, WENSS at 330\,MHz, SUMSS at 840\,MHz, and FIRST at 1400\,MHz to estimate the systematic flux scale offset. Assuming their corrected flux scale to be the true flux scale, we have re-scaled our VLASS epoch 2 values to match the corrected VLASS epoch 1 flux scale. For this, we first compared the flux densities of epoch 2 to epoch 1 for our sources. The comparison was done only for point sources with a deconvolved major axis size less than 2\arcsec~in both epoch data. We found a mean ratio of 0.95 for peak flux densities from epoch 1 to epoch 2, showing that the uncorrected epoch 2 flux densities were larger than the uncorrected epoch 1 values. Finally, to match the corrected scale from \citet{Gordon2021}, we divided our epoch 2 values by $\approx$\,0.87/0.95 = 0.92. We scaled all our VLASS peak flux density values by this factor. Although this is a large correction factor and a more accurate flux scale correction probably requires a deeper analysis, this correction is sufficient for our purpose. Changes in this factor do not affect our results significantly as we discuss later. \par

\begin{figure}
\centering
  \includegraphics[width=0.8\columnwidth]{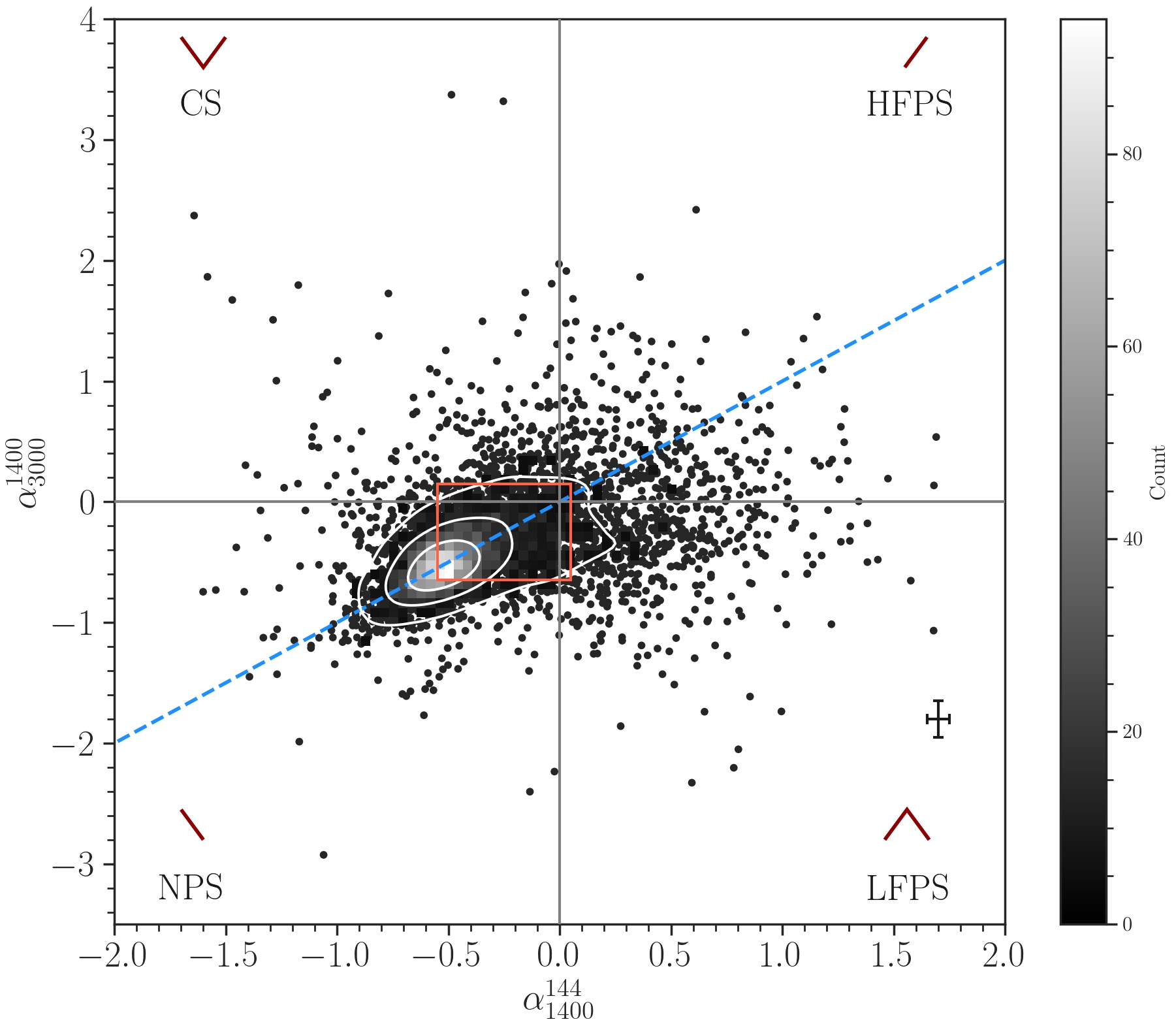}
  \caption{Colour-colour plot of spectral indices for the radio AGN sample. The different quadrants are labelled with the spectral shapes of the sources lying in that region. The spectral shapes are also shown above the labels. The red square marks the region where flat-spectrum (FS) sources lie. A 1:1 line is shown in blue. While studying the link of spectral shape with gas kinematics, only HFPS, LFPS and NPS sources are considered. }
  \label{colcolplot}    
\end{figure}

The next step was to match the resolution of VLASS ($\sim$\,$2.5-3\arcsec$) images to FIRST ($\sim$\,5.4\arcsec).
We smoothed the VLASS cutouts around our sources to match the FIRST resolution, using the same approach as in K23. We used the task \texttt{IMSMOOTH} in CASA to smooth all VLASS cutouts to the resolution of FIRST. We re-extracted the flux densities from these images using pyBDSF \citep{Mohan2015}. Finally, the spectral indices were estimated using the formula $\alpha = \textrm{log}(F_{\textrm{1,peak}}/F_{\textrm{2,peak}})/\textrm{log}(\nu_{1}/\nu_{2})$. We plot all the sources on a colour-colour diagram of spectral indices from 144 to 3000\,MHz, shown in Fig.~\ref{colcolplot}. Spectral classification of the sources was then done using this diagram. \par

As discussed in the introduction, the radio spectrum is absorbed during the young phase in the radio AGN life cycle. If the observed spectral shape is fully inverted in our frequency range ($\alpha^{144}_{1400}>0,\alpha^{1400}_{3000}>0$), we classify the source as high frequency peaked spectrum (HFPS), except for a source in the FS region, as discussed below. They lie in the top right quadrant of the colour-colour plot in Fig.~\ref{colcolplot}. These sources likely have a peak frequency larger than 3000\,MHz. Some HFPS sources may have a peak frequency between our frequency range, however, we expect this group to be dominated by the inverted spectrum sources. If the observed spectral shape has a turnover within our frequency range ($\alpha^{144}_{1400}>0,\alpha^{1400}_{3000}<0$), we label the source as low-frequency peaked spectrum (LFPS), shown in the bottom right quadrant of the colour-colour plot. These sources likely have a peak frequency within our frequency range. Again, sources lying in the FS region are not included, as discussed below. In the radio AGN life cycle, HFPS sources represent a younger phase than LFPS sources, since the peak moves to lower frequencies as the source evolves. We first combine these two groups and call them peaked spectrum (PS) sources, to study their \OIII properties. But we also study these two groups separately in more detail. Thanks to the low-frequency point provided by LoTSS, we can detect PS sources down to 144\,MHz.\par

As radio AGN evolve, the peak in their spectrum shifts to lower frequencies, and eventually the spectrum becomes fully optically thin ($\alpha^{144}_{1400}<0,\alpha^{1400}_{3000}<0$). We label these as non-peaked spectrum (NPS) sources, and they lie in the bottom left quadrant of Fig.~\ref{colcolplot}, except for the sources in the FS region. The host galaxy properties of PS and NPS sources are compared in Fig.\ref{sourceproperties},  \ref{lumvsz} and \ref{sampleproperties_highz}. It can be clearly seen that they have similar $L_\mathrm{1.4GHz}$, $L_{\textrm{\OIII}}$, $M_{\star}$, $D_\mathrm{{n}}(4000)$ and $\sigma_{\star}$ distributions. The $\sigma_{\star}$ values used throughout the paper have been corrected for the SDSS instrumental resolution ($\approx$\,70\kms) and the resolution of the base spectra used for stellar continuum modelling ($\approx$\,150\kms). The distributions show that there are no intrinsic differences between the host galaxies of PS and NPS sources. A similar conclusion for the optical properties of GPS/CSS and Megahertz-Peaked spectrum sources was also reached by \citet{Nascimento2022}.\par

Sources where free-free emission from the core dominates the radio emission, or where the emission is relativistically beamed due to a small jet viewing angle, can have a flat radio spectrum. These flat spectrum sources are usually defined to have a spectral index $\alpha>-0.5$ \citep{ODea1998} and are marked with the red box in Fig.~\ref{colcolplot}. We classify sources with $-0.55<\alpha^{144}_{1400}<0.05$ and $-0.65<\alpha^{1400}_{3000}<0.15$ as flat spectrum (FS) sources. The spectral index and radio luminosities of these sources could affected by relativistic beaming, leading to 'false' high luminosity sources. Therefore we do not use them in our analysis of the \OIII kinematics. Some sources also have a convex-shaped spectrum, which is optically thin at lower frequencies and optically thick at higher frequencies. Although their nature is uncertain, such spectra could be a result of multiple epochs of jet activity. We label these as convex spectrum (CS) sources and they lie in the top left quadrant of Fig.~\ref{colcolplot}. \par

Out of the 5\,720 radio AGN, flux densities with a signal-to-noise ratio greater than 5 were available for 5\,686 sources in LoTSS and 5\,569 sources in VLASS. The spectral shape classifications and number of sources in each group are summarised in Table~\ref{spectralshapetable}. We note that the flux scale correction of VLASS data could overestimate the flux densities at 3000\,MHz, artificially flattening the spectral index between $1400-3000$\,MHz. This could cause some NPS sources to move into the FS and CS region of Fig.~\ref{colcolplot}, and lower the number of NPS sources. However, this will not affect our results since we are still able to select a clean sample of NPS sources. \par

\begin{figure}
  \centering
  \includegraphics[width=0.55\columnwidth]{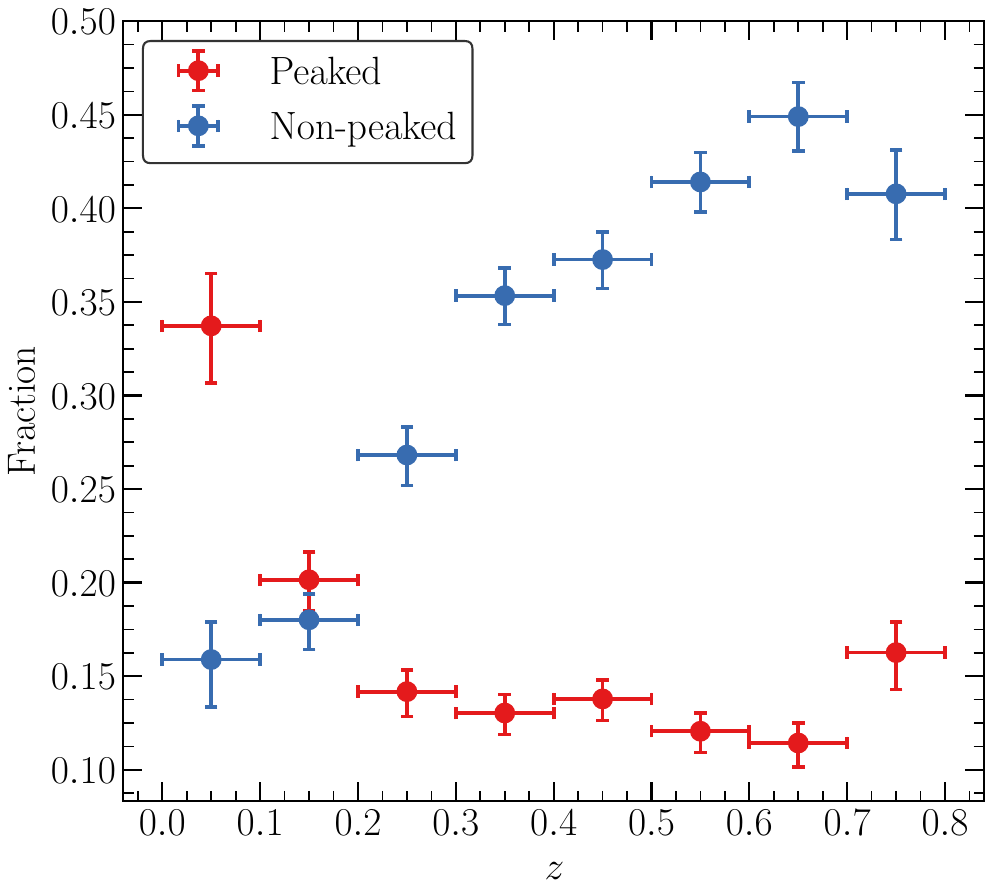}
  \caption{Fraction of PS and NPS sources in redshift ($z$) bins for the full radio AGN sample. The fraction is the number of PS (or NPS) sources divided by the total number of sources in the bin (of all spectral shapes). The horizontal error bars show widths of the redshift bins used to estimate the fraction and the vertical error bars show uncertainties on the fraction of PS and NPS sources in the bin.}
  \label{specshapefraction}    
\end{figure}

Although we use only the peak flux densities in the central region to calculate the spectral indices, the physical scale covered by a telescope beam increases with redshift. This leads to a larger contribution from any steep spectrum diffuse emission around the PS source, if present, to the observed peak flux density. This resolution effect means that genuine PS sources could be classified as NPS. Therefore the fraction of PS sources should decrease and the fraction of NPS sources should increase with redshift, as can be seen in Fig.~\ref{specshapefraction}. This contamination by PS sources makes it harder to select a clean sample of NPS sources at high redshifts. We discuss this in the context of our results in Sect.~\ref{specshape discussion}.\par

Errors in the spectral indices were estimated using a quadrature combination of the RMS noise and the flux scale errors of the surveys. We used a flux scale error of 10\% for LoTSS and 5\% for FIRST. Since we have also performed scaling of the VLASS flux scale, we used a conservative value of 10\% for VLASS data. The median error for the spectral index from $144-1400$\,MHz was 0.05, and for $1400-3000$\,MHz was 0.15. To test the impact of these errors on the spectral shape classification, we took the following approach. We added a random value $\Delta\alpha_{i}$, to each spectral index $\alpha_{i}$, to obtain a spectral index with an offset, given by $\alpha^\textrm{{offset}}_{i}=\alpha_{i}+\Delta\alpha_{i}$. This $\Delta\alpha_{i}$ was drawn randomly from a Gaussian distribution with unit area, zero mean and standard deviation equal to the error in $\alpha_{i}$. The spectral indices with offsets are shown in Fig.~\ref{lowspecindex_offset} and \ref{highspecindex_offset}. We then reclassified all the sources using the colour-colour plot with the new offset spectral indices. All results presented in this paper were then tested with the classifications obtained with these spectral index offsets. \par

\subsection{Radio source sizes}
\label{radio morphology}

\begin{figure}
\centering
  \includegraphics[width=0.7\columnwidth]{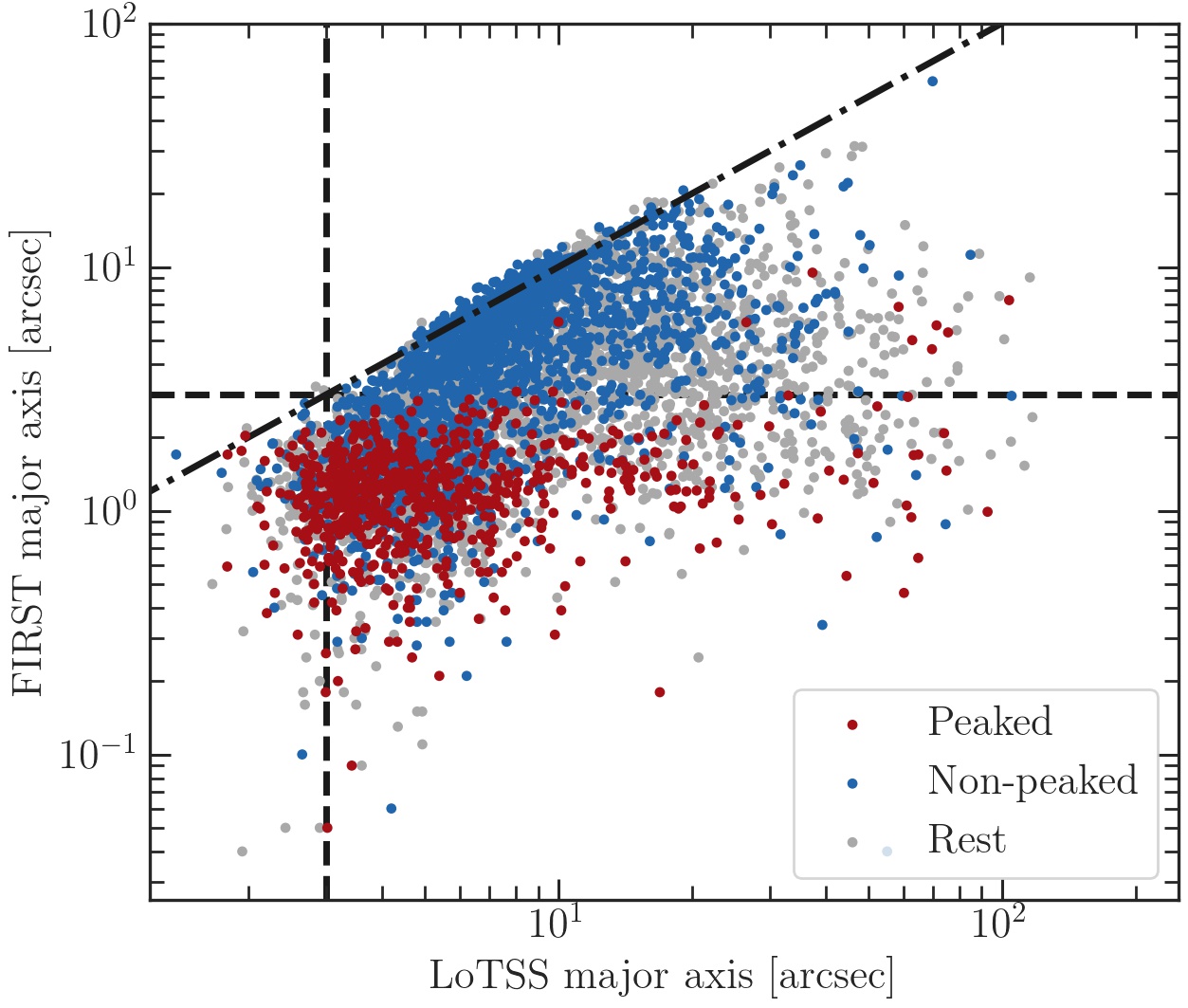}
  \caption{ Deconvolved major axis sizes from LoTSS vs FIRST for the full radio AGN sample up to $z\sim0.8$. The red, blue and grey points mark the PS, NPS and the rest of the sources in the sample. The horizontal and vertical dashed lines mark the 3\arcsec size. A 1:1 ratio is marked by the dashdotted line.}
  \label{sizecomp}    
\end{figure}

Morphologically compact radio AGN are considered to be young sources in the radio AGN life cycle. Therefore the presence of more disturbed ionised and neutral gas in these sources has been used to suggest a link between the radio AGN life cycle and feedback (e.g. \citealt{Gereb2015,Maccagni2017a,Molyneux2019,Murthy2019,Santoro2020}). However, in this study, we aim to go further and use the radio spectral properties to characterise young radio AGN. Indeed, it can be seen in Fig.~\ref{sizecomp} that compact sizes are not always the most reliable indicator for a young source. We find that PS sources span a wide range of sizes in LoTSS, but are almost all smaller than 3$\arcsec$ in FIRST \footnote{Although VLASS images have a higher angular resolution ($\sim$\,$3\arcsec$) than LoTSS and FIRST, their lower image fidelity makes it harder to determine reliable source sizes.}. This difference is due to the significantly higher sensitivity of LoTSS to extended emission than FIRST. This can also be seen in the fact that LoTSS sizes for our sources are consistently greater than FIRST sizes. NPS sources on the other hand span a wide range of sizes in both surveys. Most importantly, this shows that using only a size threshold in a single frequency continuum survey does not guarantee the selection of a clean sample of young radio AGN. \par

As mentioned before, studies have found evidence for more disturbed ionised gas in compact sources. However, there is no universal definition of a compact radio AGN. For comparison with the literature, we also explore this link in our sample. Different studies use either a ratio of major and minor axis sizes, total and peak flux densities or a combination of the two to classify a source as compact or extended. Using angular sizes for this can lead to sources of very different physical sizes being classified as compact (or extended) at low and high redshifts. To avoid this, we use the deconvolved major axis sizes from LoTSS as a proxy for the radio sizes of our sources and convert them to physical sizes using the source redshifts. Although the majority of our sample is unresolved ($\sim$\,$75\%$) in LoTSS according to the criteria of \citet{Shimwell2022}, at $z<0.4$ these sizes can still give reasonable upper limits. We explore the relation between \OIII kinematics and source sizes in Sect.~\ref{radio sizes} and discuss them in Sect.~\ref{sizes discussion}. To quantify the link between the two, we use the fraction of disturbed \OIII sources above and below 20\,kpc. We choose this limit because GPS and CSS sources, which are young radio AGN, typically have sizes varying between $1-20$\,kpc (e.g. \citealt{ODea1998,ODea2021a,Fanti1990,Dallacasa2002}). \par

\section{Results}
\label{results}

\begin{table}
\centering
\caption{Number of sources in each spectral shape group}
         \label{spectralshapetable}
\renewcommand{\arraystretch}{1.15}
\setlength{\tabcolsep}{4pt}
\begin{tabular}{cccc}
    \hlineB{3}
    \noalign{\vspace{0.05cm}}
    \hline
    \noalign{\smallskip}
     Group & Total no. of sources & $z<0.4$ & $z>0.4$  \\
          & \multicolumn{3}{c} {(Number of \OIII detections)} \\     
    \noalign{\smallskip}
    \hlineB{3}
    \noalign{\smallskip}
    \noalign{\smallskip}
  Full RAGN sample & 5720 & 2663 & 3057 \\ 
   & (1693) & (827) & (866) \\
    \noalign{\smallskip}
  Peaked & 850 & 455 & 395 \\
   & (365) & (217) & (148) \\
    \noalign{\smallskip}
  Non-peaked & 1961 & 711 &  1250 \\ 
   & (580) & (206) & (374) \\
      \noalign{\smallskip}
  Flat spectrum & 2529 & 1313 & 1216 \\
   & (592) & (352) & (240) \\
      \noalign{\smallskip}
  Concave & 242 & 104 & 138 \\ 
   & (133) & (42) & (91) \\
      \noalign{\smallskip}
  Unclassified & 138 & 80 & 58 \\ 
   & (23) & (10) & (13) \\
    \noalign{\smallskip}
    \noalign{\smallskip}    
    \hline
    \noalign{\vspace{0.05cm}}  
    \hlineB{3}
 \end{tabular}
    \flushleft
    \textbf{Note.} This table summarises the number of sources in each spectral shape group in the full radio AGN sample up to $z=0.8$. It also lists the number of sources of different spectral shapes in the low redshift ($z<0.4$) and high redshift ($z>0.4$) groups respectively. The number of \OIII detections in each group is shown in brackets. See Sect.\ref{spectral properties of radio agn}.    
\end{table}

 \begin{figure*}
    \centering
    \begin{subfigure}{0.66\columnwidth}
      \includegraphics[width=\columnwidth]{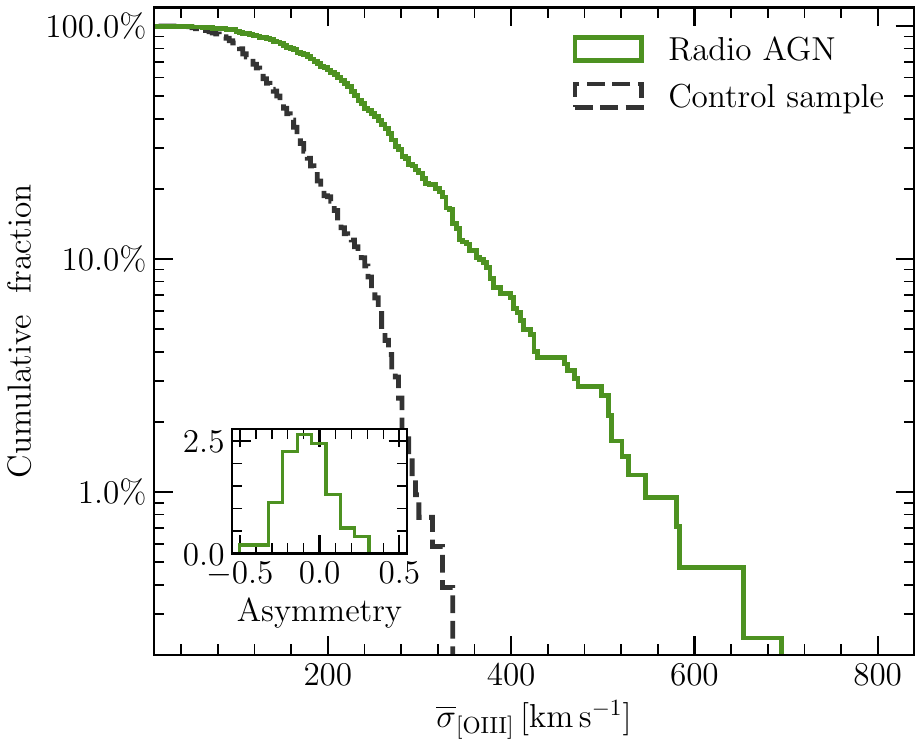}
      \caption{$0.02<z<0.4$}
      \label{cumdist_lowz_all}
    \end{subfigure}
    \hfill
    \begin{subfigure}{0.66\columnwidth}
      \includegraphics[width=\columnwidth]{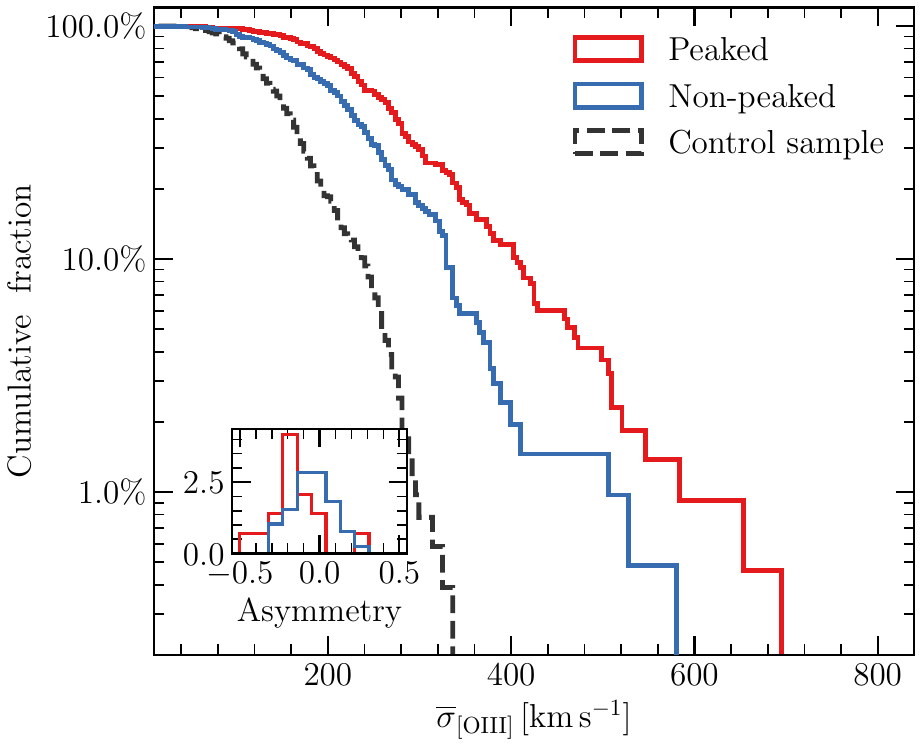}
      \caption{$0.02<z<0.4$}
      \label{cumdist_specshape_lowz}
    \end{subfigure}
    \hfill
    \begin{subfigure}{0.66\columnwidth}
      \includegraphics[width=\columnwidth]{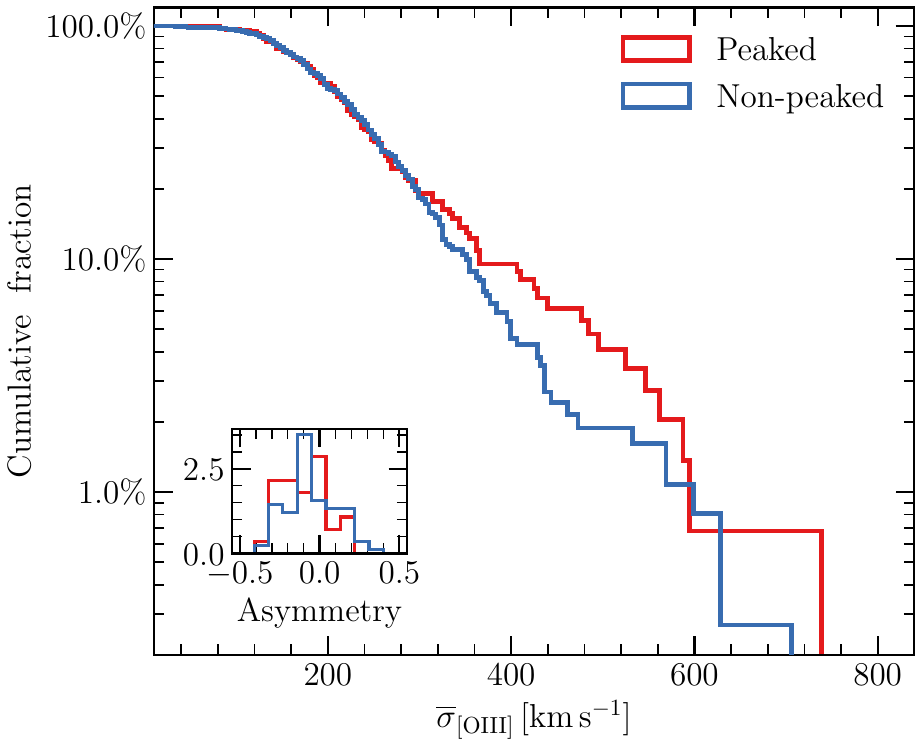}
      \caption{$0.4<z<0.8$}
      \label{cumdist_specshape_highz}
    \end{subfigure}
    \caption{Cumulative distributions of the flux-weighted average \OIII velocity dispersions for \OIII detections. The inset plots show the asymmetry distributions for \OIII profiles with multiple kinematic components in every group.\textbf{(a)} Low redshift radio AGN sample (PS + NPS) and the control sample of optical AGN up to $z\sim0.4$ from \citet{Mullaney2013}, described in Sect.~\ref{selectragn}. \textbf{(b)} Low redshift PS, NPS sources and the control sample. \textbf{(c)} High redshift PS and NPS sources.}
    \label{cmldist_specshape}
 \end{figure*}

We can now investigate the link between the radio spectral shape (a proxy for the evolutionary stage) and \OIII gas kinematics in our sample of 5\,720 radio AGN up to $z\sim0.8$. As we want to trace the changes in \OIII gas kinematics in different stages of the radio AGN life cycle, we restrict the sample to PS and NPS sources. This gives a final sample size of 2\,811 radio AGN, out of which 945 have \OIII detections ($\approx$\,$34\%$). We compare PS and NPS source properties in Fig.~\ref{sourceproperties} and ~\ref{sampleproperties_highz}, which shows that they have quite similar host galaxies. We remind the reader that PS sources represent a young phase of activity ($\sim$\,$0.1-1$ Myr) whereas NPS sources represent more evolved sources ($\sim$\,$10-100$ Myr). As mentioned in Sect.~\ref{sample contruction}, we divide the sample into two groups about $z=0.4$ and discuss them separately (see Table~\ref{spectralshapetable}). We first discuss the role of radio spectral shape in driving \OIII kinematics, followed by $L_\mathrm{1.4GHz}$, $L_{\textrm{\OIII}}$ and source sizes.\par

\subsection{Radio spectral shape}
\label{specshape}

Although our radio AGN host galaxies are selected to harbour radio jets, that does not imply a lack of radiation from the AGN that can also disturb the \OIII gas. Therefore, we first test whether the kinematical disturbances are dominantly driven by jets or radiation pressure. For this, we compare the flux-weighted average \OIII velocity dispersions ($\overline{\sigma}_{\textrm{\OIII}}$) of our radio AGN sample with the control sample of optical AGN selected in Sect.~\ref{sample contruction}, that have a radiatively efficient AGN. This control sample has been selected to not have any radio jets, but is matched in other host galaxy properties to the radio AGN sample. Since the control sample only goes up to $z=0.3$, we use the low redshift group for comparison. Here and throughout the paper, $\overline{\sigma}_{\textrm{\OIII}}$ has been corrected for the SDSS instrumental resolution. Fig.~\ref{cumdist_lowz_all} shows the cumulative distributions of $\overline{\sigma}_{\textrm{\OIII}}$ for the two samples (see Sect.~\ref{modelling optical spectra} for the definition of $\overline{\sigma}_{\textrm{\OIII}}$). The profiles for radio AGN are consistently broader than optical AGN with similar $L_{\textrm{\OIII}}$ and stellar mass ranges. Selecting only type 1 or 2 AGN in the control sample, or changing the $L_{\textrm{\OIII}}$ or $M_{\star}$ range did not affect this result. This shows that \OIII gas is more disturbed in radio AGN host galaxies and highlights the impact of jets on disturbing the surrounding ionised gas.\par

Next, we split the low redshift sample into PS and NPS sources, shown in Fig.~\ref{cumdist_specshape_lowz}. There are 217 PS and 206 NPS sources with \OIII detections in the low redshift sample. We find that $\overline{\sigma}_{\textrm{\OIII}}$ is consistently larger for PS than NPS sources. A 2-sample KS test shows a significant difference between the two groups (KS statistic = 0.23, p-value = 2.9$\times$10$^{-5}$). Therefore we can reject the null hypothesis at a >$99\%$ confidence level that the two $\overline{\sigma}_{\textrm{\OIII}}$ samples were drawn from the same distribution. The asymmetry parameter distributions, shown in the inset of the plot, are also significantly different (KS statistic = 0.47, p-value = 0.007). PS sources have significantly more blueshifted profiles (median asymmetry = -0.17) in comparison to NPS sources (median asymmetry = -0.05). These trends show that \OIII is more disturbed and blueshifted in PS sources than NPS sources. \par

To quantify the fraction of sources with `disturbed' \OIII gas, we use a threshold of $\overline{\sigma}_{\textrm{\OIII}}=350$\kms. This limit is the 3$\sigma$ deviation from the average stellar velocity dispersion of the low redshift sample. Sources in the low redshift sample with $\overline{\sigma}_{\textrm{\OIII}}$ greater than this value are classified as kinematically disturbed, throughout the paper. We find that out of the 217 PS sources, 40 are disturbed according to this threshold, which corresponds to a fraction of $18.4^{+3.0}_{-2.4}$\,\% \footnote{Throughout the paper, the 
errors on the proportion of disturbed sources are estimated at a 68\% confidence interval using a Bayesian approach for proportion uncertainties of a binomial population, outlined in \citet{Cameron2011}.}. On the other hand, out of the 206 NPS sources, only 12 are disturbed, which is a fraction of $5.8^{+2.1}_{-1.3}$\,\%. Therefore we find that \OIII gas in PS sources is $\sim$\,3 times more likely to be kinematically disturbed than NPS sources. Using a more conservative threshold of $\overline{\sigma}_{\textrm{\OIII}}>425$\kms (which corresponds to an FWHM>1000\,\kms), we obtain a fraction of $8.3^{+2.3}_{-1.5}$\,\% for PS sources and $1.5^{+1.4}_{-0.5}$\,\% for NPS sources. Therefore our results are robust to the chosen threshold value.\par

We perform a similar comparison of PS and NPS sources in the high redshift sample, shown in Fig.~\ref{cumdist_specshape_highz}. A 2-sample KS test shows that there is no significant difference in the $\overline{\sigma}_{\textrm{\OIII}}$ distributions (KS statistic = 0.06, p-value = 0.86), although PS sources have slightly broader \OIII profiles than NPS sources. The asymmetry distributions too do not show any significant difference (KS statistic = 0.21, p-value = 0.21). Using the same method to define a $\overline{\sigma}_{\textrm{\OIII}}$ threshold as for the low redshift sample, we estimate a value of 440\,\kms for the high redshift sample, above which an \OIII profile is classified as disturbed. This value is used for the high redshift sample throughout the paper. We find that 10 out of the 148 PS sources (6.8$^{+2.7}_{-1.6}$\,\%) and 11 out of the 374 NPS sources (2.9$^{+1.2}_{-0.7}$\,\%) are kinematically disturbed. Again, the disturbed proportion is larger in PS sources, by a factor of $\approx$\,2. However, the difference between the two groups is only of $\approx$\,$2\sigma$ significance. Therefore, we find only marginal evidence for more disturbed \OIII in PS sources than NPS sources at $z>0.4$.\par

We also find that the results for low and high redshift samples are the same when we use spectral shape classifications after random offsets in spectral indices (as described in Sect.~\ref{spectral properties of radio agn}). This can be seen in the $\overline{\sigma}_{\textrm{\OIII}}$ distributions shown in Fig.~\ref{cumdist_specshape_lowz_offset} and \ref{cumdist_specshape_highz_offset}. \par
Since both the low and high redshift samples cover a wide range in $L_\mathrm{1.4GHz}$ and $L_{\textrm{\OIII}}$, there could be underlying trends that drive the differences in the \OIII~profiles of PS and NPS sources. To disentangle their role, we first investigate the role of $L_\mathrm{1.4GHz}$ and $L_{\textrm{\OIII}}$ in driving the \OIII kinematics below. We then study the differences in the PS and NPS sources while controlling for $L_\mathrm{1.4GHz}$ and $L_{\textrm{\OIII}}$. \par
 \begin{figure*}
    \centering   
    \begin{subfigure}{0.8\columnwidth}
      \includegraphics[width=\columnwidth]{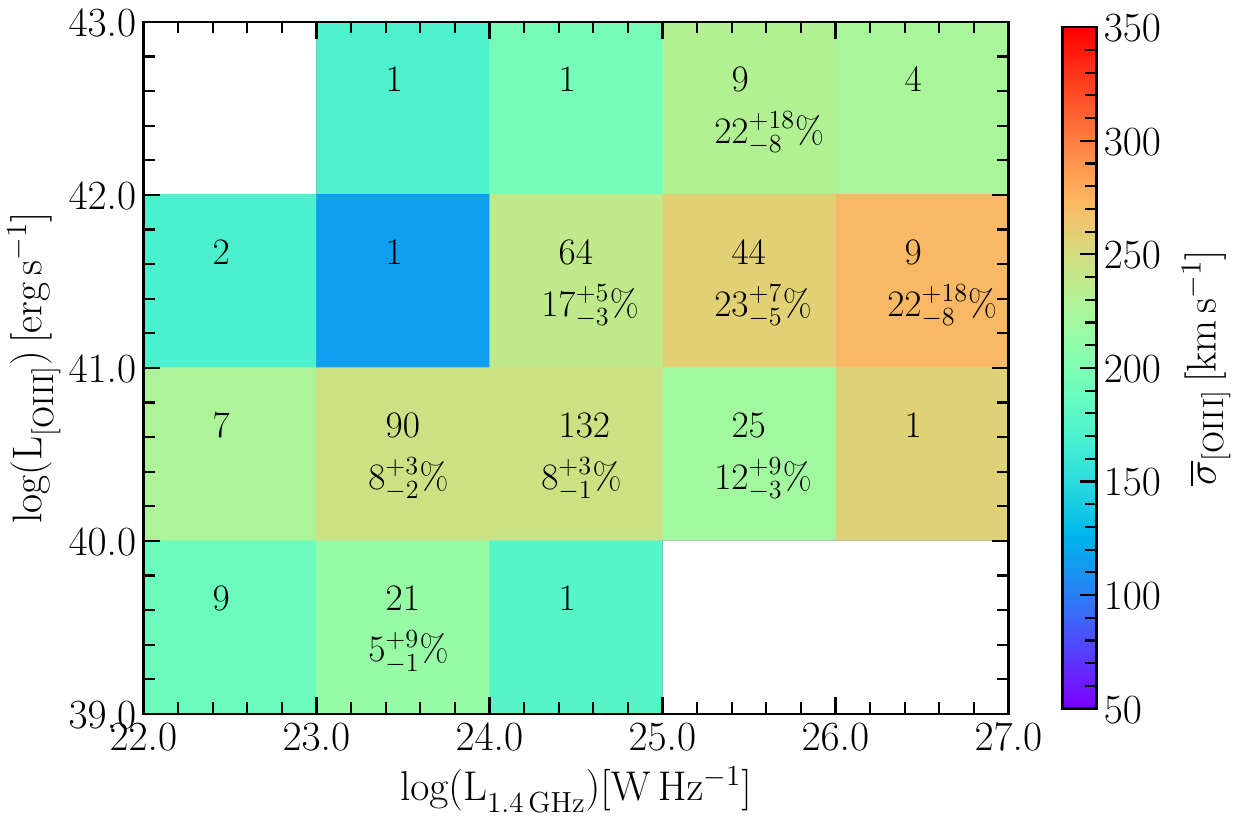}
      \caption{$0.02<z<0.4$}
      \label{lowz_radoptlum}
    \end{subfigure}
    \hspace{1cm}
    \begin{subfigure}{0.66\columnwidth}
      \includegraphics[width=\columnwidth]{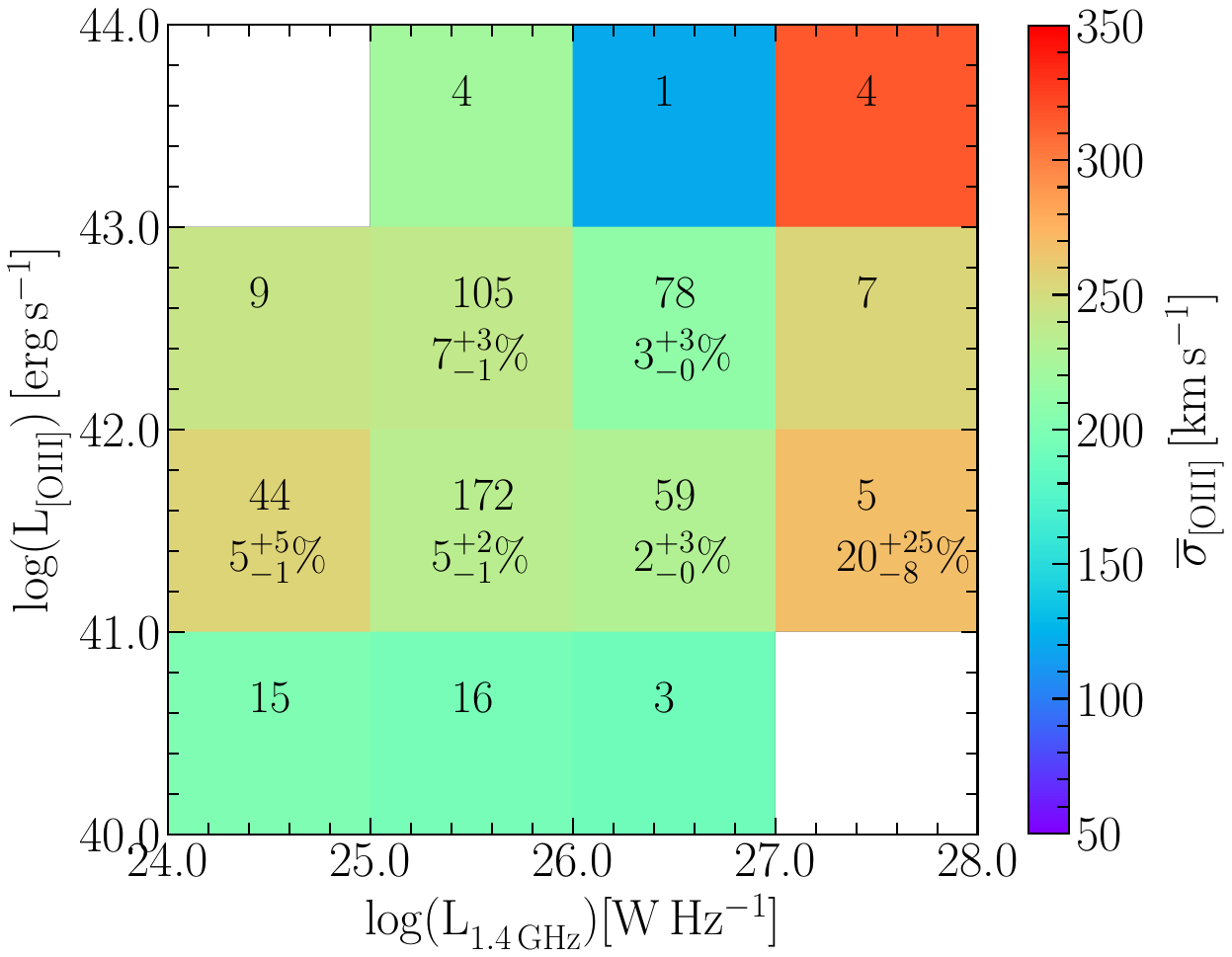}
      \caption{$0.4<z<0.8$}
      \label{highz_radoptlum}
    \end{subfigure}
    \caption{Colour maps showing the average $\overline{\sigma}_{\textrm{\OIII}}$ values in bins of $L_\mathrm{1.4GHz}$ and $L_{\textrm{\OIII}}$
    for the \textbf{(a)} low redshift and \textbf{(b)} high redshift sample. The colour of each bin represents the average velocity dispersion in that bin. The number of sources (bin size) is mentioned at the top of each bin. Fraction of sources with disturbed \OIII are mentioned below the bin size, only for bins that had at least one disturbed source.} 
    \label{radoptlum}
\end{figure*}
 
\subsection{Radio and \OIII luminosities}
\label{radiolumvsoiii}


The total energy output of an AGN is related to $L_\mathrm{1.4GHz}$ and $L_{\textrm{\OIII}}$. Therefore, it is important to understand their role in driving the \OIII kinematics in our sample. As shown in Fig.~\ref{sample_properties_distribution}, there is a correlation between $L_\mathrm{1.4GHz}$ and $L_{\textrm{\OIII}}$ of our sources, i.e. sources with high $L_\mathrm{1.4GHz}$ also tend to have large $L_{\textrm{\OIII}}$values. Since these two properties are not independent, it is important to control for $L_{\textrm{\OIII}}$ while studying the impact of varying $L_\mathrm{1.4GHz}$ on \OIII profiles, and vice versa. To achieve this, we constructed a 2D histogram where the sources are binned in $L_\mathrm{1.4GHz}$ and $L_{\textrm{\OIII}}$ simultaneously, shown in Fig.~\ref{radoptlum}. This allows us to trace the impact of varying one luminosity while controlling for another. The colour map in this plot shows the average $\overline{\sigma}_{\textrm{\OIII}}$ value in each bin. The number of sources and fraction of disturbed \OIII sources are also mentioned in each bin. \par

In the plot for the low redshift sample in Fig.~\ref{lowz_radoptlum}, we see a dependence of the \OIII profile widths on $L_\mathrm{1.4GHz}$ and $L_{\textrm{\OIII}}$. We find that the highest fraction of disturbed sources and largest $\overline{\sigma}_{\textrm{\OIII}}$ values are found at $L_\mathrm{1.4GHz}>10^{25}$\whz and $L_{\textrm{\OIII}}>10^{41}$\ergpers, i.e the high luminosity region. While controlling for  $L_{\textrm{\OIII}}$, the average $\overline{\sigma}_{\textrm{\OIII}}$ value and fraction of disturbed sources increases with $L_\mathrm{1.4GHz}$. The opposite is also true i.e. increasing $L_{\textrm{\OIII}}$ while controlling for $L_\mathrm{1.4GHz}$ also gives more disturbed \OIII. Although the differences in the fraction of disturbed sources are within $1-2\sigma$, we can see a systematic trend with increasing $L_\mathrm{1.4GHz}$ and $L_{\textrm{\OIII}}$. To avoid the effect of small number statistics, we only use bins with more than 10 sources for this result. Although not a large bin size, it is a reasonable limit given the low and high redshift sample sizes.\par

In the high redshift sample shown in Fig.~\ref{highz_radoptlum}, the trend with luminosities is less significant. Although we see more disturbed profiles in the $L_{\textrm{\OIII}}>10^{41}$\ergpers region, there seems to be no clear trend with $L_\mathrm{1.4GHz}$ or $L_{\textrm{\OIII}}$. The fraction of disturbed sources between any two consecutive bins is within $1\sigma$ of each other. The sources with the large $\overline{\sigma}_{\textrm{\OIII}}$ values also do not cluster in one region of the plot, but they do lie above $L_{\textrm{\OIII}}>10^{41}$\ergpers. The fraction of disturbed sources in bins overlapping with the low redshift sample is lower for the high redshift sample, however, this is a result of the larger threshold value for defining an \OIII profile as disturbed in the high redshift sample.\par

\subsection{Radio spectral shape and source luminosities}
\label{specshape and lum}

\begin{figure*}
  \centering
    \begin{subfigure}{0.8\columnwidth}
      \includegraphics[width=\columnwidth]{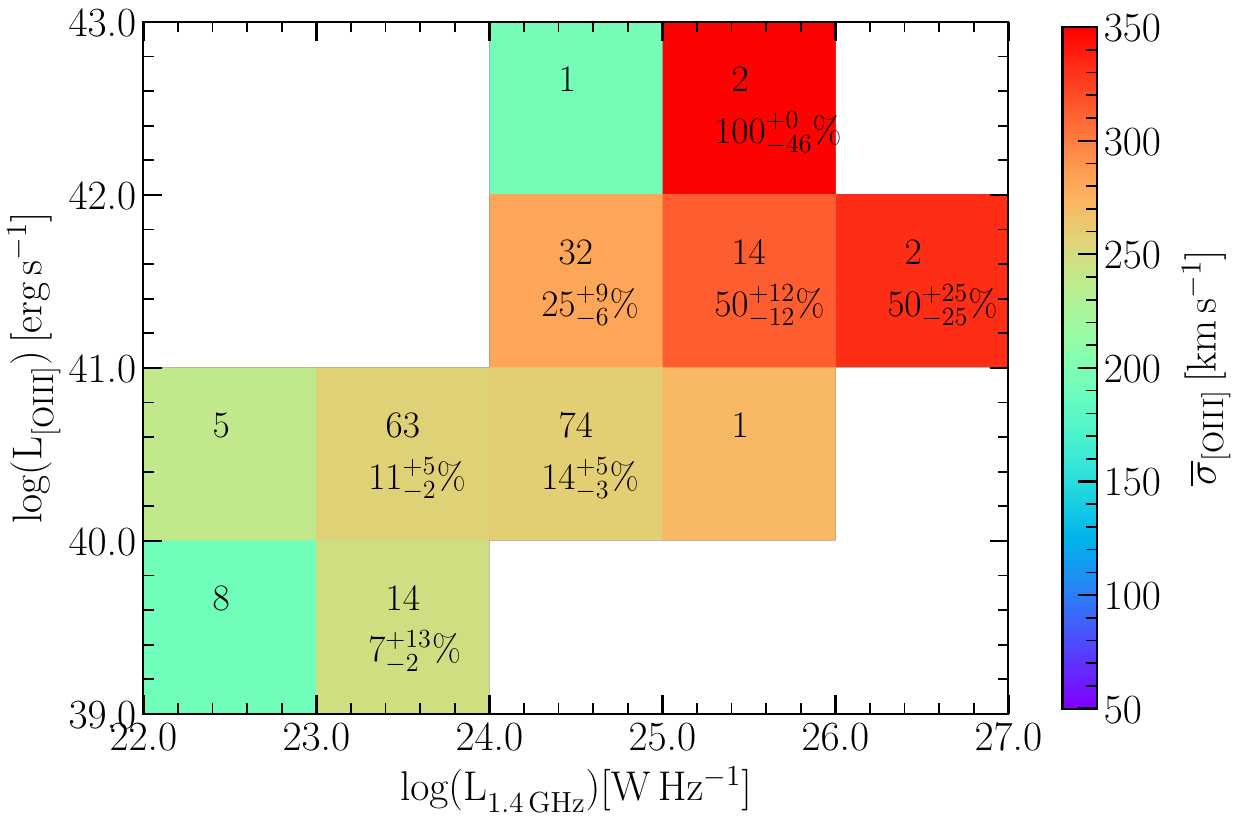}
      \caption{Peaked spectrum $0.02<z<0.4$}
    \end{subfigure}
    \hspace{1cm}
    \begin{subfigure}{0.8\columnwidth}
      \includegraphics[width=\columnwidth]{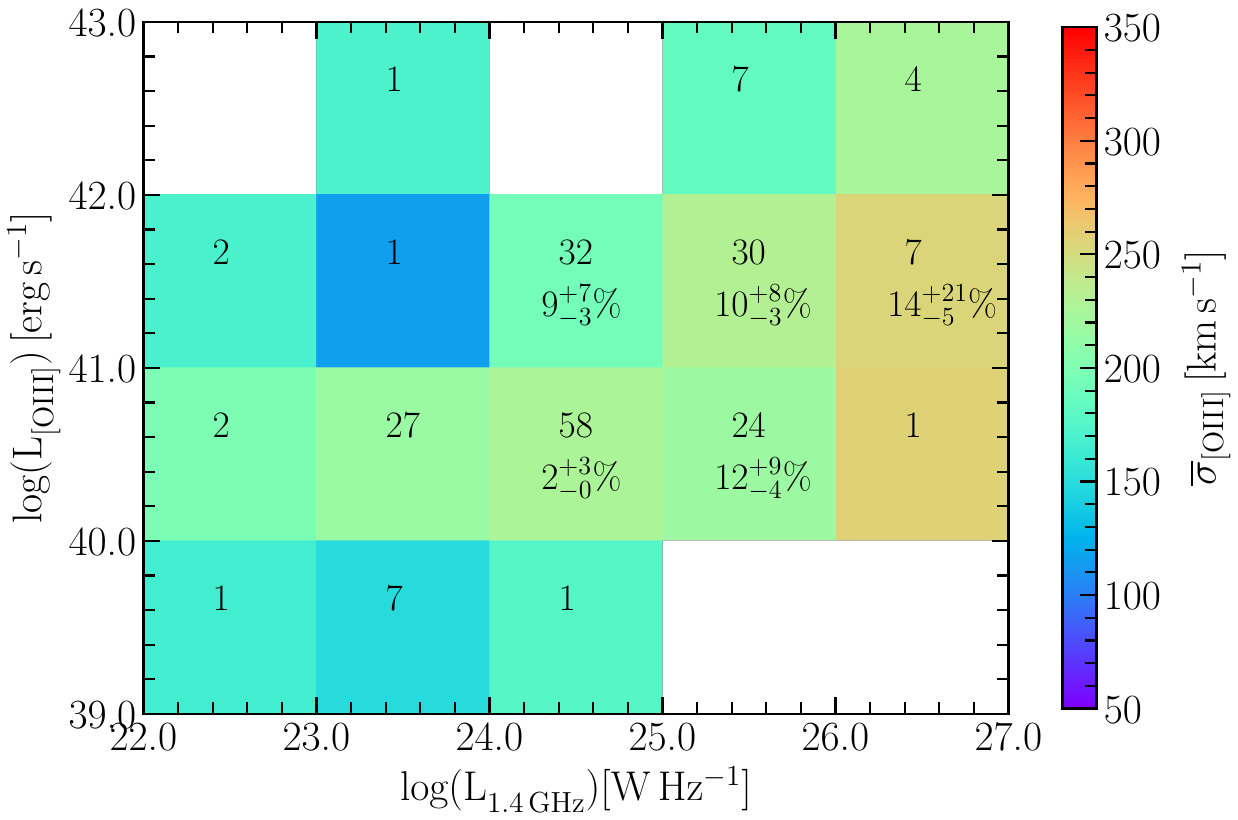}
      \caption{Non-peaked spectrum $0.02<z<0.4$}
    \end{subfigure}\\
    \vspace{0.5cm}
    \begin{subfigure}{0.7\columnwidth}
      \includegraphics[width=\columnwidth]{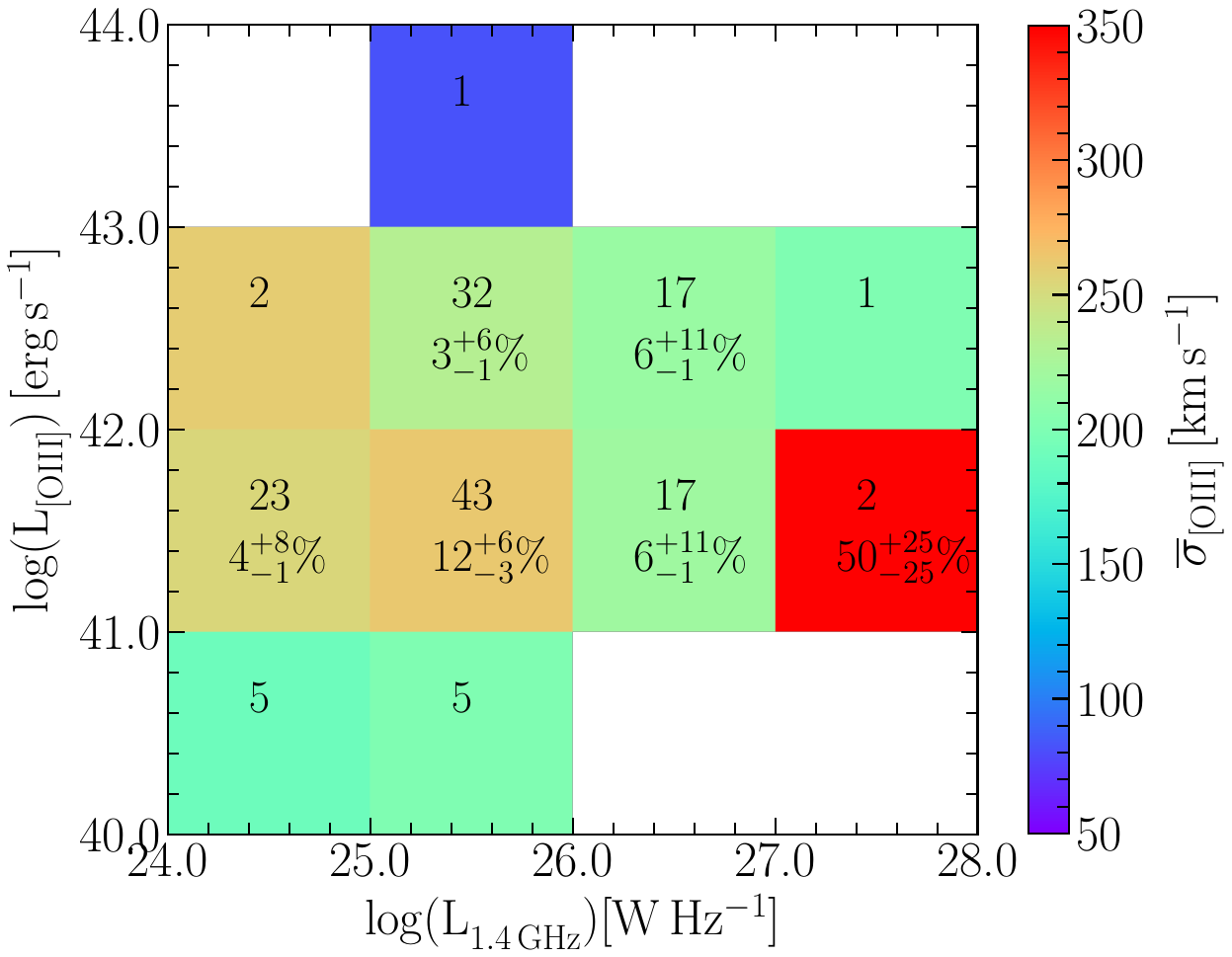}
      \caption{Peaked spectrum $0.4<z<0.8$}
    \end{subfigure}
    \hspace{2cm}
    \begin{subfigure}{0.7\columnwidth}
      \includegraphics[width=\columnwidth]{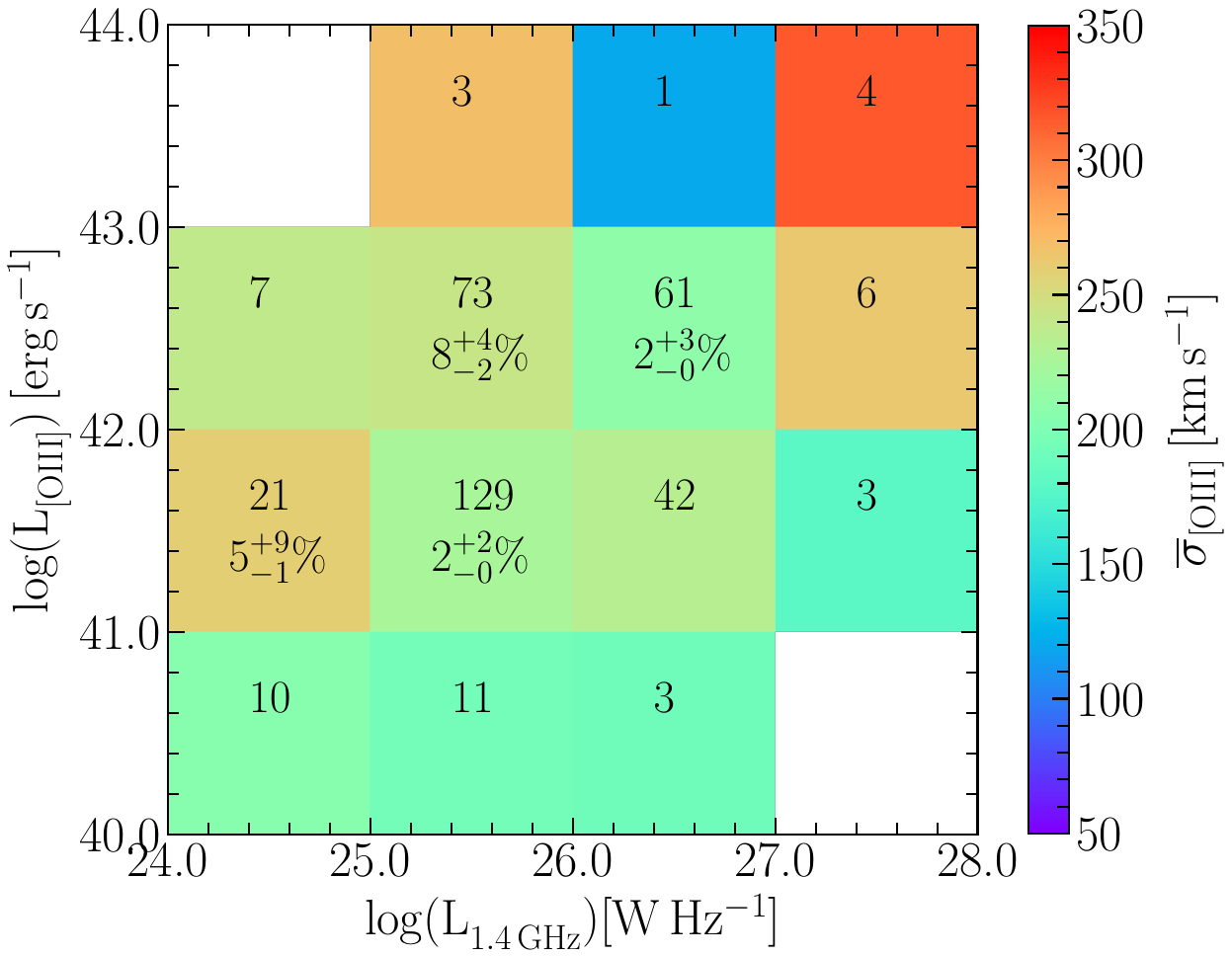}
      \caption{Non-peaked spectrum $0.4<z<0.8$}
      
    \end{subfigure}
    \caption{Same plot as Fig.~\ref{radoptlum}. The top row shows the plots for \textbf{(a)} peaked spectrum and \textbf{(b)} non-peaked spectrum sources in the low redshift sample. The bottom row shows the plots for \textbf{(c)} peaked spectrum and \textbf{(d)} non-peaked spectrum sources in the high redshift sample.}
    \label{radoptlum_specshape}
  \end{figure*}

In the sections above, we found that the radio spectral shape, and the source luminosities ($L_\mathrm{1.4GHz}$ and $L_{\textrm{\OIII}}$) are linked to the widths of the \OIII profiles. Given that the PS and NPS sources discussed in Sect.~\ref{specshape} cover a wide range of luminosities (see Fig.~\ref{sample_properties_distribution}), it is important to disentangle the role of $L_\mathrm{1.4GHz}$ and $L_{\textrm{\OIII}}$  for these groups. We control for these luminosities and construct similar plots as Fig.~\ref{radoptlum} for the PS and NPS sources in the low and high redshift samples, shown in Fig.~\ref{radoptlum_specshape}.\par

Comparing the PS and NPS sources in the low redshift sample, we see that the \OIII profiles of PS sources are consistently wider at any $L_\mathrm{1.4GHz}$ and $L_{\textrm{\OIII}}$ bin. For bins with more than 10 sources, PS sources always have a larger fraction of disturbed \OIII than NPS sources, even if we control for $L_\mathrm{1.4GHz}$ and $L_{\textrm{\OIII}}$. This shows that the relation with radio spectral shape, discussed in Sect.~\ref{specshape}, is not driven by source luminosities. The \OIII profiles of these sources are indeed linked to their radio spectral shape. In the colour maps, the difference between the \OIII profile widths of PS and NPS sources also seems to be larger in the high luminosity region. This suggests also a link between the difference in \OIII profile widths and the source luminosities. In Sect.~\ref{stacking_specshape_lum}, we assess this relationship using a stacking analysis. Similar to Sect.~\ref{specshape}, these results are robust to the errors in spectral indices. We find similar trends in PS and NPS sources with spectral index offsets, as shown in Fig.~\ref{radoptlum_lowz_peak_offset} and \ref{radoptlum_lowz_nonpeak_offset}. This shows that the relation between \OIII widths and radio spectral shape is a real trend and not a result of erroneous spectral shape classification. \par

We construct a similar plot for the high redshift PS and NPS sources, in Fig.~\ref{radoptlum_specshape}. Here we find only marginal differences in the widths of the \OIII profiles of PS and NPS sources in any bin. The fraction of disturbed sources, is almost always larger for PS than NPS sources, at any $L_\mathrm{1.4GHz}$ or $L_{\textrm{\OIII}}$ bin. This suggests that in comparison to NPS sources, more disturbed \OIII in PS sources can also be found at $0.4<z<0.8$, however the difference is only marginal. Similar to the low redshift sample, our results are consistent with the classification after spectral index offsets, shown in Fig.~\ref{radoptlum_highz_peak_offset} and \ref{radoptlum_highz_nonpeak_offset}. We do not present a stacking analysis below for the high redshift sample, since we did not find any significant difference between the profiles. But discuss this result further in Sect.~\ref{discussion}. \par

\subsection{Stacking analysis of PS and NPS sources}
\label{stacking_specshape_lum}

 \begin{figure*}
 \centering
    \begin{subfigure}{0.5\columnwidth}
      \includegraphics[width=\columnwidth]{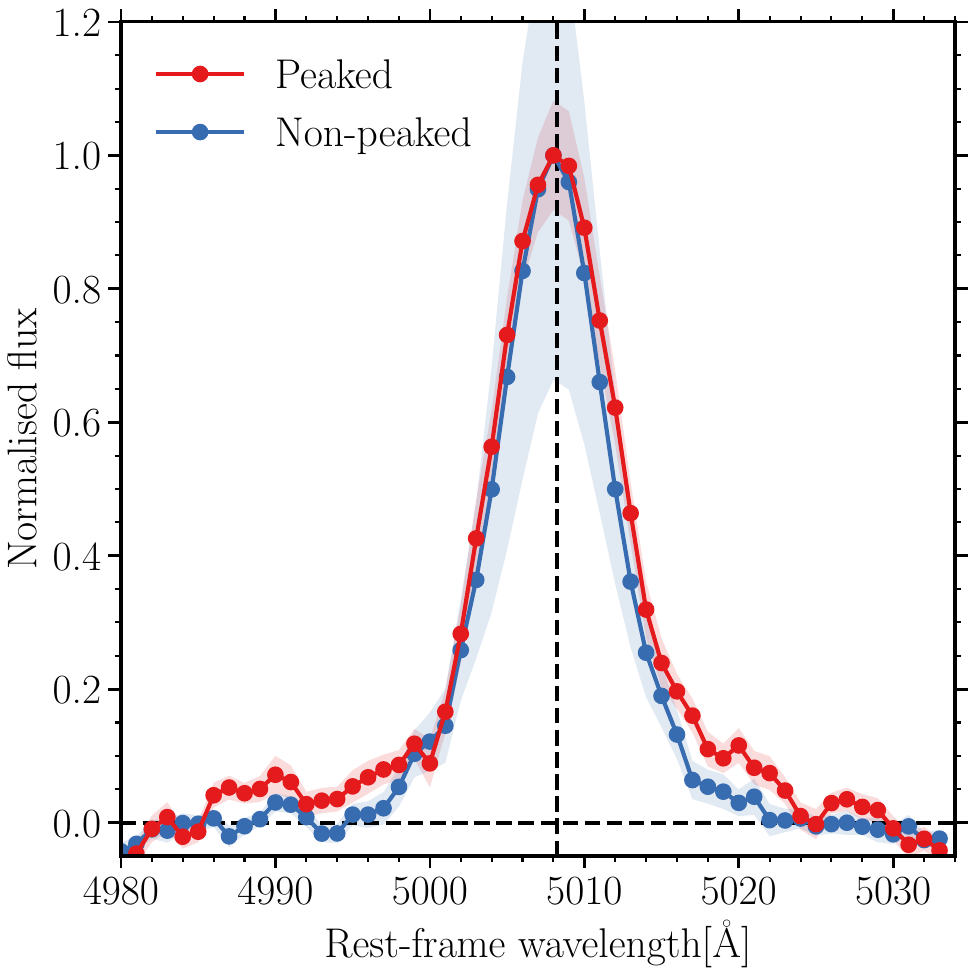}
      \caption{$L_\mathrm{1.4GHz}=10^{23}-10^{24}$\whz}
    \end{subfigure}
    \hspace{1cm}
    \begin{subfigure}{0.5\columnwidth}
      \includegraphics[width=\columnwidth]{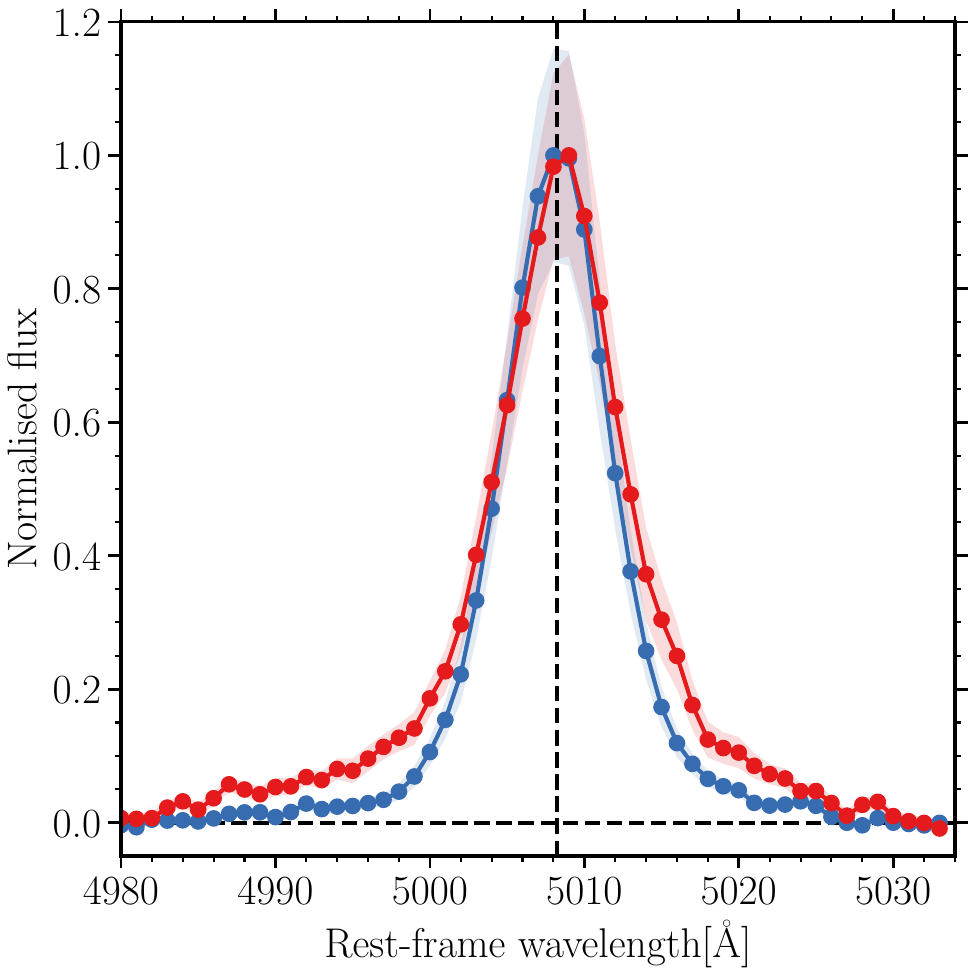}
      \caption{$L_\mathrm{1.4GHz}=10^{24}-10^{25}$\whz}
    \end{subfigure}
    \hspace{1cm}
    \begin{subfigure}{0.5\columnwidth}
      \includegraphics[width=\columnwidth]{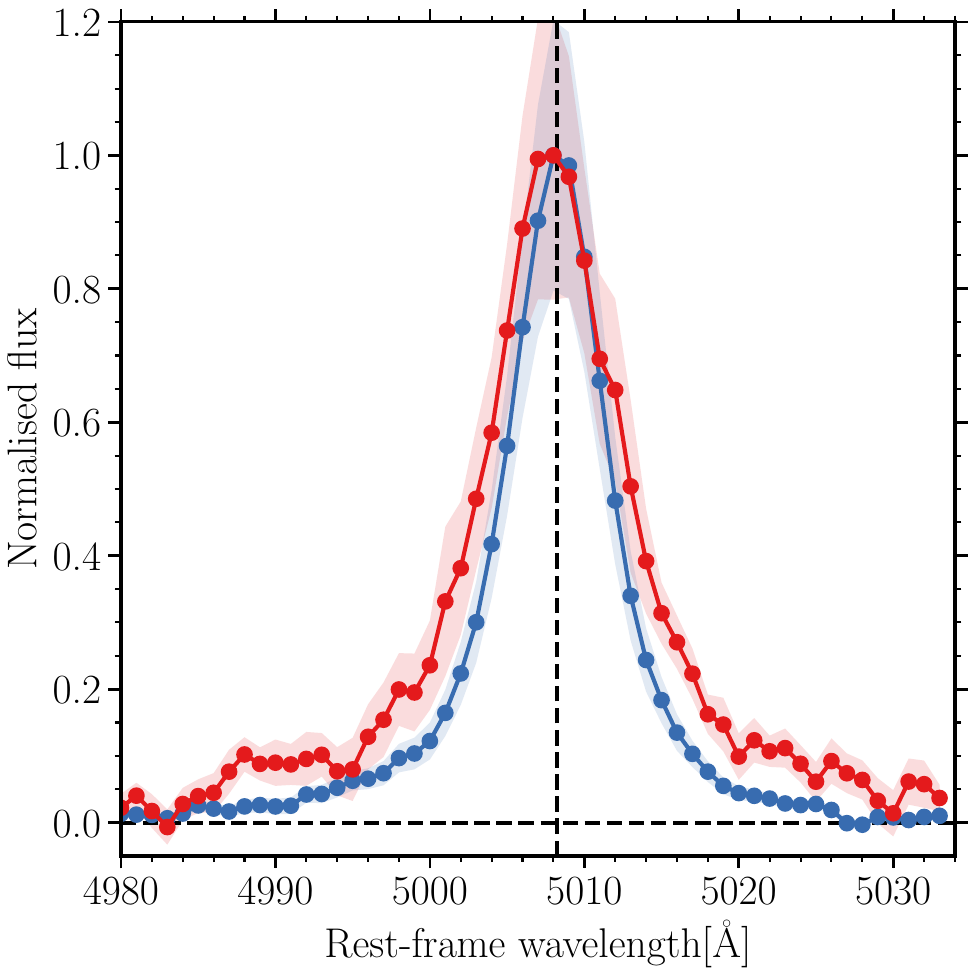}
      \caption{$L_\mathrm{1.4GHz}=10^{25}-10^{26}$\whz}
    \end{subfigure}\\
    \vspace{0.3cm}
    \begin{subfigure}{0.5\columnwidth}
      \includegraphics[width=\columnwidth]{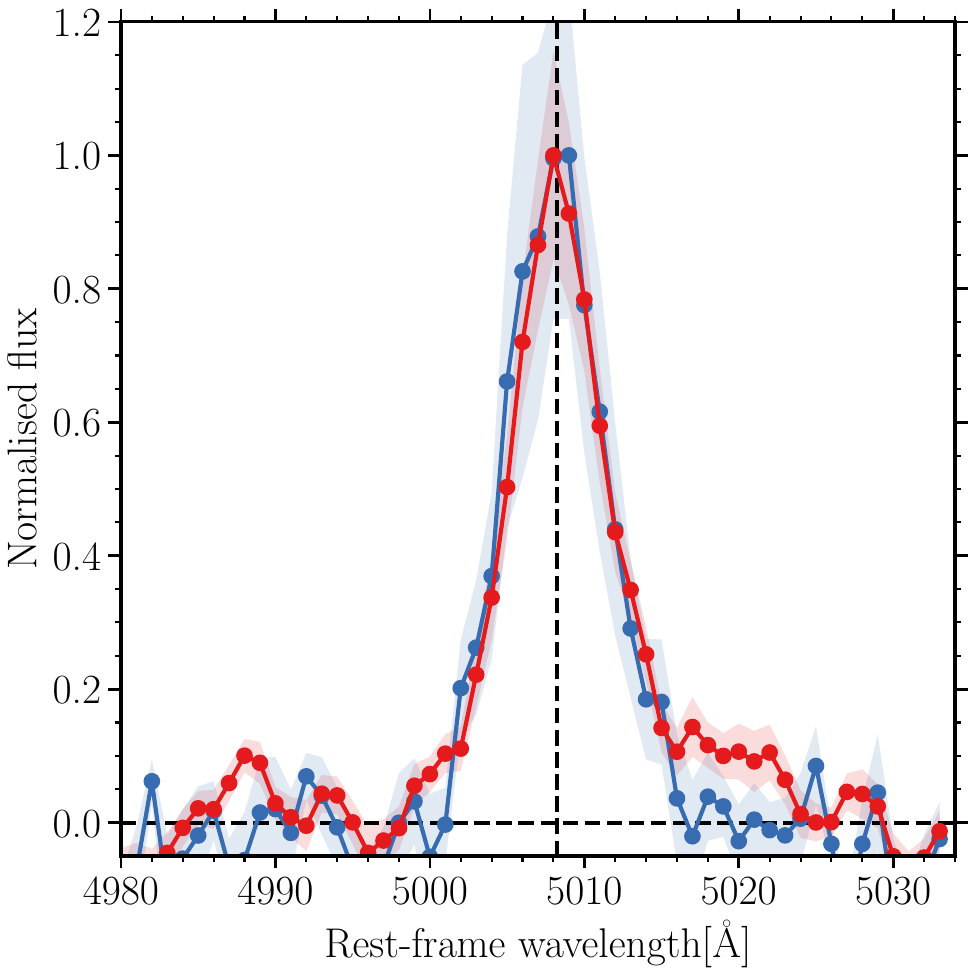}
      \caption{$L_{\textrm{\OIII}}=10^{39}-10^{40}$\ergpers}
    \end{subfigure}
    \hspace{1cm}
    \begin{subfigure}{0.5\columnwidth}
      \includegraphics[width=\columnwidth]{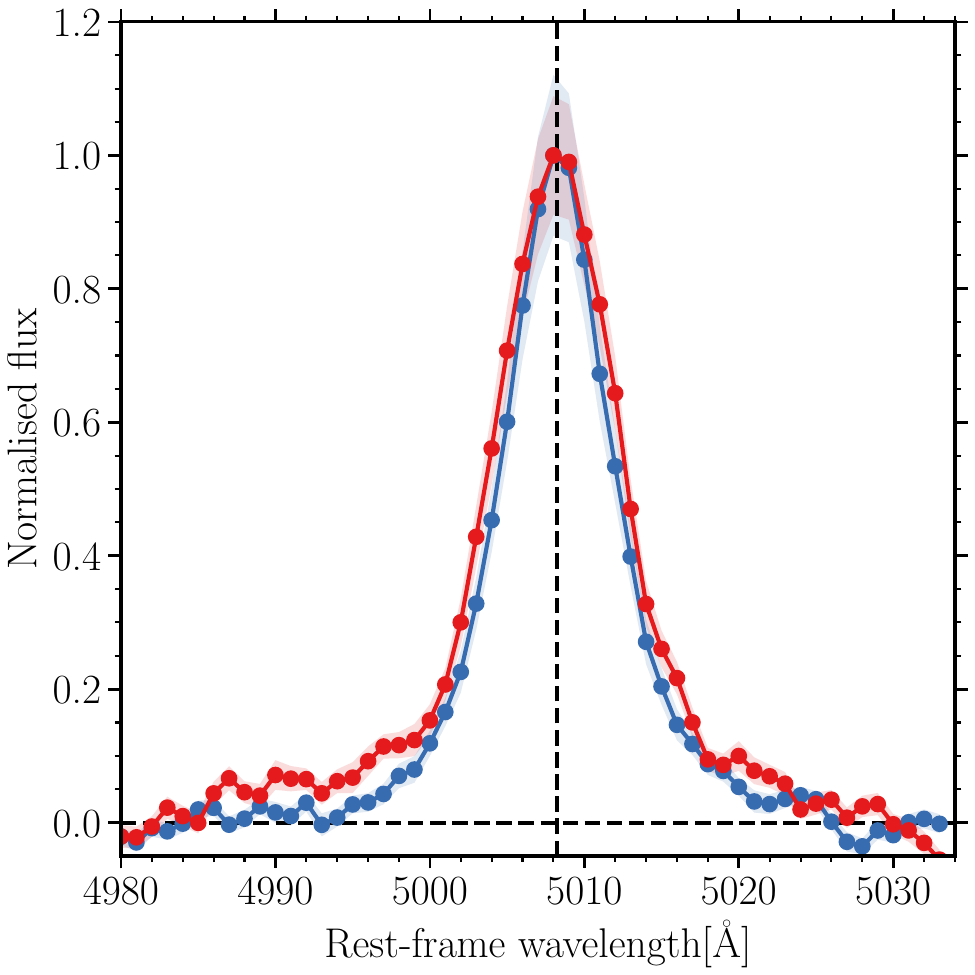}
      \caption{$L_{\textrm{\OIII}}=10^{40}-10^{41}$\ergpers}
    \end{subfigure}
    \hspace{1cm}
    \begin{subfigure}{0.5\columnwidth}
      \includegraphics[width=\columnwidth]{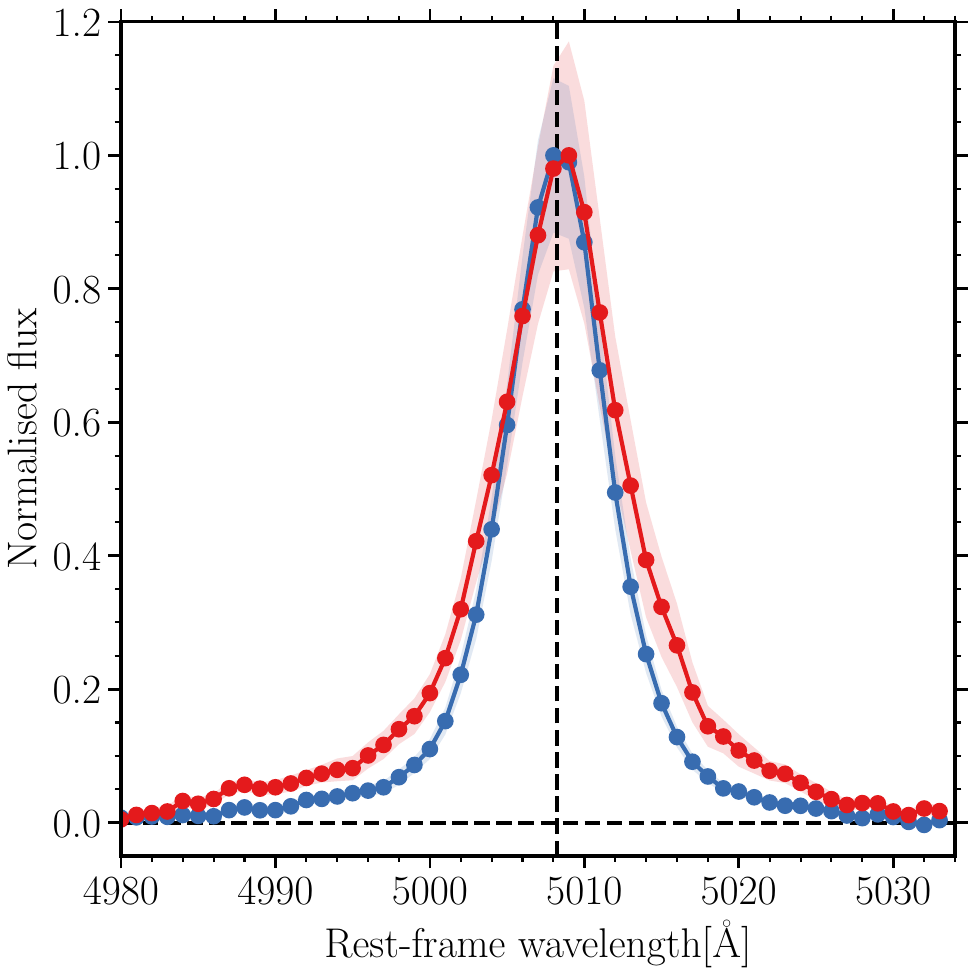}
      \caption{$L_{\textrm{\OIII}}=10^{41}-10^{42}$\ergpers}
    \end{subfigure}
    
    \caption{Stacked \OIII profiles for the peaked and non-peaked spectrum sources in the low redshift sample, while controlling for $L_\mathrm{1.4GHz}$ and $L_{\textrm{\OIII}}$. The shaded regions show the $1\sigma$ errors on the stacked profiles, estimated by stacking 1000 bootstrapped samples. The best-fit model parameters for each profile are summarised in Table~\ref{model fit}. \textbf{(a)$-$(c)} Stacked profiles for PS and NPS sources while controlling for $L_\mathrm{1.4GHz}$. \textbf{(d)$-$(f)} Stacked profiles for PS and NPS sources while controlling for $L_{\textrm{\OIII}}$.}
    \label{stacked_radlumbins}
 \end{figure*}

In the section above, we find that the difference in the widths of \OIII profiles of low redshift PS and NPS sources is related to $L_\mathrm{1.4GHz}$ and $L_{\textrm{\OIII}}$. But this is hard to judge from Fig.~\ref{radoptlum_specshape} alone. We also want to observe these differences in the widths of average \OIII profiles of PS and NPS sources and quantify them. For this purpose, we use a stacking analysis. We stack the \OIII profiles for low redshift PS and NPS sources while controlling for $L_\mathrm{1.4GHz}$ and $L_{\textrm{\OIII}}$ individually.\par
To cover most of the source luminosity range discussed in Sect.~\ref{radiolumvsoiii} while keeping a significant number of sources in each group, we decided to split the sample into three groups of $L_\mathrm{1.4GHz}$ from $10^{23}$ to $10^{26}$\,\whz and $L_{\textrm{\OIII}}$ from $10^{39}$ to $10^{42}$\,\ergpers. For consistency with Sect.~\ref{radiolumvsoiii} and \ref{specshape and lum}, we keep the width of each luminosity group the same, that is 1 dex on the log scale. Since we want to study the average \OIII profiles of these sources and have a small number of objects in the highest luminosity groups, we have carefully tested whether the results are reliable for the stacked profile. We have removed three sources with high S/N spectra and extremely disturbed and broad \OIII profiles ($v_\textrm{broad}$ up to $-600$\kms and FWHM$_\textrm{broad}$ up to 1450\kms) that dominate the stacked profile in the $10^{25}-10^{26}$\,\whz group. We do this to avoid the results being driven by a few individual extremely disturbed sources, and to get a stacked profile more representative of the true average. \par

The stacked \OIII profiles are now shown in Fig.~\ref{stacked_radlumbins}. For the groups in $L_\mathrm{1.4GHz}$, we find that the differences in the widths of stacked profiles become more prominent with increasing $L_\mathrm{1.4GHz}$. In the $10^{23}-10^{24}$\,\whz group, the stacked profile of PS sources is marginally broader than the NPS sources. Although in Fig.~\ref{radoptlum_specshape} we see that PS sources in this $L_\mathrm{1.4GHz}$ range have wider profiles and more disturbed \OIII than NPS sources. The difference is much more significant at $10^{24}-10^{25}$\,\whz and the most at $10^{25}-10^{26}$\,\whz. We measure the widths of the profiles by fitting models made up of Gaussian components as discussed in Sect.~\ref{modelling optical spectra}. Although the figures only show the 5007\,\AA~component of the \OIII doublet line, while fitting the models we also use the 4958\,\AA~component for accurate characterisation of the stacked profile. The best-fit model parameters are summarised in Table~\ref{model fit}. Here, we focus on the widths of the low-level broad components. In the $L_\mathrm{1.4GHz} = 10^{24}-10^{25}$\,\whz group, FWHM of the broad component is 1338\,\kms for PS and 1179\,\kms for NPS sources. At $L_\mathrm{1.4GHz} = 10^{25}-10^{26}$\,\whz, the FWHM of the broad component is 1639\,\kms for PS and 1091\,\kms for NPS sources. Therefore, the difference between the FWHMs of PS and NPS sources increases from $\approx$\,160\,\kms to $\approx$\,550\,\kms with $L_\mathrm{1.4GHz}$. This suggests that the `difference' in the impact of young and evolved radio AGN on \OIII gas also increases with $L_\mathrm{1.4GHz}$.\par

Next, we investigate the role of  $L_{\textrm{\OIII}}$, shown in the bottom panel of Fig.~\ref{stacked_radlumbins}. Similar to $L_\mathrm{1.4GHz}$, we find no difference in the stacked profiles of the lowest luminosity group from $10^{39}-10^{40}$\,\ergpers. At higher luminosities, the difference in the broad component widths decreases marginally with increasing $L_{\textrm{\OIII}}$. In the stacked profiles of $L_{\textrm{\OIII}} = 10^{40}-10^{41}$\,\ergpers group, FWHM of broad component is 1353\,\kms for PS and 981\,\kms for NPS sources. However in the highest luminosity group of $L_{\textrm{\OIII}} = 10^{41}-10^{42}$\,\ergpers, FWHM of the broad component is 1434\,\kms for PS and 1202\,\kms for NPS sources. The difference between the FWHMs decreases from $\approx$\,370\,\kms to $\approx$\,230\,\kms with $L_{\textrm{\OIII}}$. Therefore, we find no significant change in the average profiles of PS and NPS sources with increasing $L_{\textrm{\OIII}}$ beyond $10^{40}$\,\ergpers. Including the \OIII non-detections while stacking did not affect any results. \par

\begin{figure}
\centering
  \includegraphics[width=0.6\columnwidth]{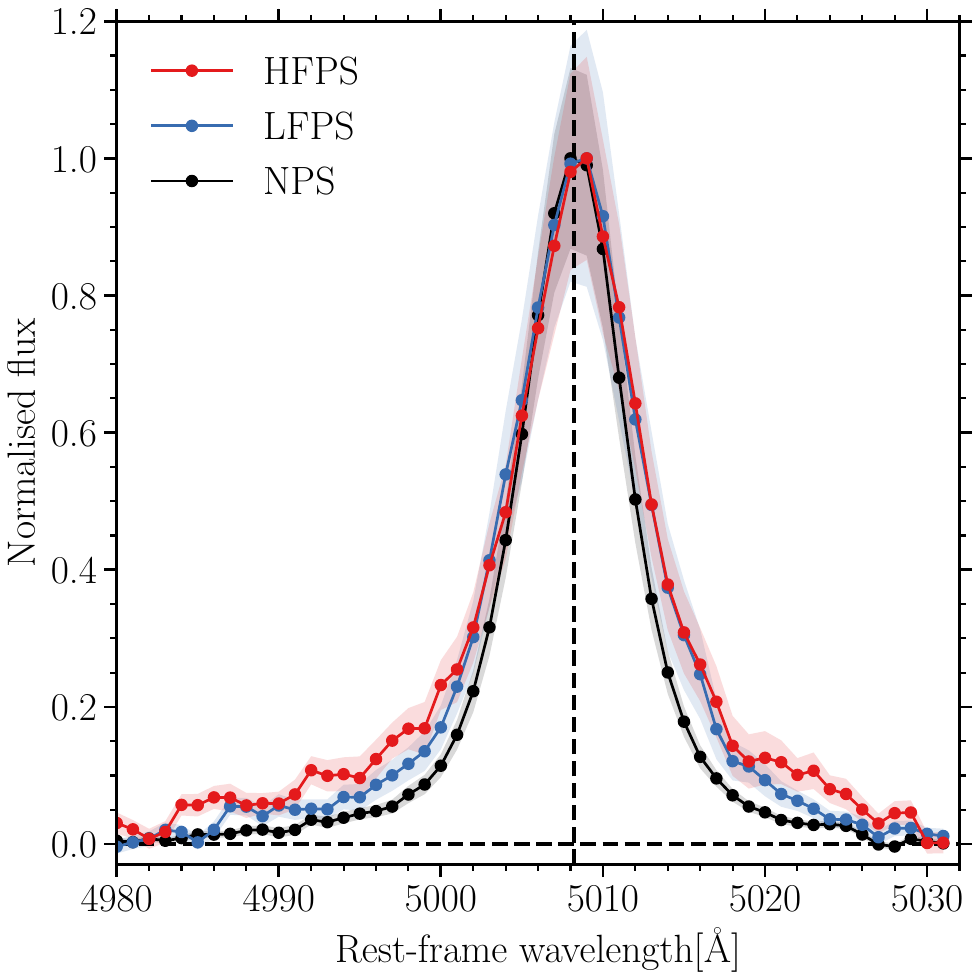}
  \caption{Stacked profile for HFPS, LFPS and NPS sources in the low redshift sample covering a $L_\mathrm{1.4GHz}$ range of $10^{24}-10^{26}$\,\whz. The shaded regions show the $1\sigma$ errors estimated using the bootstrapped samples. The best-fit model parameters and group sizes are summarised in Table.~\ref{model fit}.}
  \label{stacked_specshape_3groups}    
\end{figure}

\subsection{Stacking analysis of HFPS, LFPS and NPS sources}
\label{stacking_specshape_3bins}
\begin{figure*}
    \centering
    \begin{subfigure}{0.65\columnwidth}
      \includegraphics[width=\columnwidth]{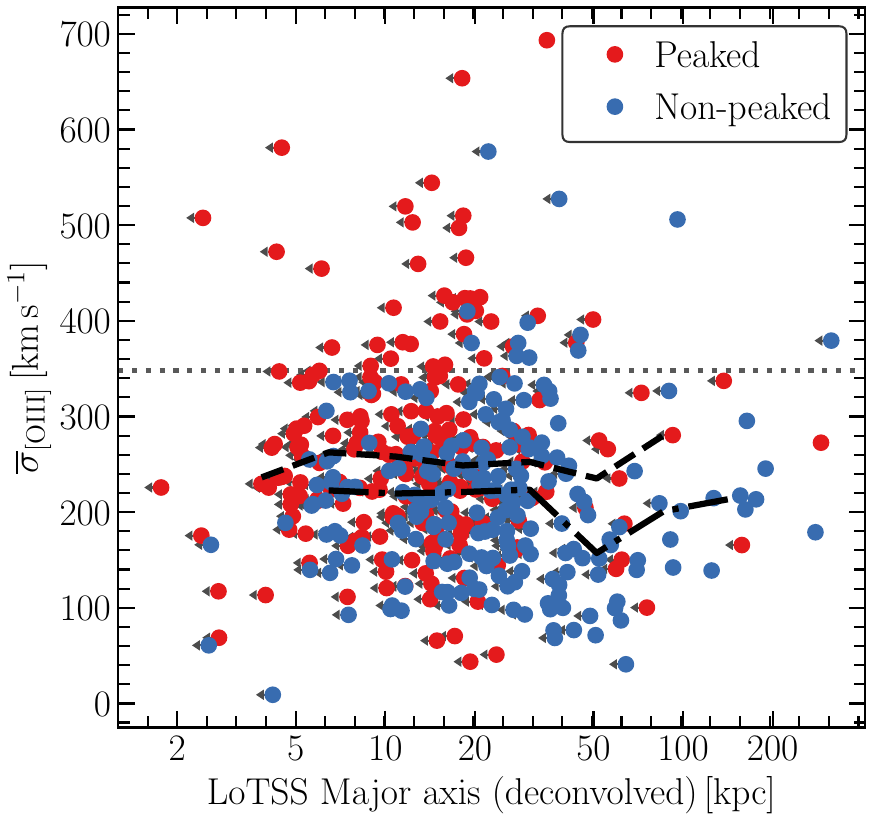}
      \caption{$0.02<z<0.4$}
    \end{subfigure}\hspace{2cm}
    \begin{subfigure}{0.65\columnwidth}
      \includegraphics[width=\columnwidth]{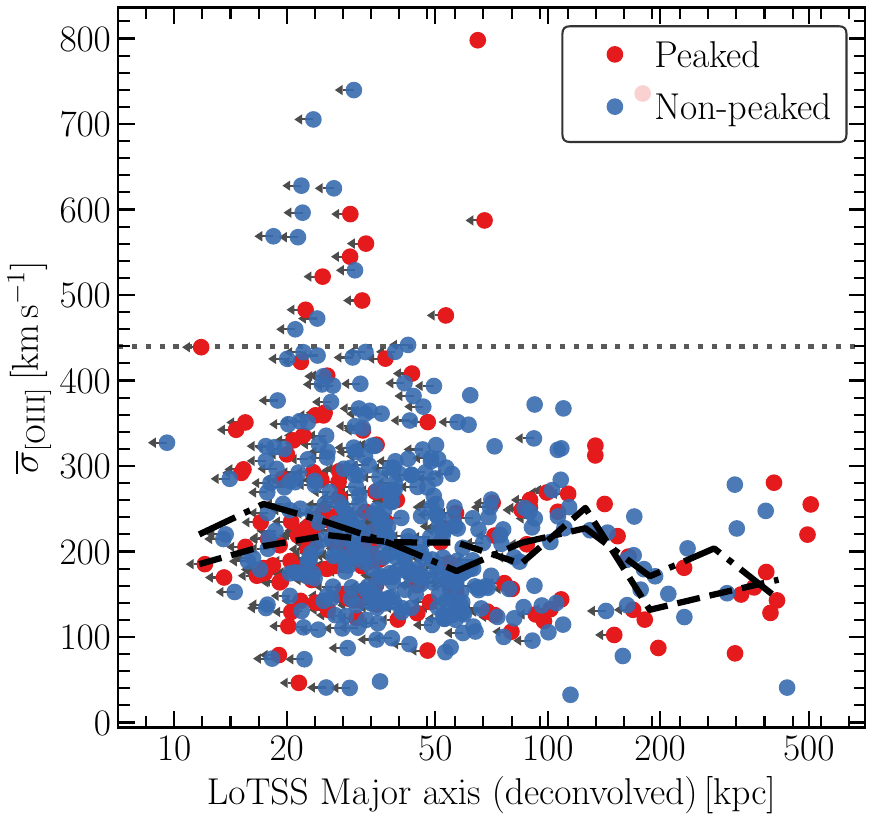}
      \caption{$0.4<z<0.8$}
    \end{subfigure}
    \caption{LoTSS major axis physical sizes (deconvolved) vs $\overline{\sigma}_{\textrm{\OIII}}$ for the \textbf{(a)} low redshift  and \textbf{(b)} high redshift sample. The different colours show the peaked and non-peaked spectrum sources. Sources that are unresolved in LoTSS are marked with a horizontal arrow since the sizes are upper limits in these cases. Dashed and dashdotted lines show the median values for the peaked and non-peaked sources respectively, in bins of major axis sizes. The dotted horizontal line marks the $\overline{\sigma}_{\textrm{\OIII}}$ threshold for classifying the \OIII in a source as disturbed. }
    \label{size_vdisp}
\end{figure*}
In the sections so far, we have found significant evidence to show that \OIII gas in PS sources is more disturbed than NPS sources in the low redshift sample, even while controlling for $L_\mathrm{1.4GHz}$ and $L_{\textrm{\OIII}}$. This supports the picture of young radio AGN having a stronger impact on the surrounding \OIII gas than their evolved counterparts. However, as discussed in Sect.~\ref{spectral properties of radio agn}, PS sources are made up of HFPS sources, which would be the youngest in the sample, and LFPS sources which would be "less" young. Although we make a sharp distinction between HFPS and LFPS sources here, it is worth noting that the peaked sources have a continuous distribution in the colour-colour plot of Fig.~\ref{colcolplot}. The distinction used here is only to make bins for stacking. If the impact on \OIII kinematics in radio AGN is linked to their evolutionary stage, in principle we should be able to trace the differences in the \OIII profiles of HFPS, LFPS and NPS sources in our low redshift sample. Since the number of HFPS and LFPS sources is small, 105 and 112 respectively, we have used a stacking analysis to test whether we can recover the difference in the average \OIII profiles of these groups. \par

In the stacking analysis of the previous section, we find that the profiles show significant differences in the $L_\mathrm{1.4GHz}$ range of $10^{24}-10^{26}$\,\whz. Therefore, we have restricted the sample to this luminosity range. We also exclude the four strong PS sources mentioned in the section above, out of which one is HFPS and three are LFPS sources. The stacked profiles are shown in Fig.~\ref{stacked_specshape_3groups} and the best-fit parameters and bin sizes are summarised in Table~\ref{model fit}. We observe a clear difference in the stacked profiles of the three groups. HFPS sources show the broadest profile at the base, on both blue and red sides, and a broad component FWHM of 1642\,\kms. The profiles then get narrower gradually as we move to LFPS and NPS sources, which have a broad component FWHM of 1333\,\kms and 1145\,\kms, respectively. This shows that the average \OIII profile of our sources gets narrower as we go from the youngest to the most evolved sources, showing the changing impact on \OIII gas. It also shows that it is possible to trace the changing impact on the average \OIII profiles, using a radio spectral shape classification. Similar to the stacking analysis in the section above, including the \OIII non-detections while stacking did not affect our results. \par

\subsection{Radio sizes}
\label{radio sizes}

In this section, we investigate the relation between $\overline{\sigma}_{\textrm{\OIII}}$ and the total radio sizes (physical) from LoTSS, plotted in Fig.~\ref{size_vdisp} for the low and high redshift samples. We do not observe any clear trend between the two quantities. In the low redshift sample, 30 out of the 256 sources (11.7$^{+2.3}_{-1.7}$\,\%) smaller than 20\,kpc and 19 out of the 167 sources (11.4$^{+3.0}_{-2.0}$\,\%) larger than 20\,kpc are disturbed. Similarly in the high redshift sample, 5 out of the 50 sources (10.0$^{+6.0}_{-2.9}$\,\%) smaller than 20\,kpc and 54 out of the 472 sources (11.4$^{+1.7}_{-1.3}$\,\%) larger than 20\,kpc are disturbed. However, it is interesting that the median $\overline{\sigma}_{\textrm{\OIII}}$ at different sizes is consistently larger for PS sources than NPS sources in the low redshift sample, as shown by the dashed and dashdotted lines in Fig.~\ref{size_vdisp}. This shows that irrespective of the radio size, a PS source at the centre typically has more disturbed \OIII gas than an NPS source. We discuss the role of radio sizes and morphology further in Sect.~\ref{sizes discussion}. \par

At this stage, we remind the reader that the radio spectral shape is determined using only the peak flux densities in the central region of the sources, therefore PS or NPS does not classify the spectrum over the entire extent of the radio source. This method allows us to identify a PS at the centre of large sources, which could be candidates for restarted radio AGN. We identify 138 out of 850 PS sources up to $z=0.8$ that have extended emission in their LoTSS images on scales >\,50\,kpc. This is done by visual inspection of the LoTSS images and also includes the \OIII non-detections. Since the young source is likely $\sim$\,$0.1-1$\,Myr old, this phase of activity could not have formed the extended emission observed on >\,50\,kpc scales. This suggests that the PS source and the extended emission represent two distinct phases of activity, and these sources are good restarted candidates. This sample could be useful to study the cumulative impact of AGN feedback on the host galaxies over multiple epochs of activity and will be followed up in detail in a future project.

\subsection{Ionised gas densities of PS and NPS sources}
\label{gas properties}

Although the PS and NPS host galaxies are similarly massive, the difference in the \OIII kinematics of these sources could also be due to a richer medium available for interaction with the young jets in PS sources. To test this, we estimated the electron densities for these sources using the \Sii$\mathrm{\lambdaup\lambdaup}$6717,30\,\AA~doublet method. For sources that have a $\geq$3\,$\sigma$ \Sii~detection, the 6717 to 6730\,\AA~line ratio is related to the electron density as \citep{Sanders2015}:

\begin{ceqn}
\begin{align}
      n_{e}(R) = \frac{cR-ab}{a-R},
\end{align}
\end{ceqn}

where $n_{e}$ is the electron density, $R$ is the flux ratio of the 6717 to 6730\,\AA~\Sii~line, $a=0.4315$, $b=2,107$ and $c=627.1$. These values are estimated for a gas with an electron temperature of 10\,000\,K. Since this relation saturates at electron densities below $\sim$\,10\,cm$^{-3}$ ($R<1.3$) and above $\sim$\,10\,000\,cm$^{-3}$ ($R>0.45$), we only used sources with $0.45<R<1.3$, to stay within the theoretical line ratio limits of this relation. We note that this method has several limitations as it assumes a fully ionised medium whereas the gas could be partially ionised. It is also not sensitive to high-density gas (see a discussion in \citealt{Davies2020IonizedRates}). Studies using auroral and transauroral lines have found significantly higher densities of the ionised gas than those estimated using the \Sii$\mathrm{\lambda\lambda}$6716,6731\,\AA~ratio \citep{Holt2011, Santoro2020, Rose2018, Davies2020IonizedRates}.\par
\begin{figure}
\centering
  \includegraphics[width=0.55\columnwidth]{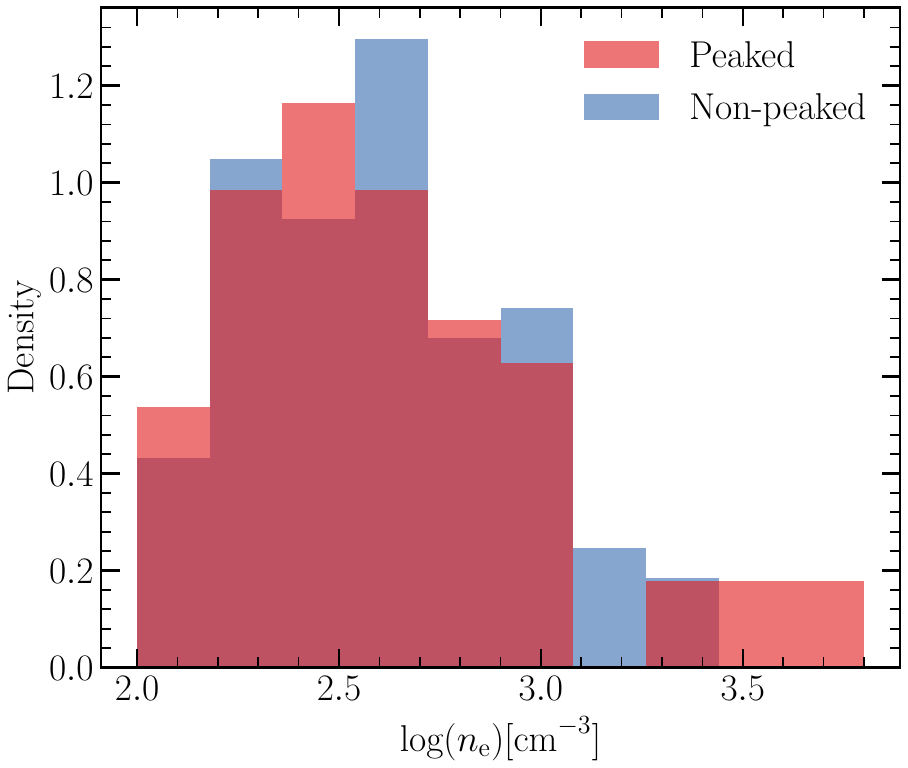}
  \caption{Electron density distributions for PS and NPS sources, that have $\geq$\,3\,$\sigma$ \Sii detections.}
  \label{gasprop}    
\end{figure}

 
Out of the 566 \Sii~detections in the radio AGN sample, electron densities can be estimated for 301 sources using this method, out of which 66 are PS and 90 are NPS sources up to $z\approx0.5$. In Fig.~\ref{gasprop}, we plot distributions of the estimated electron densities. We find no significant difference in the PS and NPS distributions, confirmed by a 2-sample KS test (statistic=0.14, p value=0.40). Although the densities could only be estimated for a subset of the PS and NPS sources, it gives some insight into the ionised gas medium. The lack of any difference between the two suggests that the ionised medium is similarly dense in both groups. We discuss this in the context of jet-driven feedback in the next section.\par

 \section{Discussion}
\label{discussion}

In this paper, we have investigated the \OIII gas kinematics for a large sample of 5\,720 radio AGN, over a $L_\mathrm{1.4GHz}$ range of $10^{22.5}-10^{28}$\,\whz, out to redshift $z=0.8$. Our careful selection of radio AGN allows us to probe the feedback from jets on the \OIII gas kinematics. Using the spectral shape over a wide frequency range ($144-3000$\,MHz) we have separated young and evolved radio AGN, making it possible to study their impact in detail. Below, we discuss our results from Sect.~\ref{results} in the context of jet-driven feedback in radio AGN host galaxies.

\subsection{Impact of jets over the radio AGN life cycle}
\label{specshape discussion}
In our analysis, we have been able to disentangle the effects of radio jets and radiation on disturbing the ionised gas. Comparison between the widths of \OIII profiles of our radio AGN sample and the control sample in Fig.~\ref{cumdist_lowz_all} shows that the presence of jets in AGN host galaxies causes significantly more disturbed \OIII gas. Evidence for a positive correlation between the presence of radio emission and disturbed \OIII kinematics has been found before, for instance, by \citet{Mullaney2013} and recently by \citealt{Escott2024UnveilingEmission}. However, the radio emission in their sources could have different origins: star formation, shocks, or low-power jets. On the other hand, our sample is selected to have radio emission dominated by jets since we study jet-driven feedback on \OIII. \par

With the spectral shape classification, we were able to go one step further and study the changes in the impact on \OIII at different evolutionary stages of radio AGN. This link between the radio spectral shape and the \OIII kinematics is the most important result of this paper. In Sect.~\ref{specshape}, we found that \OIII in PS sources is $\sim$\,3 times more likely to be disturbed than NPS sources in the low redshift sample, even though both groups cover similar ranges of host galaxy properties. Since the radio emission in our sample is dominated by jets, this gives insight into the link between jets and \OIII kinematics. Our results show that the impact of radio jets on surrounding \OIII gas changes as they grow, with the strongest impact on the gas kinematics occurring when they are young. At evolved stages, \OIII is less disturbed. 
These trends support previous evidence in this context by K23, now confirmed with a sample that is $\sim$\,44 times larger. Since the young phase of activity typically lasts for $\sim$\,$0.001-1$ Myr (see Sect.~\ref{introduction}), this suggests that the impact of radio AGN on the warm ionised gas is most extreme in the initial stages, which are a small fraction of their total lifetime ($\sim$\,$10-100$ Myr). At later stages of evolution, the impact on the ionised gas is more `gentle'.\par

In the high redshift ($z>0.4$) sample, we find only marginal evidence for more disturbed \OIII in PS sources ($\sim$\,2 times more likely). Although less significant ($2\sigma$), the observed trend is still in agreement with the low redshift sample. We propose that this is likely due to contamination of the NPS sources by PS sources in the high redshift sample, as mentioned in Sect.~\ref{spectral properties of radio agn}. This contamination would lead to broader \OIII in NPS sources at high redshift. This can be seen in Fig.~\ref{cmldist_specshape}, where see that the fraction of sources with $\overline{\sigma}_{\textrm{\OIII}}>350$\,\kms is larger for NPS sources in the high redshift than the low redshift sample.\par

Radio and \OIII luminosity of an AGN are crucial parameters to determine the energy output of the system which couples with the gas in the host galaxy. Although PS and NPS sources in our sample cover a wide range in $L_{\textrm{\OIII}}$ and $L_\mathrm{1.4GHz}$ (see Fig.~\ref{sample_properties_distribution}), we find that controlling for these parameters does not affect our results. At any $L_{\textrm{\OIII}}$ or $L_\mathrm{1.4GHz}$, PS sources show wider and larger fraction of disturbed \OIII profiles than NPS sources. This is most evident in the low redshift sample. This shows that the difference we find in \OIII profiles of PS and NPS sources is not driven by source luminosities, but is indeed a result of the changing impact of jets in radio AGN along their life cycle. \par 

Controlling for $L_{\textrm{\OIII}}$ in the stacking analysis provides insight into the role of radiation in our sample (Fig.~\ref{stacked_radlumbins}). We find that above $L_{\textrm{\OIII}}=10^{40}$\ergpers, \OIII luminosity does not play a significant role in determining the difference between widths of stacked profiles of PS and NPS sources. PS sources are consistently broader than NPS in different $L_{\textrm{\OIII}}$ groups. However, below $10^{40}$\ergpers, we do not observe any difference between PS and NPS stacked profiles, likely due to the low S/N of the spectra. The strength of radiation is likely not significant for determining the difference in the impact of PS and NPS sources. \par

Similarly, from the stacked profiles in Fig.~\ref{stacked_radlumbins}, we find that the 'difference' between the profile widths of PS and NPS sources, increases with $L_\mathrm{1.4GHz}$. That is, there is a positive link between the varying impact of radio AGN and the 1.4\,GHz luminosity. Since $L_\mathrm{1.4GHz}$ is a proxy for jet power \citep{Cavagnolo2010,McNamara2012,Hardcastle2018}, this points towards young and powerful jets being more effective at pushing the ionised gas clouds in their path to faster velocities. This is broadly in agreement with the results of some jet-ISM interaction simulations which predict that powerful jets would drill through the surrounding medium with ease while driving fast outflows \citep{Mukherjee2016,Mukherjee2018}. These simulations also indicate that low-power jets, on the other hand, would struggle to break through the ISM and dissipate their energy by inducing turbulence in the ambient medium. It is not possible to make a direct comparison of these simulations with our results, and that requires spatially resolved optical spectra. However, the trends we observe for the sample appear to agree with these predictions.  \par

The larger sample in this study compared to K23 means that we can also trace the changing impact in different regions of the colour-colour plot in Fig.~\ref{colcolplot}. Our stacked profiles of HFPS, LFPS and NPS sources (see Fig.~\ref{stacked_specshape_3groups}) show a gradual narrowing as we move from the youngest to the "less" young and the most evolved source. This can also be observed by tracking the change in the broad component FWHM of the stacked profiles. On average, within the PS source group, the impact of HFPS sources on \OIII is stronger than LFPS sources. This further supports our results on the varying impact on \OIII over the radio AGN life cycle. Such a detailed tracing of the impact at different stages of the life cycle has been possible due to the addition of low-frequency data (144\,MHz) with LoTSS, which allowed the detection of a large number of PS sources. Although evidence for disturbed ionised gas has been found in powerful young AGN before ($L_\mathrm{1.4GHz}\gtrsim$\,$10^{26}$\whz, e.g. \citealt{Gelderman1994,Holt2008,Holt2011,DeVries2009,Santoro2020}), our study expands this to a much larger sample of typical radio AGN, down to $L_\mathrm{1.4GHz}\sim$\,$10^{23}$\,\whz. We find that the impact on ionised gas and the difference between young and evolved radio AGN is significant even down to $L_\mathrm{1.4GHz}=10^{24}$\,\whz~(confirming the findings for single objects like \citealt{Murthy2022,Oosterloo2017,Audibert2019,Ruffa2022}). \par

We further check that the differences we observe in \OIII profiles of PS and NPS sources are not due to any selection effects. PS sources are selected based on absorption in their radio spectra. Other than synchrotron self-absorption, this could be caused by free-free absorption from a highly dense ionised medium surrounding the radio jets. Therefore PS sources could also have more high-density ionised gas. AGN with such a dense medium in the central region would have a larger possibility of jet-ISM interactions. They would be more likely to show prominent broad wings in their spectra than NPS sources. However, as we show in Fig.~\ref{gasprop} for a subset of sources, the ionised gas densities for PS and NPS sources are similar. Therefore, the result of \OIII kinematics with radio spectral shape is not due to a selection effect, but due to genuine differences in their interaction with the ionised gas. \par

\subsection{Disturbed \OIII gas in NPS sources}

It is intriguing that NPS sources consistently show less disturbed \OIII gas than PS sources, even though the jets are still active. Since the disturbed gas is likely driven by shocks due to jet-ISM interaction \citep{Sutherland2007,Mukherjee2018}, one possibility is that by the time jets have grown to larger scales, the shocked gas in the central region has cooled down, and is not as disturbed as before. In this case, the strength of the broad component could be reduced, leading to narrower observer profiles in NPS sources. Indeed, using hydrodynamical simulations of jet-ISM interaction, \citet{Meenakshi2022c} found that as the jets start and propagate, they shock ionise the clouds, pushing them to high velocities, and clear the region which is then photoionised by the AGN radiation. In their scenario, once the jets have evolved to larger scales and their direct interaction with the gas clouds has stopped, the shocked gas clouds would cool down. Another possibility is that on the time scale of the NPS sources (a few to tens of Myr), the disturbed \OIII gas clouds have been pushed by jet-ISM interaction to distances beyond the coverage of the 3\arcsec SDSS fibre. In that case, as well, the gas within the SDSS fibre would appear less disturbed. \par

Determining the exact nature of the disturbed gas in NPS sources is beyond the scope of this paper. Testing these possibilities would require high-resolution spatially resolved optical spectra. Indeed, spatially resolved molecular gas kinematics has provided evidence for changing effects of jet-ISM interaction as the jets expand in radio-loud (for example PKS\,0023-26 by \citealt{Morganti2021a}) and radio-quiet (for example \citealt{Girdhar2024QuasarQuasars}) AGN (see \citealt{Morganti2023YoungImpact} for an overview). We are currently performing a study with integral field unit (IFU) data for a subset of these sources to examine the size and shape of the disturbed \OIII gas emission in more detail and understand its nature. 

 \begin{figure*}
    \centering
    \begin{subfigure}{0.75\columnwidth}
      \includegraphics[width=1\columnwidth]{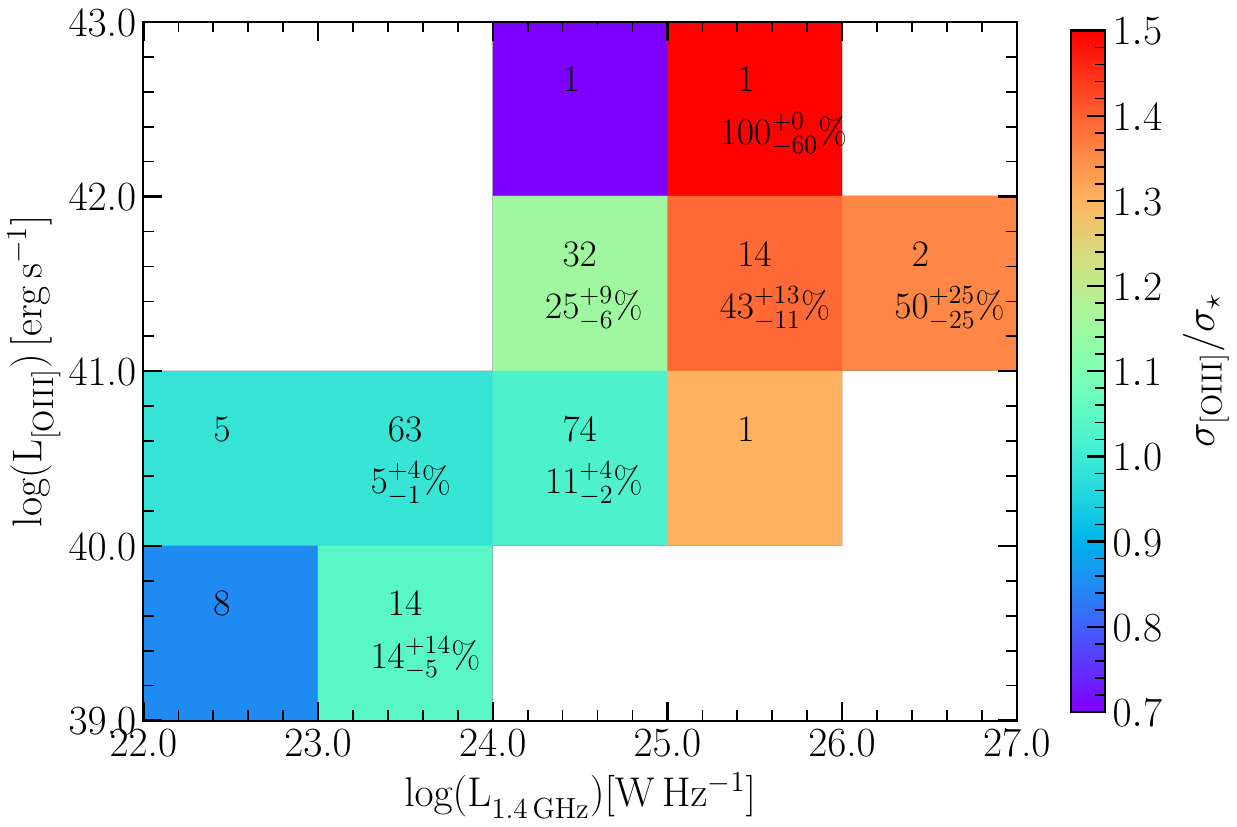}
      \caption{Peaked ($0.02<z<0.4$)}
      \label{peaked_normalised}
    \end{subfigure}
    \hspace{0.7cm}
    \begin{subfigure}{0.75\columnwidth}
      \includegraphics[width=1\columnwidth]{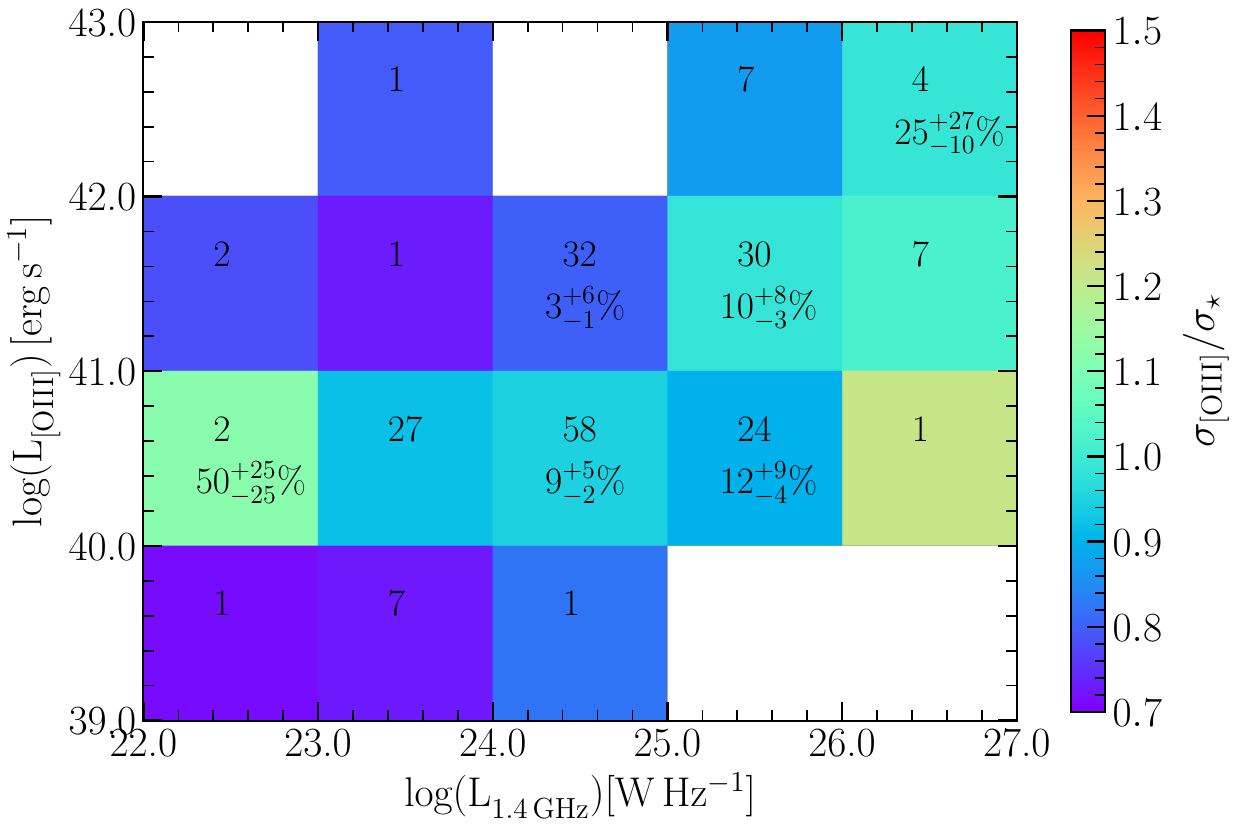}
      \caption{Non-peaked ($0.02<z<0.4$)}
      \label{nonpeaked_normalised}
    \end{subfigure}\\
    \vspace{0.2cm}
    \begin{subfigure}{0.6\columnwidth}
      \includegraphics[width=\columnwidth]{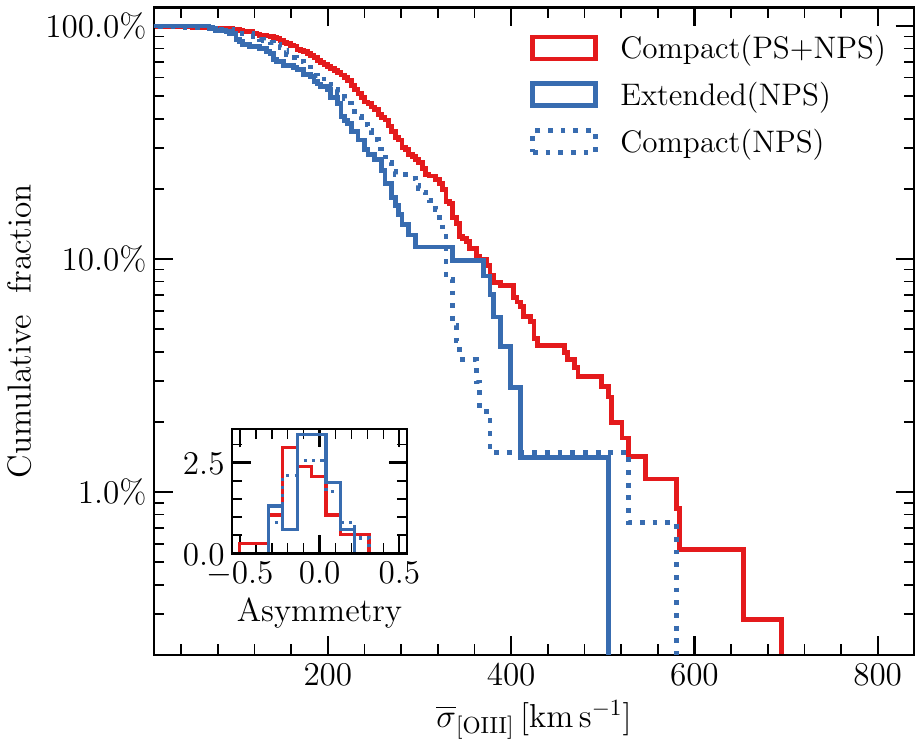}
      \caption{FIRST compact and extended}
      \label{firstsizes}
    \end{subfigure}
    \hspace{1.7cm}
    \begin{subfigure}{0.6\columnwidth}
      \includegraphics[width=\columnwidth]{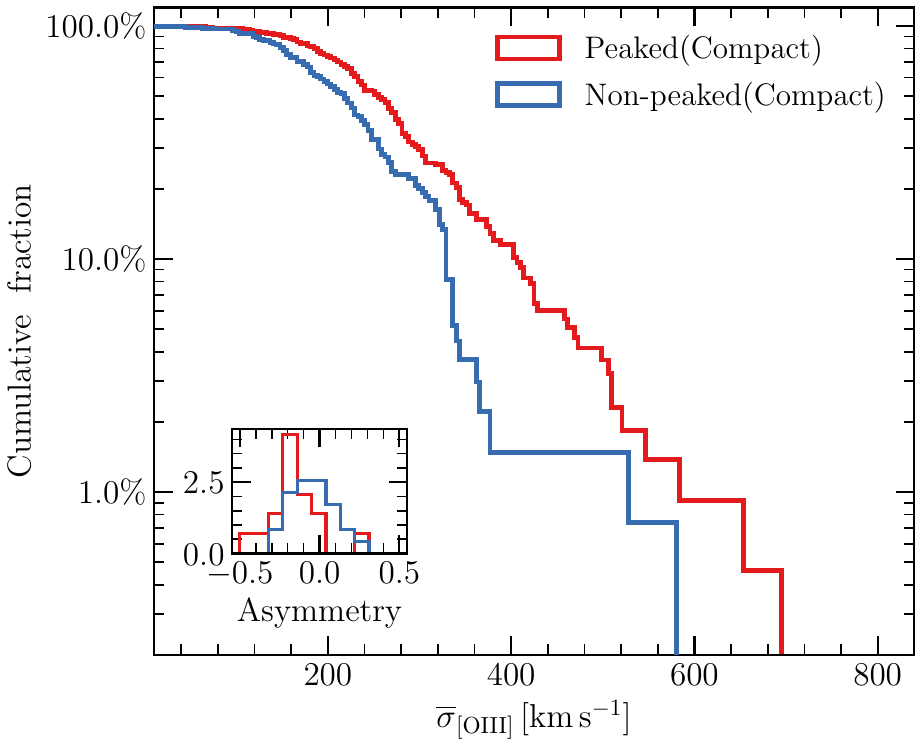}
      \caption{FIRST compact PS+NPS sources}
      \label{firstcompact}      
      
    \end{subfigure}
    \caption{Distributions of $\sigma_{\textrm{\OIII}}/\sigma_{\star}$ and $\overline{\sigma}_{\textrm{\OIII}}$. \textbf{(a) \& (b)} Colour map showing the average $\sigma_{\textrm{\OIII}}/\sigma_{\star}$ values for peaked and non-peaked sources in the low redshift sample, in bins of $L_\mathrm{1.4GHz}$ and $L_{\textrm{\OIII}}$, similar to Fig.~\ref{lowz_radoptlum}. \textbf{(c)} Cumulative distributions of the average \OIII velocity dispersions. Comparison for compact ($<3\arcsec$) and extended ($>3\arcsec$) sources according to FIRST sizes in the low redshift sample. We also plot the compact NPS sources for comparison. The inset plot shows the asymmetry distributions for the \OIII profiles with multiple kinematic components. \textbf{(d)} Cumulative distributions for low redshift PS and NPS sources that are compact in FIRST.}
    \label{cmldist_firstsize}
 \end{figure*}

\subsection{Role of $L_\mathrm{1.4GHz}$ and $L_{\textrm{\OIII}}$}
\label{luminosities discussion}

In Sect.~\ref{radiolumvsoiii}, we find that the width and fraction of disturbed \OIII profiles increase with both $L_\mathrm{1.4GHz}$ and $L_{\textrm{\OIII}}$. This trend is most evident for the low redshift sample, as shown in Fig.~\ref{lowz_radoptlum}. With our approach of separating the PS and NPS sources, we can observe an even stronger trend between $L_\mathrm{1.4GHz}$ and \OIII profile widths. This can be seen for the low redshift sample in Fig.~\ref{radoptlum_specshape}. Therefore  $L_\mathrm{1.4GHz}$, and hence the radio jet power, plays an important role in determining the impact on \OIII kinematics. The highest luminosity radio AGN have the most disturbed gas. However, this trend is not very clearly observed in the high redshift sample.\par

It is worth noting that some studies of AGN samples have found slightly different trends between $L_\mathrm{1.4GHz}$ and \OIII profile widths. For instance, \citet{Mullaney2013} found that the average widths of \OIII profiles were largest between $L_\mathrm{1.4GHz}=10^{23}-10^{25}$\,\whz, and therefore \OIII was less disturbed at luminosities below and above this range. Our analysis differs in the sample selection criteria, as mentioned before. Therefore, our sample is more suited for studying the role of radio luminosity in jet-driven feedback. Our additional selection criteria of spectral shape (PS and NPS sources) means that our sample is also not contaminated by flat-spectrum sources, which could affect the results from \citet{Mullaney2013} at $L_\mathrm{1.4GHz}>$\,10$^{25}$\,\whz.\par

Previous studies have reported a correlation between the \OIII and stellar velocity dispersions ($\sigma_{\textrm{\OIII}}$ vs $\sigma_{\star}$), suggesting that the \OIII gas kinematics are determined by the bulge gravitational potential (for example \citealt{Nelson1996StellarBulge,Boroson2002DoesAGN,Sexton2020BayesianProperties}). Although there is a large scatter in this correlation, it is still important to be cautious about the role of the bulge in determining \OIII kinematics \citep{Bennert2018Studying,Le2023OAGNs}. Based on this correlation, some studies have used $\sigma_{\star}$ to account for the gravitational kinematics component in $\sigma_{\textrm{\OIII}}$. For instance, \citet{Woo2016} have used the quantity $\sigma_{\textrm{\OIII}}/\sigma_{\star}$ to study the impact of AGN feedback for a sample of 39\,000 type 2 AGN out to $z$\,$\sim$\,0.3. They find that the correlation between $L_\mathrm{1.4GHz}$ and $\sigma_{\textrm{\OIII}}$ vanishes when $\sigma_{\textrm{\OIII}}$ is normalised using   $\sigma_{\star}$. Using similar techniques for a sample of $\sim$\,$6000$ AGN out to $z$\,$\sim$\,0.4, \citet{Ayubinia2023InvestigatingAGNs} have found the same result between $L_\mathrm{1.4GHz}$ and $\sigma_{\textrm{\OIII}}/\sigma_{\star}$. They also do not find any dependence of $\sigma_{\textrm{\OIII}}/\sigma_{\star}$ on the radio-loud/radio-quiet, compact/extended groups of radio AGN. These results have suggested that radio AGN have a limited impact on \OIII kinematics, and radio AGN-driven outflows are much weaker than accretion-driven outflows (although radio AGN can provide an extra boost to the disturbed \OIII gas). \par

For comparison, we test our results for PS and NPS sources in the low redshift sample using $\sigma_{\textrm{\OIII}}/\sigma_{\star}$, as shown in Fig.~\ref{peaked_normalised} and \ref{nonpeaked_normalised}. We again find that PS sources have consistently more disturbed \OIII kinematics. In this case, we define \OIII to be disturbed if $\sigma_{\textrm{\OIII}}/\sigma_{\star}>1.5$, however the specific value of this limit does not affect our broad result. This shows that there is an intrinsic difference in the \OIII kinematics of PS and NPS sources. We also note that using this approach to select disturbed \OIII sources does not affect our results. The fraction of disturbed sources at any source luminosity is in agreement with the result presented in Fig.~\ref{radoptlum_specshape}. Another important result is the dependence of this ratio on $L_\mathrm{1.4GHz}$ and $L_{\textrm{\OIII}}$. We find that PS (and to some extent NPS) sources with higher luminosities have larger $\sigma_{\textrm{\OIII}}/\sigma_{\star}$. This also supports our result for the dependence of \OIII kinematics on source luminosity. However, we did not use the $\sigma_{\textrm{\OIII}}/\sigma_{\star}$ approach for the entire paper because the large scatter between $\sigma_{\textrm{\OIII}}$ and $\sigma_{\star}$ makes it less reliable. The difference in the relation between $\sigma_{\textrm{\OIII}}/\sigma_{\star}$ and $L_\mathrm{1.4GHz}$ for our sample compared to \citet{Woo2016} can be attributed to sample selection effects, as they study type 2 AGN. Although \citet{Ayubinia2023InvestigatingAGNs} also study radio AGN, most of their sources lie below $L_\mathrm{1.4GHz}=10^{24}$\whz whereas most of our sources lie above it. Indeed, in Fig.~\ref{peaked_normalised} and \ref{nonpeaked_normalised}, we do not observe any significant dependence on radio luminosity below this limit. Our separation of PS and NPS sources also helps emphasise the role of $L_\mathrm{1.4GHz}$.\par

From Fig.~\ref{lowz_radoptlum}, we also conclude that increasing only $L_{\textrm{\OIII}}$ (which is proxy for the total radiative power of an AGN;  \citealt{Heckman2004}) also increases the \OIII profile widths and fraction of disturbed sources. Most of the radio AGN are low excitation radio galaxies (LERGs), i.e. systems with radiatively inefficient accretion and low accretion rates \citep{Best2012}. Our results suggest that even in such systems, radiation strength plays an important role in driving feedback on \OIII gas kinematics. For a sample of $\sim$\,800 LERGs up to $z$\,$\sim$\,0.3, \citet{Singha2021} also found that the majority of sources with an \OIII outflow (only $\sim$\,1.5\% of their sample) had high accretion rates. They concluded that outflows in LERGs are radiative pressure driven. We note a crucial difference between our studies regarding the identification of disturbed \OIII. \citet{Singha2021} focus only on the outflow detections, which are the most extreme form of feedback, whereas we study the average velocity dispersion for the entire radio AGN sample. However, there seems to be a general agreement between the results on the role of radiation pressure. \par

\subsection{\OIII and radio source sizes}
\label{sizes discussion}

Finally, we also studied the dependence of $\overline{\sigma}_{\textrm{\OIII}}$ on the radio sizes from LoTSS in Sect.~\ref{radio sizes}, and found no significant relation between the two. We find that sources larger and smaller than 20\,kpc in LoTSS are equally likely to have disturbed \OIII in both low and high redshift samples. However, it is interesting that we find larger $\overline{\sigma}_{\textrm{\OIII}}$ values for PS than NPS sources, at all sizes in the low redshift sample (Fig.~\ref{size_vdisp}). This suggests that in radio AGN host galaxies, the impact on \OIII is not determined by the extent of the radio emission, but by the emission properties in the central region. \par

Many studies in literature have found links between the radio morphology of AGN and the multiphase gas kinematics. In neutral gas, \Hi~absorption studies have found broader and more asymmetric line profiles in compact radio sources compared to extended sources (for e.g. \citealt{Gereb2014,Gereb2015,Maccagni2017a}). Since young radio AGN usually have a compact radio morphology, this suggested the presence of more disturbed \Hi~gas in young radio AGN. In the ionised phase, \citet{Mullaney2013} have found evidence for more disturbed \OIII in AGN with compact radio morphology. \citet{Molyneux2019} use the type 1 and 2 AGN sample up to $z\sim0.2$ from \citet{Mullaney2013}, and also find more disturbed \OIII in compact radio emission sources. They conclude that the SDSS spectra of their AGN are twice more likely to have broad \OIII profiles (FWHM\,>\,1000\,\kms) if the radio emission is confined within 3\arcsec in their FIRST images. It is worth noting that the typical radio luminosity of their sources is lower by $\sim$\,2 orders of magnitude compared to ours, as they do not select for radio AGN. The radio emission in their sources can also be due to AGN-driven winds that drive outflows and shock the ISM to produce radio emission. This could also explain the correlation between radio size and the prevalence of outflows. Another major difference in our study is the way we select young radio AGN, using a spectral shape classification down to low frequencies of 144\,MHz (thanks to the addition of LoTSS data) instead of source extent. This allows us to identify young phases of activity even at the centres of extended sources. \par

Still, it is worth comparing their results with our low redshift sample in the context of radio morphology and spectral shape. For this, we split our low redshift sample using the FIRST deconvolved major axis sizes into compact ($<3\arcsec$) and extended ($>3\arcsec$) sources. It is important to note from Fig.~\ref{sizecomp} that all the PS sources in the low redshift sample are smaller than $3\arcsec$ in FIRST. Therefore the compact group consists of both PS and NPS sources but the extended group consists only of NPS sources. We compare their $\overline{\sigma}_{\textrm{\OIII}}$ distributions in Fig.~\ref{firstsizes} and find that they are significantly different (KS statistic = 0.22, p-value = 4.8$\times$10$^{-3}$). We also find that the fraction of sources with FWHM\,>\,1000\,\kms is 4.5$^{+1.4}_{-0.9}$\,\% for compact and 1.4$^{+3.1}_{-0.4}$\,\% for extended FIRST sources. This is in agreement with the results of \citet{Molyneux2019}.\par

However, we propose that this result in our sample is driven by the PS sources present in the compact group. We test for this by controlling for spectral shape and only using the NPS sources for the compact group, shown in Fig.~\ref{firstsizes}. We find that the distributions of compact and extended sources are now not significantly different (KS statistic = 0.13, p-value = 0.35). The fraction of sources with FWHM\,>\,1000\,\kms also comes down to 1.5$^{+1.9}_{-0.5}$\,\% for compact sources. Therefore, controlling for spectral shape removes the difference between the \OIII profile widths of compact and extended FIRST sources. This shows that the difference found before is indeed driven by PS sources. Further, in Fig.~\ref{firstcompact} we even compare the $\overline{\sigma}_{\textrm{\OIII}}$ distributions of only compact PS and NPS sources, and find that they are still significantly different (KS statistic = 0.20, p-value = 1.6$\times$10$^{-3}$). We conclude that controlling for FIRST sizes does not affect our results on the \OIII profiles of PS and NPS sources.\par

Therefore, we find that the \OIII kinematics in our radio AGN sample is not significantly dependent on the FIRST or LoTSS sizes. Instead, we consistently find that the \OIII kinematics is most strongly related to the radio spectral shape. We conclude that young radio AGN picked using their peaked radio spectrum are more likely to show disturbed \OIII profiles than evolved radio AGN with a non-peaked spectrum.

\section{Summary and conclusions}
\label{summary}
In this study, we have selected a sample of 5\,720 radio AGN up to $z$\,$\sim$\,0.8 where the radio emission is dominated by jets. We have shown that the presence of radio jets in AGN disturbs the \OIII gas more compared to AGN without jets. Using a wide-frequency radio spectral analysis, we have been able to identify sources in different stages of their life cycle. We have then established a link between the radio spectral shape (proxy for AGN evolutionary stage), and the \OIII kinematics. Our main findings are summarised below:

\begin{enumerate}
    \item \OIII in PS sources is $\sim$\,3 times more likely to be disturbed than in NPS sources at $z<0.4$. This shows that jets in young radio AGN have a stronger impact on the surrounding ionised gas. At $z>0.4$, we find marginal evidence for PS sources having $\sim$\,2 times more disturbed \OIII. 
    \item The fraction of disturbed sources and the average \OIII velocity dispersion is consistently higher for PS sources, even when controlling for 1.4\,GHz and \OIII luminosity. This shows that the feedback is indeed linked to the evolutionary stage of radio AGN. The most extreme impact of radio AGN is likely limited to the initial stages of evolution. 
    \item The difference between the widths of stacked \OIII profiles of PS and NPS sources increases with 1.4\,GHz luminosity. This shows that young and powerful jets have the strongest impact on \OIII gas. 
    \item A lower fraction of disturbed \OIII sources in the NPS group suggests that the shocked gas has cooled down by the time jets evolve to large scales. Although the shocked gas may have been pushed out to distances beyond the coverage of the SDSS fibre. Testing this hypothesis would require studying the spatially resolved gas kinematics.
    \item Using a stacking analysis, we find that the average \OIII profile becomes narrower as the source spectral shape changes from completely inverted, to peaked and finally to non-peaked. We propose that this traces, in detail, the changes in the impact of jets along the radio AGN life cycle.     
    \item We find that the fraction of disturbed \OIII sources and average velocity dispersion increases with both 1.4\,GHz and \OIII luminosity, even while controlling for one of the two. This highlights the role of the total energetic output in driving feedback and shows that in radio AGN, radiation pressure can also add to the effect of jets on ionised gas. 
    \item We do not find any strong dependence of \OIII kinematics on the source sizes. We find that any dependence is driven by PS and NPS sources, and controlling for spectral shape removes it. We suggest that the trends observed between disturbed \OIII and compact radio AGN are driven by PS sources.  
\end{enumerate}

Our study shows that the impact of jets on the host galaxy's gas could change as the jets evolve. Understanding the exact changes in the ionised gas kinematics and morphology over galaxy scales, and the eventual impact on star formation, requires detailed studies of the spatially resolved gas and stellar properties. We plan to perform such studies using IFU data from MaNGA \citep{Bundy2015OverviewObservatory} and the GMOS-N spectrograph on the GEMINI telescope, for a subset of sources from this sample. 

\begin{acknowledgements} 
P.K. acknowledges support through an Emmy Noether Grant (German Research Foundation) of Dominika Wylezalek (WY 179/1-1). LOFAR \citep{vanHaarlem2013} is the Low Frequency Array designed and constructed by
ASTRON. It has observing, data processing, and data storage facilities in several countries,
which are owned by various parties (each with their own funding sources), and are
collectively operated by the ILT foundation under a joint scientific policy. The ILT resources
have benefited from the following recent major funding sources: CNRS-INSU, Observatoire de
Paris and Université d'Orléans, France; BMBF, MIWF-NRW, MPG, Germany; Science
Foundation Ireland (SFI), Department of Business, Enterprise and Innovation (DBEI), Ireland;
NWO, The Netherlands; The Science and Technology Facilities Council, UK; Ministry of
Science and Higher Education, Poland; The Istituto Nazionale di Astrofisica (INAF), Italy. Funding for the Sloan Digital Sky 
Survey IV has been provided by the 
Alfred P. Sloan Foundation, the U.S. 
Department of Energy Office of 
Science, and the Participating 
Institutions. The National Radio Astronomy Observatory is a facility of the National Science Foundation operated under a cooperative agreement by Associated Universities, Inc. CIRADA is funded by a grant from the Canada Foundation for Innovation 2017 Innovation Fund (Project 35999), as well as by the Provinces of Ontario, British Columbia, Alberta, Manitoba and Quebec.
\end{acknowledgements}
  \bibliographystyle{aa-copy} 
  \bibliography{References} 
\onecolumn

\begin{appendix}

\twocolumn
\section{Modelling the stellar continuum and emission line characterisation}
\label{modelling optical spectra}

To obtain the emission line spectra, we first modelled the stellar continuum and removed it from the observed SDSS spectra. Modelling was done using the STARLIGHT spectral synthesis code\footnote{\url{http://www.starlight.ufsc.br/}} \citep{Fernandes2005,Fernandes2009}. Single stellar population models of \citet{Bruzual2003} with young (1, 3.2, 5, 10, 25, 40 and 100 Myr), intermediate (0.29, 0.64, 0.9, 1.43, 2.5 and 5 Gyr) and old (11 and 13 Gyr) age populations were used. We assumed a Chabrier initial mass function, and three different metallicities: 0.004, 0.02 and 0.05 in mass-fraction notation. Before modelling the stellar continuum, the spectra were shifted to rest-frame wavelength, and the following emission lines were masked: \OII$\mathrm{\lambdaup\lambdaup}$3726,28\AA, \hdelta, \hgamma, \hbeta, \OIII$\mathrm{\lambdaup\lambdaup}$4958,5007\AA, \OI$\mathrm{\lambdaup}$6300\,\AA, \Nii $\mathrm{\lambdaup\lambdaup}$6548,84\AA, \halpha~and \Sii$\mathrm{\lambdaup\lambdaup}$6717,30\,\AA. A visual inspection revealed that this procedure could reproduce the stellar continuum emission of the sources well, and can be used for subtraction to obtain the emission line spectra. \par

After obtaining the continuum subtracted spectra, modelling of the line profiles was done using Gaussian functions. For reference, we used the \OIII doublet which is usually the strongest emission line. We used kinematic components made up of two Gaussian functions, with a fixed amplitude ratio ($A_{\textrm{\OIII}4958}=\frac{1}{3}\times A_{\textrm{\OIII}5007}$) and equal widths ($\sigma_{\textrm{\OIII}4958}=\sigma_{\textrm{\OIII}5007}$) according to atomic physics. We started with a single kinematic component and added up to two components depending on whether adding an extra component significantly improved the fit. We judged this using an F-test, and considered an extra component to significantly improve the fit, if it improved the $\chi^{2}$ at >99\% confidence level. We considered a source as an \OIII detection if its amplitude was $\geq3\sigma$, where $\sigma$ is the standard deviation of the residuals after subtracting the best-fit model. We obtained 2\,363 \OIII detections, out of which 410 required double and 120 required triple components.\par

For sources with \OIII detections, we then fit the \hbeta~line. We used the \OIII best-fit model and allowed one more component to account for any possible broad line region (BLR) emission in the \hbeta~line from type 1 AGN. Out of all the \hbeta~detections, 173 sources required an extra BLR component in the \hbeta~line. Similarly using the best-fit \OIII model, we then fit the \Nii$\mathrm{\lambdaup\lambdaup}$6548,84\AA~and the \halpha~lines. The separation between the \Nii~and \halpha~lines, and the relative amplitudes of the two components of the \Nii~line ($A_\mathrm{\Nii 5648}=\frac{1}{3}\times A_\mathrm{\Nii 5684}$) were fixed. We again allowed for an extra broad component in the \halpha~line, whenever one was needed for the \hbeta~line. Lastly, we fit the \Sii$\mathrm{\lambdaup\lambdaup}$6717,30\,\AA~doublet, with a fixed velocity separation and the same velocity dispersion and number of components as the best fit \OIII model. The number of different emission line detections in the full sample and the radio AGN sample selected in the next section, are summarised in Table~\ref{emissionlinesummary}.\par

Next, we estimated the emission line fluxes by adding up fluxes of the Gaussian components in the best-fit model. In the case of \hbeta~and \halpha~line fits that required an extra BLR component, we neglected the flux of the BLR component as we did not want the light from the BLR region to affect our fluxes. For non-detections, we used a $3\sigma$ value to estimate an upper limit on the fluxes.\par

We removed 22 sources from the sample where the spectra were corrupted at the \OIII wavelengths. We further removed 30 sources where the \hbeta~profile required an extra broad component, and the emission was strong enough to leak into the \OIII profile. This contaminates the profile width and flux estimates. Finally, we were left with a sample of 6\,504 sources.\par

\textit{\OIII profile characterisation and stacking:} The multi-component fitting routine makes it challenging to compare \OIII profiles with different number of kinematic components in their best-fit model. To overcome this, we estimated a flux weighted average velocity dispersion $\overline{\sigma}_{\textrm{\OIII}}$, similar to \citet{Mullaney2013}, given as 

\begin{ceqn}
\begin{align}
      \overline{\sigma}_{\textrm{\OIII}} = \left(\sum\limits_{i}(f(i) \times \sigma_{\textrm{\OIII}}(i))^{2}\right)^{\frac{1}{2}},
\end{align}
\end{ceqn}

where $f(i)$ is the flux ratio of the $i$\,th component to the total flux of the line and $\sigma_{\textrm{\OIII}}(i)$ is the velocity dispersion of the $i$\,th component in the best-fit model. Using this one parameter we can compare profiles with different numbers of components. In addition, it is also useful to assess the shape of the line profiles and determine any asymmetry, if present. The asymmetry helps trace kinematically disturbed gas in the sources. We measured the asymmetry of the lines using the definition from \citet{Liu2013}, given as

\begin{ceqn}
\begin{align}
      A = \frac{(v_{90}-v_{50})-(v_{50}-v_{10})}{v_{90}-v_{10}},
\end{align}
\end{ceqn}

where $v_{90}$, $v_{50}$, and $v_{10}$ are the velocities at 90, 50 and 10 per cent of the cumulative flux. In this convention, a heavily blueshifted profile has $A<0$ whereas a heavily redshifted profile has $A>0$ and a symmetric profile has $A=0$. \par

In this paper, we also use a stacking analysis to measure the average differences between \OIII profiles of different source groups. Using a stacking approach, we can control for various source properties and disentangle their role in driving the \OIII kinematics. It also suppresses the noise (by 1/$\sqrt{N}$ for N samples) thus allowing the detection of low-level broad components, which would be otherwise missed in individual fits. Stacking requires accurate spectroscopic redshift measurements to align the rest-frame spectra, which we have for the entire sample. 
We estimated the stacked spectrum using the equation  
\begin{ceqn}
\label{stacking_equation_chap5}
\begin{align}
      S(\lambdaup) = \frac{\sum\limits_{i}\frac{1}{\sigma^{2}(i)}S(i,\lambdaup)}{\sum\limits_{i}\frac{1}{\sigma^{2}(i)}},
\end{align}
\end{ceqn}

where S$(i,\lambdaup)$ is the individual source spectrum, $\sigma(i)$ is the noise in a line-free region of the source spectrum and S($\lambdaup$) is the stacked spectrum. We estimate errors on the average stacked spectrum using a bootstrapping approach. We randomly select sources in each bin to create a bootstrapped sample and stack their spectra, repeating this process 1000 times. We use the standard deviation of the distribution at each wavelength as the error for that wavelength point. \par

We also estimate the impact of errors in redshift on the stacked \OIII profiles. To do this, we add a random value to each redshift, to obtain a redshift with an offset. This value is drawn randomly from a Gaussian distribution with unit area, zero mean and standard deviation equal to the error in reported in SDSS for that redshift. We then re-perform the stacking analysis to test the impact of redshift errors. We find that it has a negligible impact on the final stacked profiles. 

\section{Selecting radio AGN}
\label{selectragn}

\begin{figure*}[!ht]
  \centering
    \begin{subfigure}{0.8\columnwidth}
      \includegraphics[width=\columnwidth]{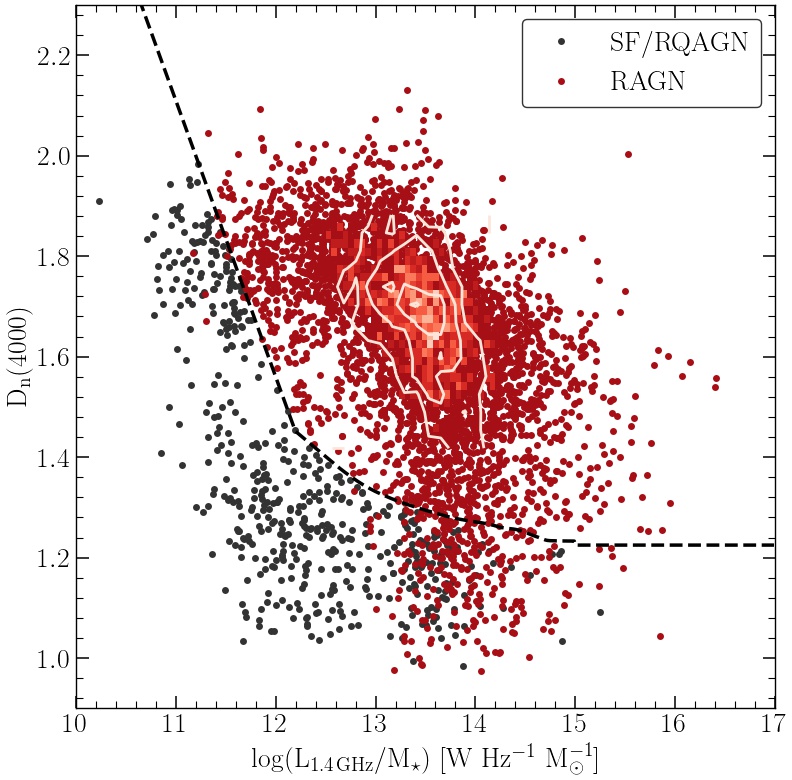}
      \caption{$D_\mathrm{{n}}(4000)$ vs $L_\mathrm{1.4GHz}$/$M_{\star}$}
      \label{d4pm}
    \end{subfigure}
    \hspace{1.5cm}
    \begin{subfigure}{0.8\columnwidth}
      \includegraphics[width=\columnwidth]{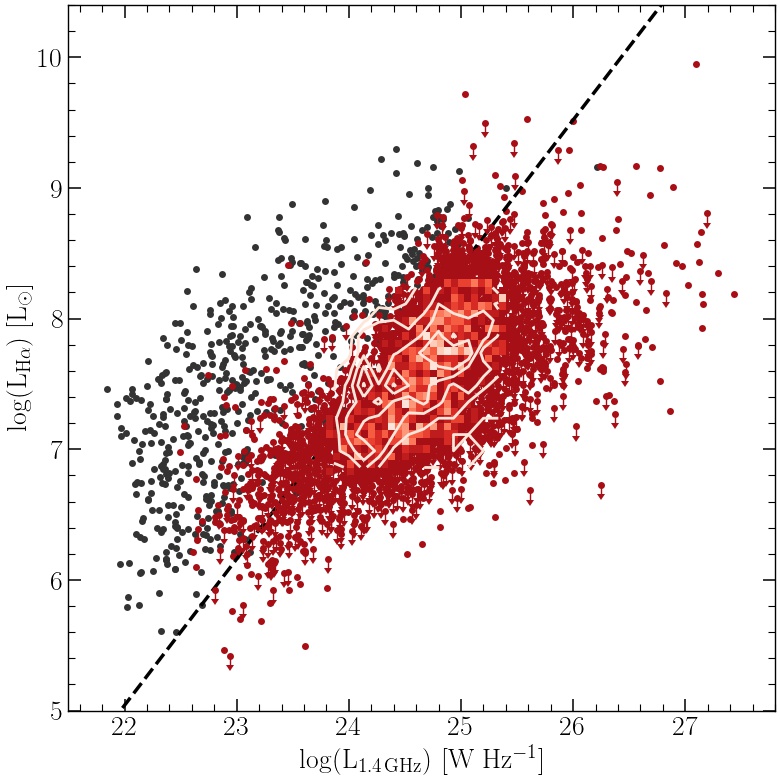}
      \caption{$L_{\textrm{H}\alpha}$ vs $L_\mathrm{1.4GHz}$}
      \label{harad}
    \end{subfigure}\\
    \vspace{0.5cm}
    \begin{subfigure}{0.8\columnwidth}
      \includegraphics[width=\columnwidth]{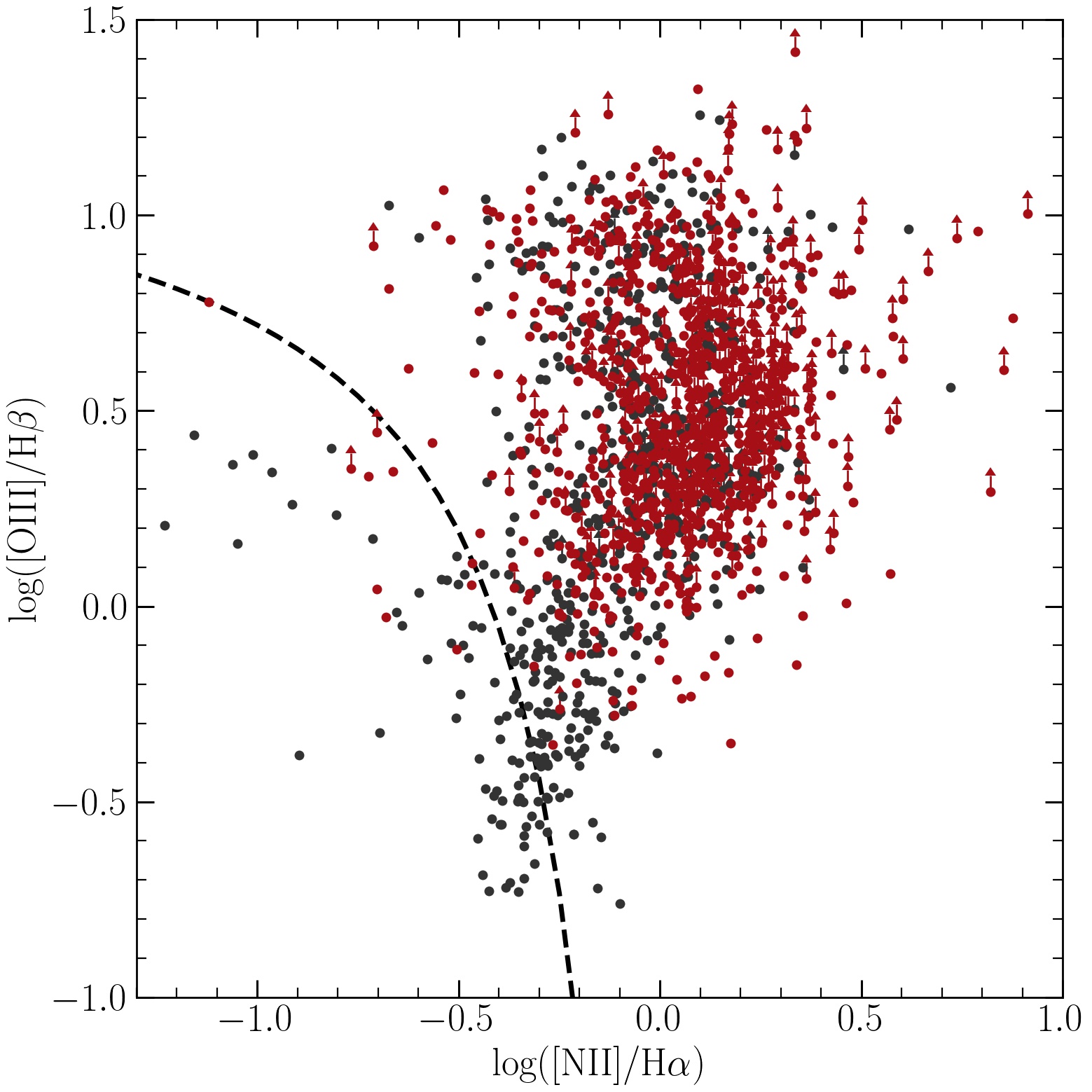}
      \caption{BPT}
      \label{bpt}
    \end{subfigure}
    \hspace{1.5cm}
    \begin{subfigure}{0.8\columnwidth}
      \includegraphics[width=\columnwidth]{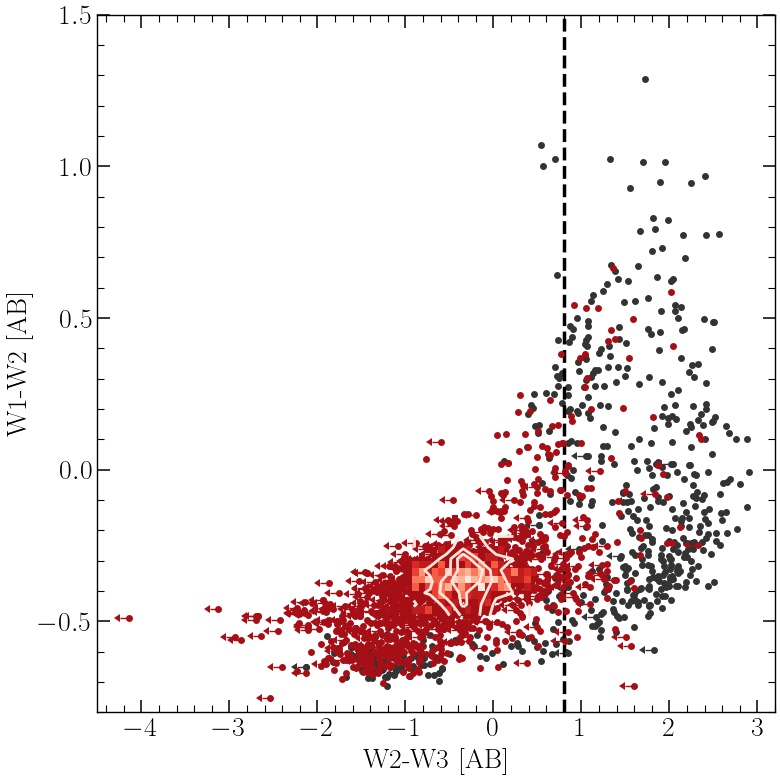}
      \caption{WISE colour-colour}
      \label{wisecolcol}
    \end{subfigure}
    \caption{Diagnostic plots for selecting radio AGN. Red points mark the sources classified as radio AGN after combining all four diagnostics, and black points show the star-forming/radio-quiet AGN (SF/RQAGN) sources. \textbf{(a)} $D_\mathrm{{n}}(4000)$ vs $L_\mathrm{1.4GHz}$/$M_{\star}$ plot. The dashed curve marks the division used for radio AGN and SF/RQAGN from \citet{Best2012}. \textbf{(b)} $L_{\textrm{H}\alpha}$ versus $L_\mathrm{1.4GHz}$ plot, with the separation line from \citet{Best2012}. \textbf{(c)} BPT diagram for the sources. The dashed curve shows the semi-empirical relation from \citet{Kauffmann2003}. \textbf{(d)} WISE colour-colour plot for sources with the division line from \citet{Sabater2019}.}
    \label{diagnostic}
  \end{figure*}

To study the relation between radio AGN life cycle and their impact on \OIII gas, we first need to select a clean sample of radio AGN. Radio emission from a galaxy can originate either from the AGN itself or star formation (SF). In the local universe, star-forming galaxies dominate a radio-selected sample up to $L_\mathrm{1.4GHz}\approx10^{23}$\,W\,Hz$^{-1}$ (e.g. \citealt{Sadler2002,Best2005}). This makes it important to separate them from radio AGN. Even AGN identified using optical emission line ratios can have radio emission dominated by star formation in the host galaxy. In this section, we discuss the approach used for selecting a clean sample of radio AGN from the 6\,504 radio selected sources in Sect.~\ref{sample contruction}. We followed the method outlined in \citet{Best2005,Best2012} and \citet{Sabater2019}, which combines four diagnostic methods. The plot for each diagnostic is shown in Fig.~\ref{diagnostic}.\par

First is the `$D_\mathrm{{n}}(4000)$ versus $L_\mathrm{1.4GHz}$/$M_{\star}$' method, shown in Fig.~\ref{d4pm} and developed by \cite{Best2005}. $D_\mathrm{{n}}(4000)$ is the 4000\,\AA~break in the spectrum of the galaxy (sensitive to the presence of a young stellar population), and $L_\mathrm{1.4GHz}$/$M_{\star}$ is the ratio of the radio luminosity to the stellar mass. This method relies on the idea that star-forming galaxies with various star formation histories, would lie in a similar region on a plot of these two quantities. This is because both of these quantities are dependent on the specific star formation rate. However, radio AGN have jets that add to the radio emission ($L_\mathrm{1.4GHz}$), allowing us to separate the two populations. \citet{Best2005} obtained the track of a galaxy with an exponentially decreasing star formation rate on this diagram, which can be used to separate the two sources of radio emission. They then added 0.225 to the $D_\mathrm{{n}}(4000)$ values of this track to construct a conservative division. This division was later modified by \citet{Kauffmann2008}, who used a straight line division at $L_\mathrm{1.4GHz}$/$M_{\star}<12.2$. For our sample, we estimated $D_\mathrm{{n}}(4000)$ as the flux ratio between 4000-4100\,\AA~and 3850-3950\,\AA. We used the stellar masses from the `Portsmouth SED-fit' catalogue by \citet{Maraston2013}, which were available for 5\,535 sources. The stellar masses were estimated from broad-band spectral energy distribution (SED) fitting of stellar population models using $u$, $g$, $r$, $i$ and $z$ band photometry and spectroscopic redshifts from SDSS. We classified the sources as radio AGN if they lie above the criterion from \citet{Kauffmann2008} \footnote{Provided by Philip Best (private communication)}, and as star-forming (i.e. galaxies where the radio emission is dominated by star formation) if they lie below it.\par

The second diagnostic uses $L_{\textrm{H}\alpha}$ versus $L_\mathrm{1.4GHz}$ to classify sources, shown in Fig.~\ref{harad}. It is based on the idea that for star-forming galaxies, both \halpha~and radio luminosities are related to the star formation rate of massive stars. Therefore sources with excess radio emission (coming from jets) can be separated from sources where both \halpha~and radio luminosities come from star formation. We used the division from \citet{Best2012}, which is a conservative division designed to produce clean radio AGN samples, and given as log($L_{\textrm{H}\alpha}$/L$_{\odot}) = 1.12 \times$ (log($L_\mathrm{1.4GHz}$/\whz)$ - 17.5)$. We classify sources with larger radio luminosity than predicted from this relation as radio AGN and the rest as star-forming. We use upper limits on \halpha~luminosities for non-detections. Overall, we were able to classify 4\,027 sources using this diagnostic.\par

The third diagnostic is the BPT diagram \citep{Baldwin1981} shown in Fig.~\ref{bpt}, which uses the \Nii/\halpha~and \oiiihbeta~emission line ratios to separate AGN and star-forming sources of ionisation. We used the relation from \citet{Kauffmann2003} to classify sources as AGN if they lie above the relation, and star forming if they lie below it. Although we do not have emission line detections for the majority of the sample, it can still provide useful information for sources with upper limits. We classified 1\,503 sources using the BPT diagram.\par

The fourth diagnostic is the WISE colour-colour diagram, shown in Fig.~\ref{wisecolcol}. Star-forming galaxies can be separated from radio AGN host galaxies (typically ellipticals with low levels of star formation) in their $W2-W3$ colour \citep{Yan2013}. \citet{Herpich2016} used a division of $W2-W3$ (AB) = 0.7 to separate galaxies with and without star formation, however, the division is not very clean. \citet{Sabater2019} modified it to $W2-W3$ (AB) = 0.8 for LoTSS sources following a comparison with the radio AGN in the Herschel-ATLAS North Galactic Plane field selected by \citet{Gurkan2018}. For this work, we used the same division as \citet{Sabater2019}. Since this is a crude division, it has less weight in the overall classification and we use it only to resolve cases where other diagnostics give contradictory results. Since we did not perform $k$-corrections to the WISE magnitudes, we use them only for $z<0.3$ sources. We classified 2\,019 sources using this method. \par

In each diagnostic, a source can be labelled as an AGN (or radio AGN), SF/radio-quiet AGN or unclassified, leading to 81 different possible combinations. A summary of the source classifications by each diagnostic is given in Table~\ref{diagnostictable}. We use these combinations to give a final classification to each source, following the approach of \citet{Sabater2019} and \citet{Best2012}. We refer the reader to these papers for further details on the classification. Sources that could not be classified with any diagnostic, or were classified as SF using only one diagnostic, were assigned a final classification of radio AGN if their 1.4\,GHz luminosity was greater than $10^{25}$\whz. 
The different combinations, number of sources in each, and their final classification are summarised in Table~\ref{classification}. Overall, we classified 5\,720 sources as radio AGN, 726 as SF/radio-quiet AGN and 58 were unclassified.

\begin{table}[!ht]
\centering
\caption{Number of sources classified by each diagnostic}
         \label{diagnostictable}
\renewcommand{\arraystretch}{1.15}
\setlength{\tabcolsep}{4pt}
\begin{tabular}{cccc}
    \hlineB{3}
    \noalign{\vspace{0.05cm}}
    \hline
    \noalign{\smallskip}
     Diagnostic & AGN & SF  & Unclassified\\
     & \multicolumn{3}{c} {(No. of overall radio AGN)} \\

    \noalign{\smallskip}
    \hlineB{3}
    \noalign{\smallskip}
    \noalign{\smallskip}
    
  $D_\mathrm{{n}}(4000)$ vs $L_\mathrm{1.4GHz}$/$M_{\star}$ & 4767 & 766 &971  \\ 
   & (4767) & (254) & (699)  \\ 
    \noalign{\smallskip}
  $L_{\textrm{H}\alpha}$ vs $L_\mathrm{1.4GHz}$ & 3208 & 819 & 2477 \\
   & (3172) & (248) & (2300)  \\ 
       \noalign{\smallskip}
  BPT & 1430 &  73 & 5001 \\
   & (917) & (7) & (4796)  \\
       \noalign{\smallskip}
  WISE col-col & 1559 & 460 & 4485 \\
   & (1404) & (76) & (4240)  \\ 
    \noalign{\smallskip}
    \noalign{\smallskip}
   
    \hline
    \noalign{\vspace{0.05cm}}  
    \hlineB{3}
 \end{tabular}
    \flushleft
    \textbf{Note.} This table summarises the number of sources classified by each diagnostic as AGN or star-forming (or radio-quiet AGN) and the sources that could not be classified. The numbers in brackets are the number of sources with the overall classification of radio AGN after combining the four diagnostics. See Sect.\ref{selectragn}.    
\end{table}

\section{Supplementary tables and figures}

\begin{table*}[!ht]
\centering
\begin{threeparttable}

\caption[]{Source classification}
         \label{classification}
\renewcommand{\arraystretch}{1.1}         
\begin{tabular}{cccccc}
    \hlineB{3}
    \noalign{\vspace{0.05cm}}
    \hline
    \noalign{\smallskip}
$D_\mathrm{{n}}(4000)$ vs $L_\mathrm{1.4GHz}$/$M_{\star}$ & BPT  & $L_{\textrm{H}\alpha}$ vs $L_\mathrm{1.4GHz}$ & WISE & Number &  Final classification \\
    \noalign{\smallskip}
    \hlineB{3}
    \noalign{\smallskip}
    \noalign{\smallskip}

AGN & AGN & AGN & AGN & 222 & AGN\\ 
AGN & AGN & AGN & Uncl & 257 & AGN \\
AGN & AGN & SF & AGN & 160 & AGN\\
AGN & AGN & SF & Uncl & 47 & AGN \\
AGN & AGN & SF & SF & 30 & AGN \\   
AGN & Uncl & AGN & AGN & 785 & AGN \\ 
AGN & Uncl & Uncl & AGN & 177 & AGN \\ 
AGN & Uncl & Uncl & Uncl & 1468 & AGN \\
AGN & Uncl & AGN & Uncl & 1583 & AGN \\     
AGN & Uncl & AGN & SF & 13 & AGN \\  
SF & AGN & SF & AGN & 74 & SF \\
SF & AGN & SF & SF & 208 & SF \\
SF & AGN & SF & Uncl & 54 & SF\\
SF & SF & SF & SF & 43 & SF \\
SF & Uncl & AGN & AGN & 29 & AGN\\ 
SF & Uncl & AGN & Uncl & 61 & AGN \\ 
SF & Uncl & Uncl & AGN & 45 & SF \\  
SF & Uncl & Uncl & SF & 21 & SF \\  
SF & Uncl & Uncl & Uncl & 157 (117) & SF$^{\star}$ \\ 
Uncl & AGN & AGN & SF & 15 & AGN \\
Uncl & AGN & AGN & AGN & 14 & AGN \\
Uncl & AGN & AGN & Uncl & 105 & AGN \\
Uncl & AGN & SF & SF & 100 & SF \\
Uncl & AGN & SF & AGN & 32 & SF \\   
Uncl & AGN & SF & Uncl & 62 (6) & SF$^{\star}$\\  
Uncl & Uncl & AGN & Uncl & 51 (33) & SF$^{\star}$\\
Uncl & Uncl & Uncl & Uncl & 592 (534) & Uncl$^{\star}$ \\
    \noalign{\smallskip}
    \noalign{\smallskip}
    \hline
    \noalign{\vspace{0.05cm}}  
    \hlineB{3}
\end{tabular}
 {\textbf{Note.} Combinations for classification of a source using the diagnostic diagrams, and the final classification assigned. Only groups with more than 10 sources are shown here. In groups with final classification marked with a $\star$, sources with $L_\mathrm{1.4GHz}\geq10^{25}$\whz were classified as radio AGN, and their numbers are mentioned in brackets. }   
\end{threeparttable} 
\end{table*}

 \begin{figure*}[!ht]
    \centering
    \includegraphics[width=1.5\columnwidth]{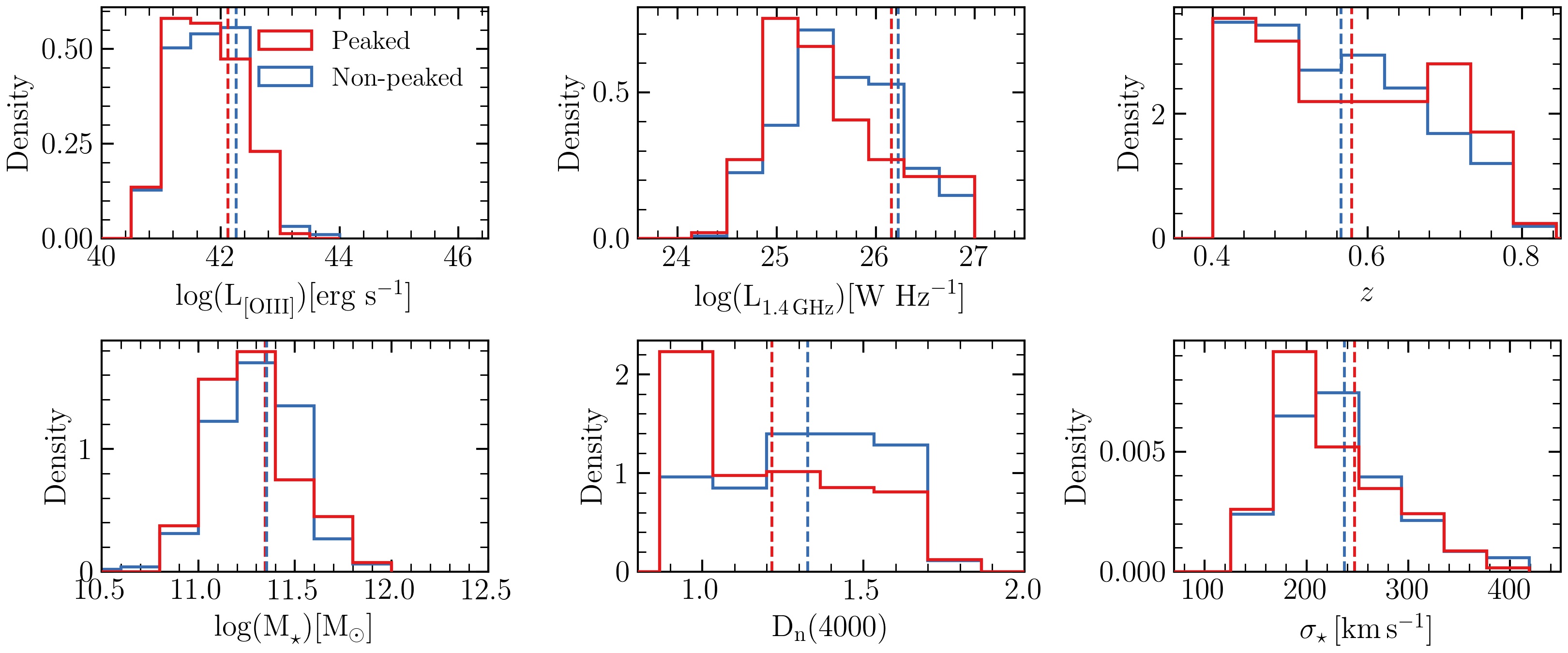}
    \caption{Distributions of host galaxy properties for the peaked and non-peaked radio AGN in the high redshift sample. Only the \OIII detections are shown. Vertical lines show the mean values of the distributions.}
    \label{sampleproperties_highz}
 \end{figure*}

\begin{table*}
\centering
\begin{threeparttable}
\caption{Model fit parameters for stacked \OIII profiles in the low redshift sample}
         \label{model fit}
\renewcommand{\arraystretch}{1}
\setlength{\tabcolsep}{4pt}
\begin{tabular}{cccccccccc}
    \hlineB{3}
    \noalign{\vspace{0.05cm}}
    \hline
    \noalign{\smallskip}
     Luminosity & Groups & $N$ & A$_\textrm{narrow}$  & FWHM$_\textrm{narrow}$ & $v_\textrm{narrow}$ & $A_\textrm{broad}$ &  FWHM$_\textrm{broad}$ &  $v_\textrm{broad}$ & FWHM$_\textrm{avg}$  \\
     & & & [$\times10^{-17}$] &  &  & [$\times10^{-17}$] &   &  &  \\
    
    \noalign{\smallskip}
    \hlineB{3}

    \noalign{\smallskip}    
    Controlling $L_\mathrm{1.4GHz}$ &&&&&&&&\\

    $10^{23}-10^{24}$\whz & PS & 77&  5.9$\pm$0.1 & 551$\pm$11 & 0$\pm$4 & - & - & - & 551$\pm$11 \\    
     & NPS &36 & 11.1$\pm$0.1 & 517$\pm$7 & 0$\pm$2 & - & - & - & 517$\pm$7 \\
    \noalign{\smallskip}
    $10^{24}-10^{25}$\whz & PS & 107 & 3.0$\pm$0.1 & 469$\pm$8 & 17$\pm$2 & 0.9$\pm$0.1 & 1338$\pm$51  & 0$\pm$12  & 667$\pm$34 \\    
     & NPS & 91&  5.3$\pm$0.1 & 441$\pm$4 & 0$\pm$1 & 0.7$\pm$0.1 & 1179$\pm$58  & 0$\pm$12  & 448$\pm$38 \\
    \noalign{\smallskip}
    $10^{25}-10^{26}$\whz & PS & 14 & 3.0$\pm$0.1 & 509$\pm$18 & 0$\pm$5 & 0.8$\pm$0.1 & 1639$\pm$132  & 0$\pm$36  & 805$\pm$88 \\    
     & NPS & 61 & 6.9$\pm$0.1 & 384$\pm$4 & 1$\pm$1 & 1.7$\pm$0.1 & 1091$\pm$36  & 0$\pm$8  & 503$\pm$25 \\
    \noalign{\smallskip}
    Controlling $L_{\textrm{\OIII}}$ &&&&&&&&\\

    $10^{39}-10^{40}$\ergpers & PS & 22 &  7.4$\pm$0.2 & 433$\pm$13 & 5$\pm$5 & - & - & - & 433$\pm$13 \\    
     & NPS & 9 &5.4$\pm$0.2 & 459$\pm$16 & 0$\pm$6 & - & - & - & 459$\pm$16 \\
    \noalign{\smallskip}
    $10^{40}-10^{41}$\ergpers & PS &143 & 2.2$\pm$0.1 & 488$\pm$16 & 0$\pm$4 & 0.5$\pm$0.01 & 1353$\pm$126  & 0$\pm$33  & 603$\pm$81 \\    
     & NPS &112 & 2.1$\pm$0.1 & 401$\pm$11 & 0$\pm$2 & 0.7$\pm$0.1 & 981$\pm$61  & 22$\pm$12  & 493$\pm$40 \\    
    \noalign{\smallskip}
    $10^{41}-10^{42}$\ergpers & PS & 47 &5.1$\pm$0.1 & 479$\pm$8 & 16$\pm$2 & 1.5$\pm$0.1 & 1434$\pm$46  & 0$\pm$13  & 718$\pm$29 \\
     & NPS &73 & 10.5$\pm$0.1 & 417$\pm$3 & 0$\pm$0 & 1.8$\pm$0.1 & 1202$\pm$35  & 0$\pm$9  & 486$\pm$23 \\
    \noalign{\smallskip}
    Spectral shape &&&&&&&&\\    

    $10^{24}-10^{26}$\whz & HFPS &54 &  2.3$\pm$0.1 & 470$\pm$11 & 17$\pm$3 & 0.7$\pm$0.1 & 1642$\pm$69  & 0$\pm$22  & 876$\pm$45 \\
     & LFPS & 67 & 3.6$\pm$0.1 & 483$\pm$9 & 11$\pm$2 & 0.9$\pm$0.1 & 1333$\pm$58  & 0$\pm$15  & 615$\pm$36 \\    
     & NPS &152 & 6.0$\pm$0.1 & 414$\pm$3 & 0$\pm$0 & 1.1$\pm$0.1 & 1145$\pm$34  & 0$\pm$8  & 473$\pm$23 \\  
    
    \noalign{\smallskip}
    \hline
    \noalign{\vspace{0.05cm}}  
    \hlineB{3}
 \end{tabular}
    {\textbf{Note.} Best-fit model parameters for the stacked \OIII profiles discussed in Sect.~\ref{stacking_specshape_lum} and~\ref{stacking_specshape_3bins}. The `Luminosity' column lists the luminosity range of the stacked bins, and the spectral shape is listed under `Groups'. Column '$N$' lists the number of sources in each group. The `A$_{\textrm{narrow}}$' and `A$_{\textrm{broad}}$' columns list the amplitudes in erg s$^{-1}$ cm$^{-2}$ $\AA^{-1}$, `FWHM$_{\textrm{narrow}}$' and `FWHM$_{\textrm{broad}}$' columns list the full width at half maximums in \kms and, `$v_{\textrm{narrow}}$' and `$v_{\textrm{broad}}$' list the velocity offsets in \kms of the Gaussian components in the best-fit model. `$FWHM_{\textrm{avg}}$' lists the flux-weighted average FWHM of the stacked profile in \kms.}
\end{threeparttable}
\end{table*}

 \begin{figure*}[!ht]
    \centering
    \begin{subfigure}{0.45\columnwidth}
      \includegraphics[width=\columnwidth]{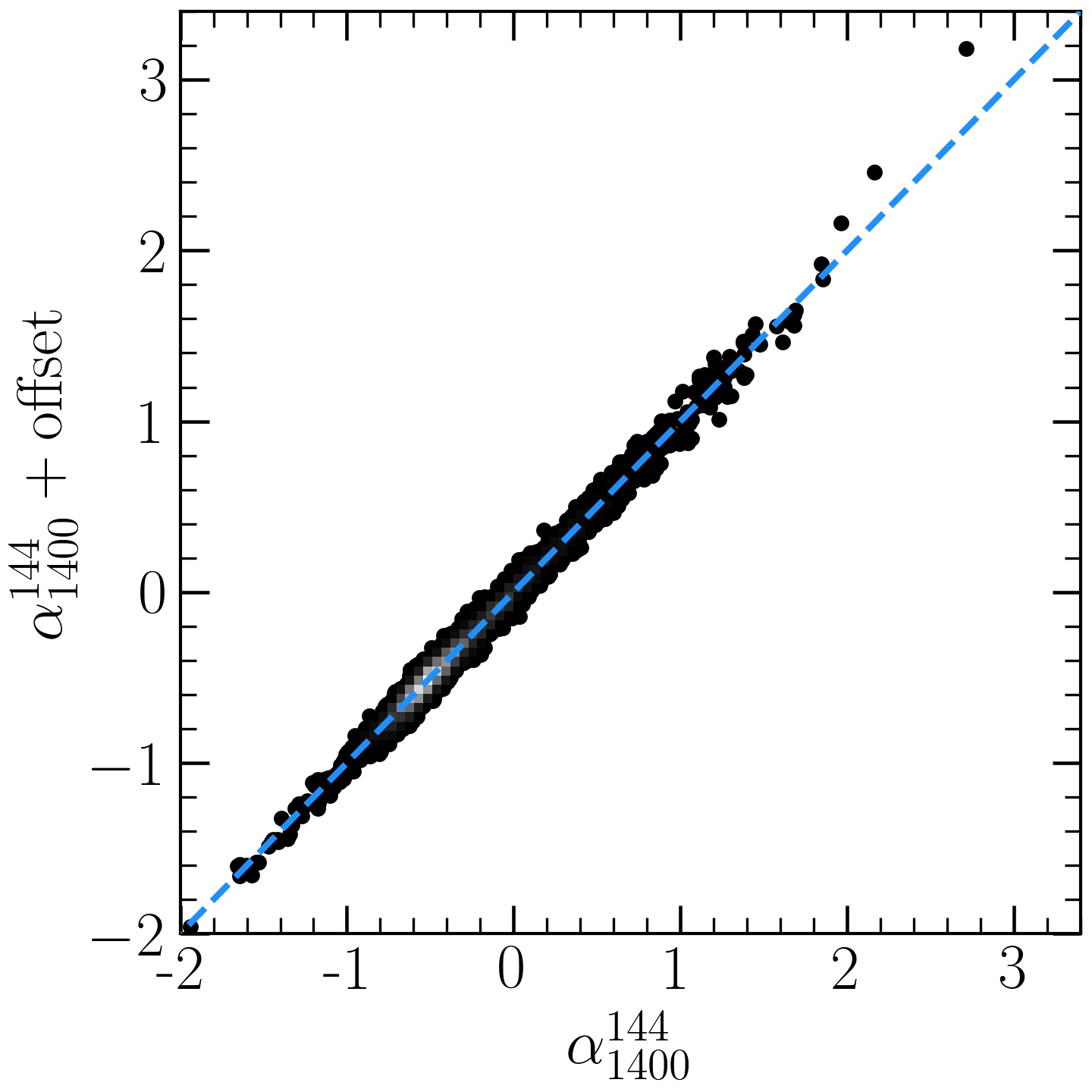}
      \caption{$\alpha^{144}_{1400}$ with offset}
      \label{lowspecindex_offset}
    \end{subfigure}
    \hfill
    \begin{subfigure}{0.45\columnwidth}
      \includegraphics[width=\columnwidth]{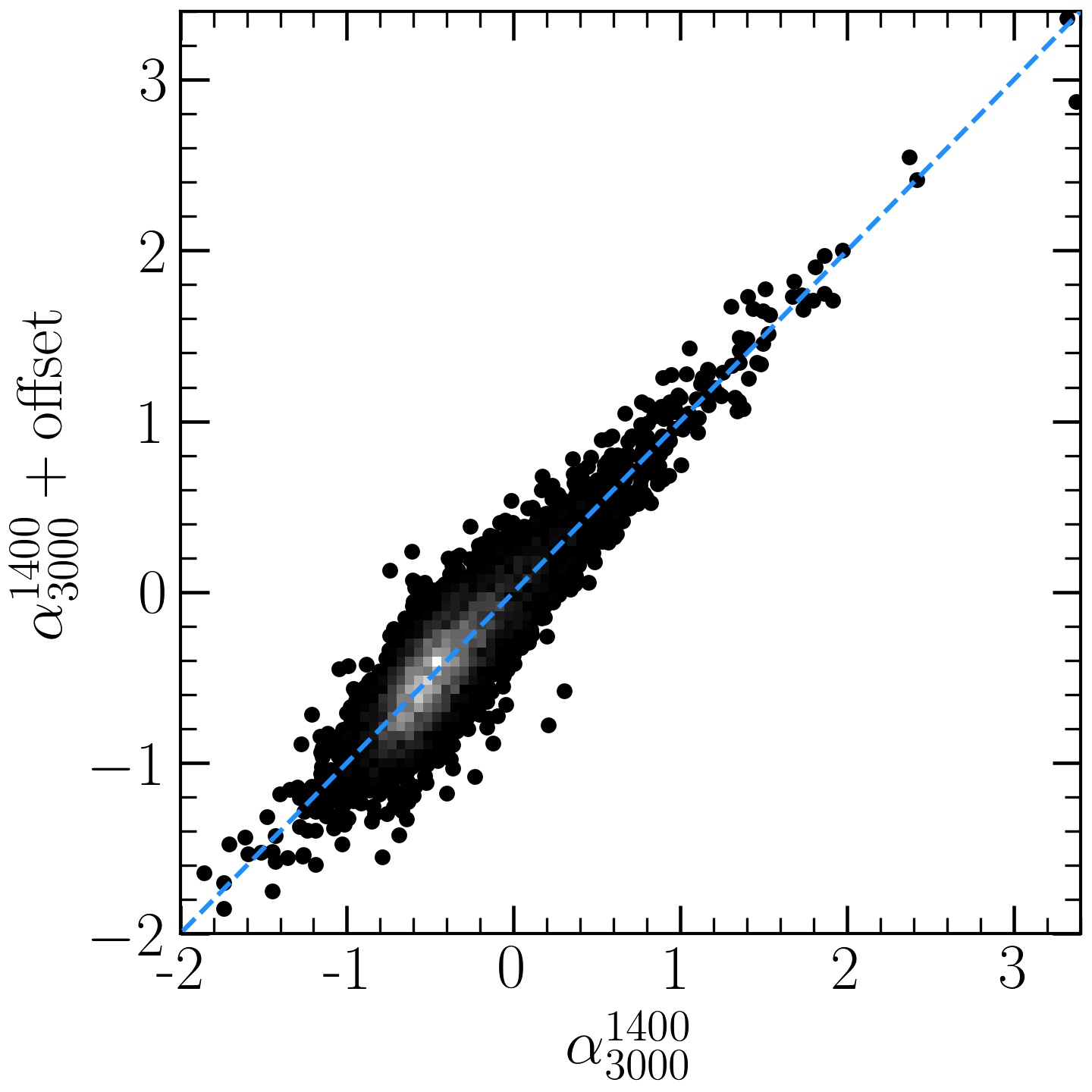}
      \caption{$\alpha^{1400}_{3000}$ with offset}
      \label{highspecindex_offset}
    \end{subfigure}
    \hfill
    \begin{subfigure}{0.53\columnwidth}
      \includegraphics[width=\columnwidth]{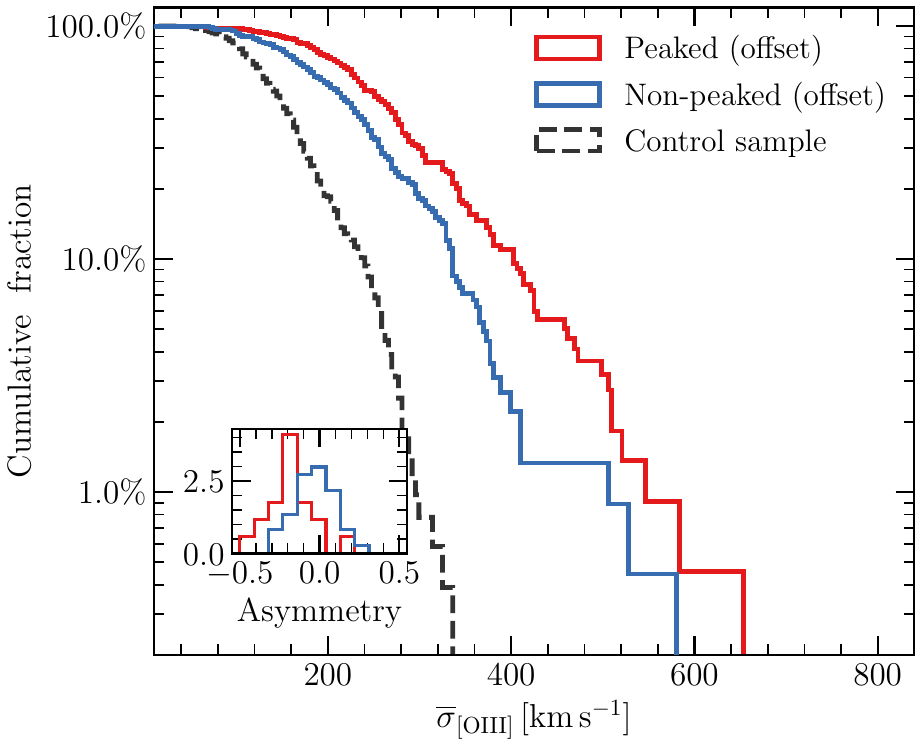}
      \caption{Cumulative $\overline{\sigma}_{\textrm{\OIII}}$ $0.02<z<0.4$}
      \label{cumdist_specshape_lowz_offset}
    \end{subfigure}
    \begin{subfigure}{0.53\columnwidth}
      \includegraphics[width=\columnwidth]{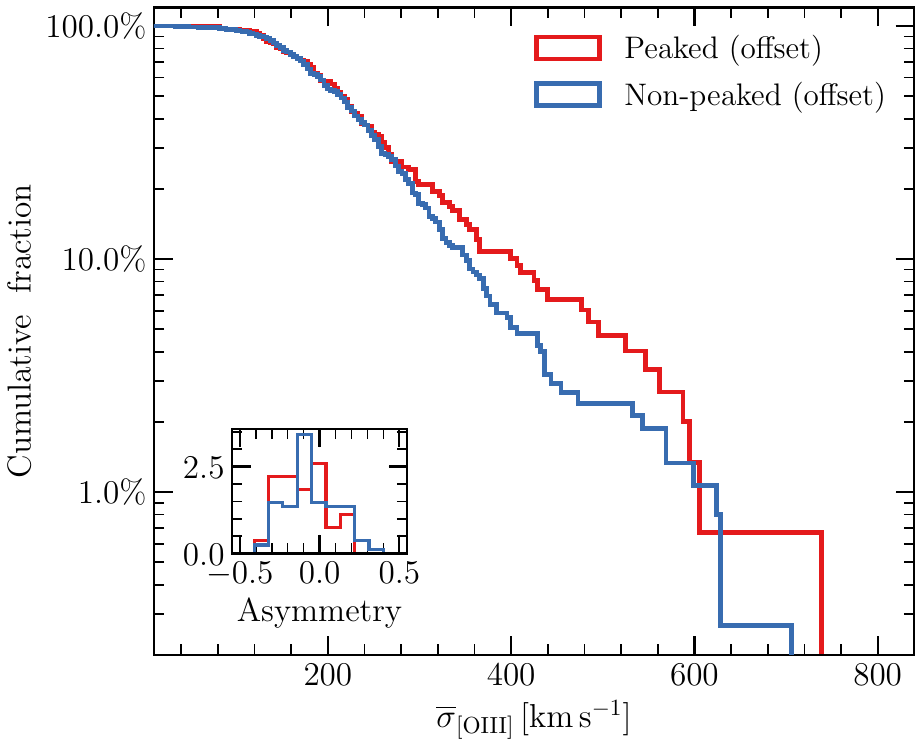}
      \caption{Cumulative $\overline{\sigma}_{\textrm{\OIII}}$ $0.4<z<0.8$}
      \label{cumdist_specshape_highz_offset}
    \end{subfigure}\\
    \vspace{0.6cm}
    \begin{subfigure}{0.54\columnwidth}
      \includegraphics[width=\columnwidth]{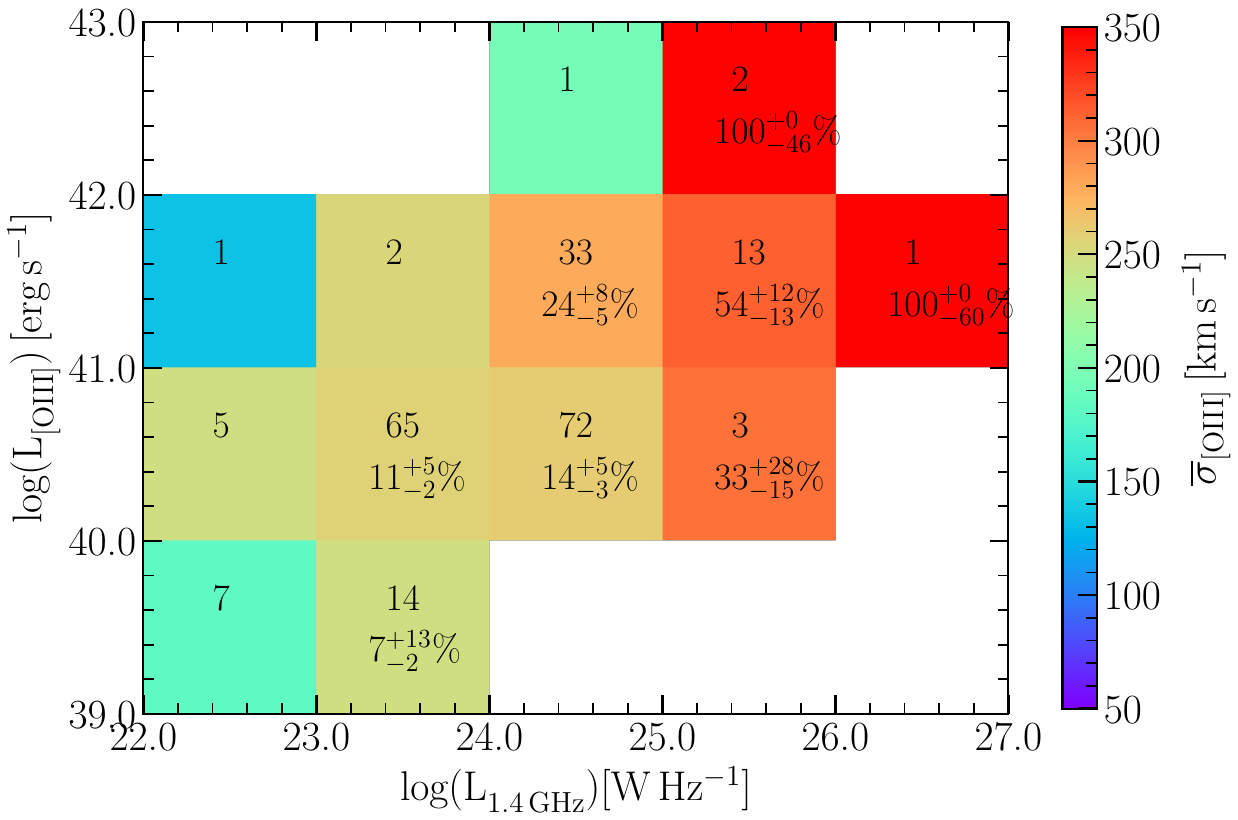}
      \caption{PS sources offset ($z<0.4$)}
      \label{radoptlum_lowz_peak_offset}
    \end{subfigure}
    \hfill
    \begin{subfigure}{0.54\columnwidth}
      \includegraphics[width=\columnwidth]{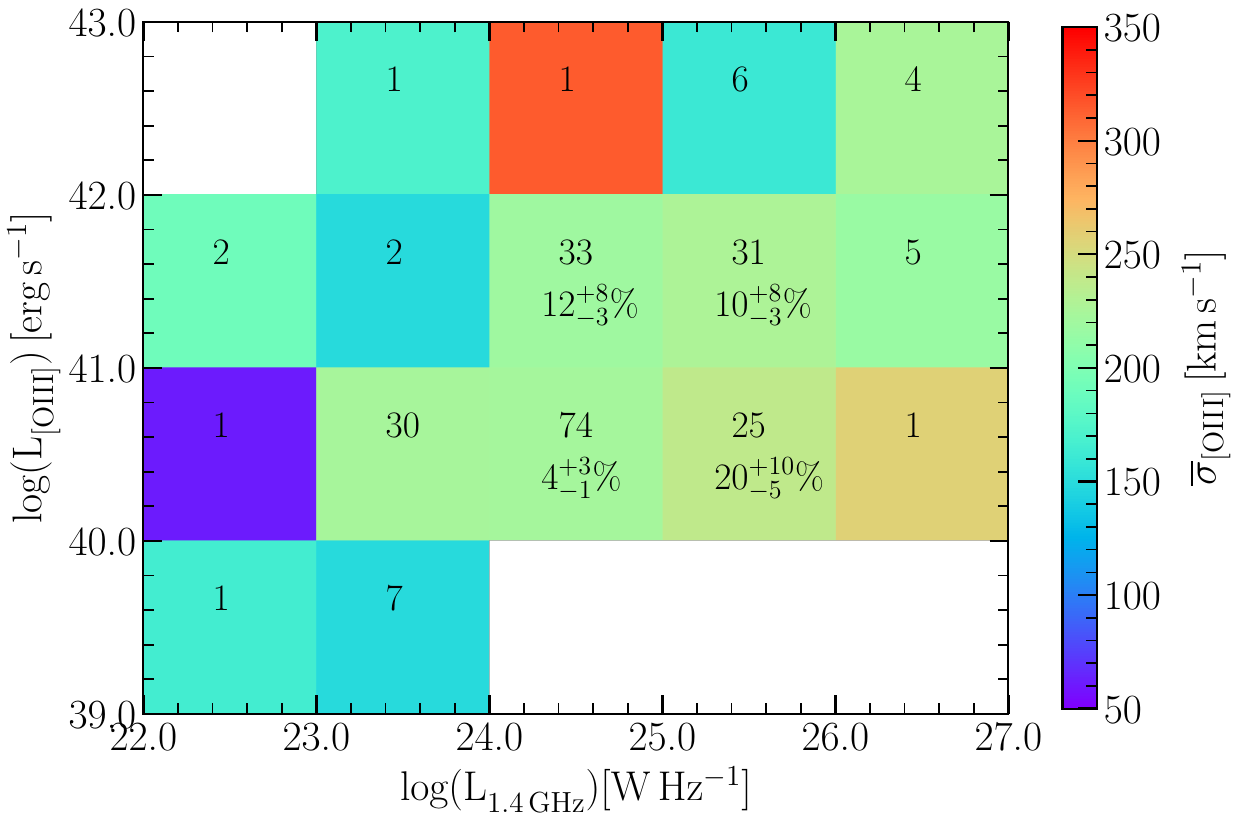}
      \caption{NPS sources offset ($z<0.4$)}
      \label{radoptlum_lowz_nonpeak_offset}
    \end{subfigure}
    \hfill
    \begin{subfigure}{0.46\columnwidth}
      \includegraphics[width=\columnwidth]{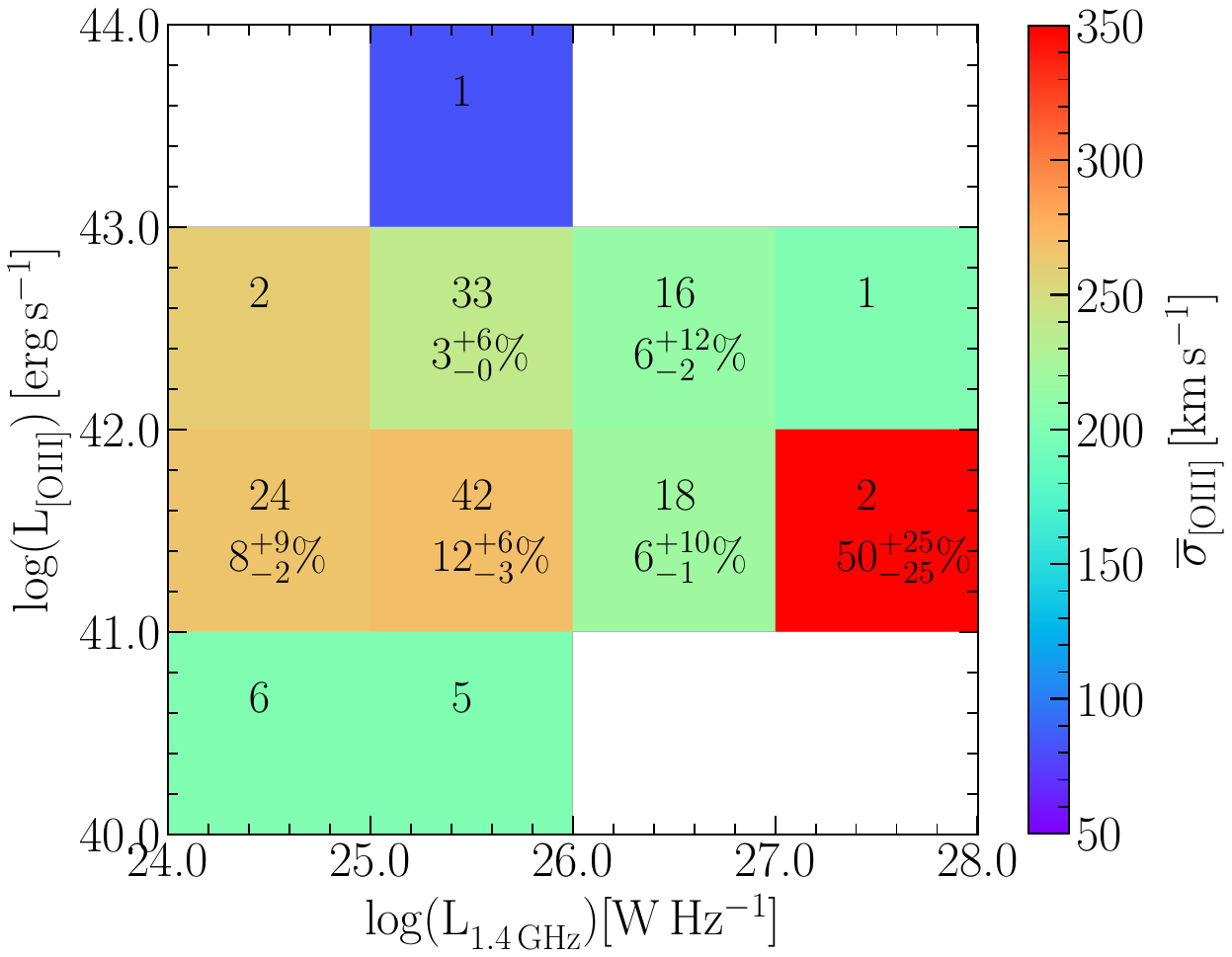}
      \caption{PS sources offset ($z>0.4$)}
      \label{radoptlum_highz_peak_offset}
    \end{subfigure}
    \hfill
    \begin{subfigure}{0.46\columnwidth}
      \includegraphics[width=\columnwidth]{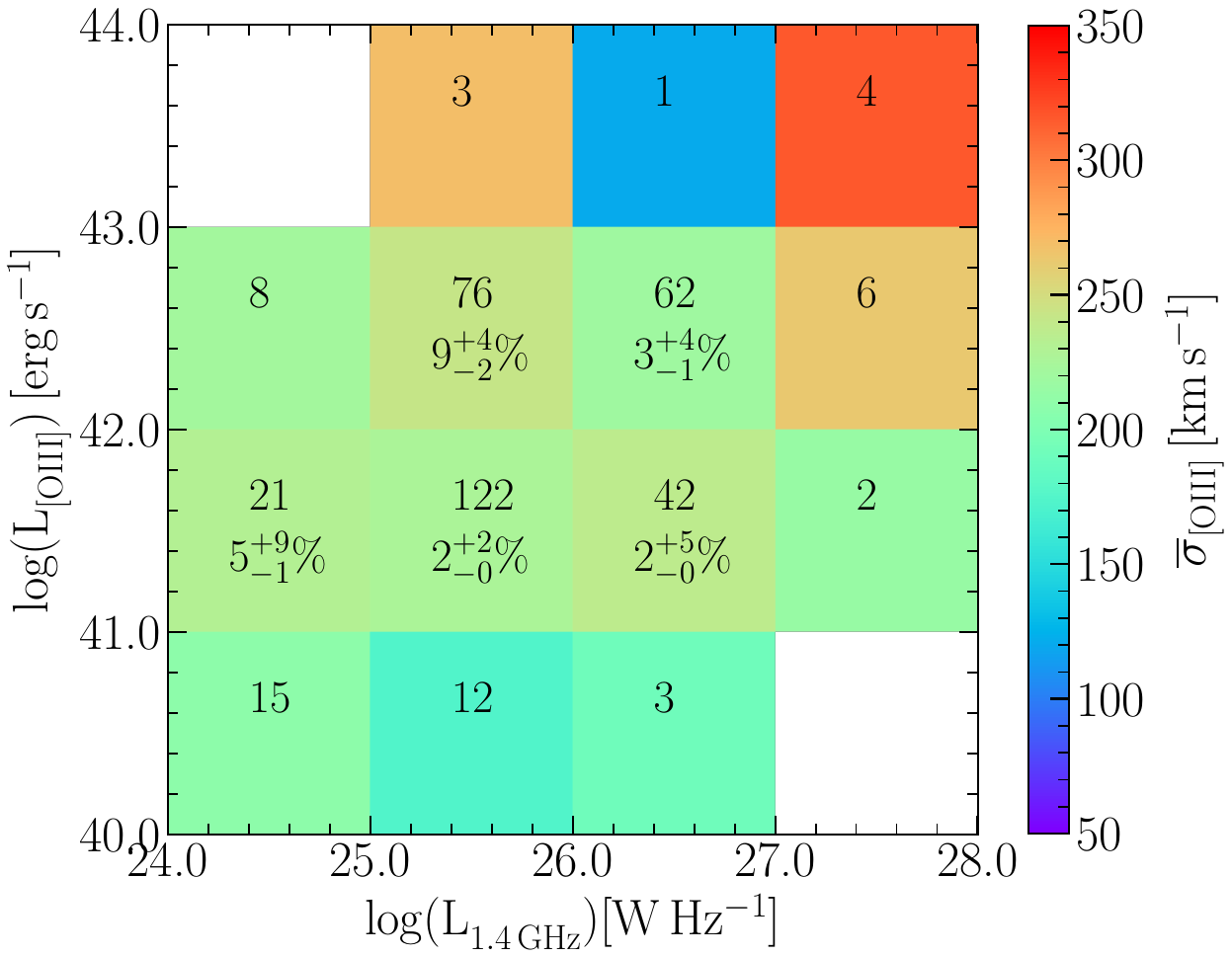}
      \caption{NPS sources offset ($z>0.4$)}
      \label{radoptlum_highz_nonpeak_offset}
    \end{subfigure}
    \caption{ Results for the radio AGN sample after adding offsets to spectral indices. \textbf{(a)$-$(b)} Comparison of spectral indices before and after adding the offsets described in Sect.~\ref{spectral properties of radio agn}. \textbf{(c)$-$(d)} Cumulative distributions of $\overline{\sigma}_{\textrm{\OIII}}$ after reclassifying sources with the spectral index offset. \textbf{(e)$-$(h)} Colour maps showing the average $\overline{\sigma}_{\textrm{\OIII}}$ values and fraction of disturbed sources in bins of $L_\mathrm{1.4GHz}$ and $L_{\textrm{\OIII}}$, again after reclassifying sources with the spectral index offset.}
    \label{specindex_offset_results}
 \end{figure*}

\end{appendix}

\end{document}